\documentclass[twocolumn,notitlepage,prmaterials,superscriptaddress,longbibliography,floatfix]{revtex4-2}

\usepackage{amsfonts}
\usepackage{textcomp}
\usepackage{times}
\usepackage{graphicx}
\usepackage{float}
\usepackage{latexsym,amsmath,amssymb,bm,euscript}
\usepackage{color}
\usepackage{subfigure}
\usepackage{epstopdf}
\usepackage[colorlinks,
linkcolor=blue,
anchorcolor=blue,
citecolor=blue,
urlcolor=blue,]{hyperref}
\usepackage{soul}
\usepackage[normalem]{ulem}
\usepackage{mathrsfs}
\usepackage{amsmath}
\usepackage{xspace}
\usepackage{natbib}
\usepackage{ulem}
\usepackage{etoolbox}
\usepackage{tikz}
\usepackage{array}[=2016-10-06]
\usepackage{longtable}
\usepackage{xcolor}
\usepackage{dcolumn}
\usepackage{multirow}
\usepackage{booktabs}

\graphicspath{{./}}

\newcommand{\ef}{E_{\mathrm{F}}}

\begin{document}
	
	\title{Binary kagome superconducting candidates \textcolor{black}{hosting} topological electronic states}
	
	\author{Xin-Wei Yi}
	\affiliation{Institute of Theoretical Physics, Chinese Academy of Sciences, Beijing 100190, China}
	
	\author{Jing-Yang You}
	\email{phyjyy@buaa.edu.cn}
	\affiliation{Peng Huanwu Collaborative Center for Research and Education, Beihang University, Beijing 100191, China}
	
	\author{Gang Su}
	\email{gsu@ucas.ac.cn}
	\affiliation{Institute of Theoretical Physics, Chinese Academy of Sciences, Beijing 100190, China} \affiliation{School of Physical Sciences, University of Chinese Academy of Sciences, Beijing 100049, China}\affiliation{Kavli Institute for Theoretical Sciences, University of Chinese Academy of Sciences, Beijing 100190, China}
	
	\begin{abstract}
		Topological superconductivity has attracted broad interest because of its connection to Majorana fermions and quantum computation. Kagome materials provide a natural setting in which interesting electronic structures, topological surface states, and superconductivity may occur in nearby energy windows. Here we carry out a systematic first-principles search for binary kagome superconducting candidates with topological electronic states across several structural families $A_mB_n$ ($m$:$n$=3:1; 3:2; 1:1). The screening combines formation-energy and phonon-stability filtering, magnetic-ground-state searches, electron-phonon-coupling calculations, and topological electronic-structure analysis. We identify 286 nonmagnetic dynamically stable candidates for electron-phonon-coupling analysis, among which 84 have estimated transition temperatures above 3 K and seven exceed the \(\sim 9\)~K reference scale of existing ambient-pressure kagome superconductors. In addition, several systems are predicted to host abundant topological surface states near the Fermi level. NbBi is identified as a \(\mathbb{Z}_2\) topological metal with clear Dirac-cone topological surface states and a kagome-derived flat band near \(\ef\). Ti$_3$Si, with the largest estimated \(T_\text{c}=15.2\)~K in this dataset, shows nodal-line-derived drumhead surface states and selected nodes that survive spin-orbit coupling near \(\ef\). \textcolor{black}{Combining superconductivity with nontrivial topological electronic structures, the binary kagome compounds predicted here represent compact and chemically tunable candidate platforms for topological superconductivity and for exploring kagome-related quantum phenomena.}
	\end{abstract}
	
	\maketitle
	
	\section{Introduction}
	
	Topological superconductivity (TSC) provides a paradigm in which superconducting phase coherence is combined with topological boundary physics. In this setting, Majorana zero modes can emerge as protected boundary or vortex excitations with non-Abelian statistics and potential applications in fault-tolerant quantum computation~\cite{RN529,RN599,RN603,RN516}. Current searches follow two complementary strategies: intrinsic superconductors with topologically nontrivial pairing, most prominently odd-parity superconductivity, and engineered proximity structures in which a conventional superconductor gaps spin-nondegenerate topological states~\cite{RN370,RN538,RN539,RN440}. Candidate odd-parity systems such as $\mathrm{Cu}_x\mathrm{Bi}_2\mathrm{Se}_3$ and $\mathrm{Sn}_{1-x}\mathrm{In}_x\mathrm{Te}$ have served as important milestones, but unambiguously identifying their pairing symmetry and isolating Majorana signatures remain difficult experimental tasks~\cite{RN532, RN441, RN535, RN584,RN515}. The proximity route has been pursued in topological-insulator/superconductor interfaces, semiconductor nanowires, and magnetic-atom chains, yet its interpretation and reproducibility can be limited by interface quality, disorder, and parameter tuning~\cite{RN572,RN606,RN574,RN575,RN577,RN578}. These limitations motivate the search for materials in which bulk superconductivity and topological surface states (TSS) occur in the same compound, as explored for Fe-based superconductors such as FeTe$_{1-x}$Se$_x$~\cite{RN533, RN592, RN594, RN536}, CaKFe$_4$As$_4$~\cite{RN548}, and LiFeAs~\cite{RN549,RN604}, as well as in high-throughput searches for topological and nodal superconductors~\cite{RN478, RN1006,RN514}.
	
	Kagome materials offer a structurally simple but electronically rich platform for this pursuit. Their corner-sharing triangular network naturally hosts geometric frustration, destructive-interference flat bands, Dirac nodes, and van Hove singularities (vHs)~\cite{RN1235}. These electronic motifs can amplify correlation effects and make kagome compounds fertile platforms for intertwined quantum phases, including electronic orders, magnetism, anomalous transport, and unconventional or multiband superconducting responses~\cite{RN1043,RN976}. The coexistence of superconductivity and intrinsic TSS in kagome materials can provide a route to topological superconductivity, provided that the superconducting order effectively gaps the spin-momentum-locked surface states. Kagome-lattice qubit proposals also show that this geometry is relevant to topological quantum computation beyond Majorana-based schemes~\cite{RN253,RN1374,RN1220}. In the \(A\mathrm{V}_3\mathrm{Sb}_5\) family (\(A\) = K, Rb, Cs), superconductivity with $T_\text{c}$ of 0.9--2.5~K coexists with \(\mathbb{Z}_2\) TSS and multiple charge-ordered or symmetry-broken states, making these compounds benchmark systems for intertwined topology, correlations, and pairing~\cite{RN1660,RN82,RN1043,RN5,RN461,RN8}. Vortex-core zero-bias conductance peaks reported on Cs-terminated CsV$_3$Sb$_5$ surfaces were discussed as possible Majorana signatures, motivating further tests of topological superconductivity in kagome metals~\cite{RN1}. Meanwhile, the Ti-based analogue CsTi$_3$Bi$_5$ was predicted by high-throughput searches and subsequently synthesized~\cite{RN985,RN967,RN795}. Later single-crystal, ARPES, and first-principles studies reported a superconducting onset near 4.8~K without long-range charge-density-wave order, together with flat bands, Dirac nodal lines, \(\mathbb{Z}_2\) TSS, and tunable superconducting and topological responses~\cite{RN967,RN966,RN1284}. These developments motivate a broader materials search for kagome compounds in which superconductivity and topological electronic structures coexist in an intrinsic setting.
	
	Here we report a high-throughput first-principles screening of binary kagome compounds across \(A_3B\), \(AB\), and \(A_3B_2\)-type structural prototypes. The search identifies 286 nonmagnetic dynamically stable systems for electron-phonon-coupling (EPC) analysis, including seven compounds with estimated \(T_\text{c}\) values above the \(\sim 9\)~K scale of the ambient-pressure kagome superconductors used as experimental references. Topological analysis further identifies distinct surface-state mechanisms: NbBi is a \(\mathbb{Z}_2\) topological metal with surface states near \(\ef\), whereas Ti$_3$Si combines the largest estimated \(T_\text{c}\) in the present dataset with nodal-line-derived surface spectral weight. These results indicate that binary kagome compounds provide a chemically compact platform for superconducting candidates with distinct topological electronic-state mechanisms.

	\section{Results}
	
	\subsection{Searching for binary kagome compounds}

	\begin{figure}[!t]
		\centering
		\includegraphics[width=\columnwidth]{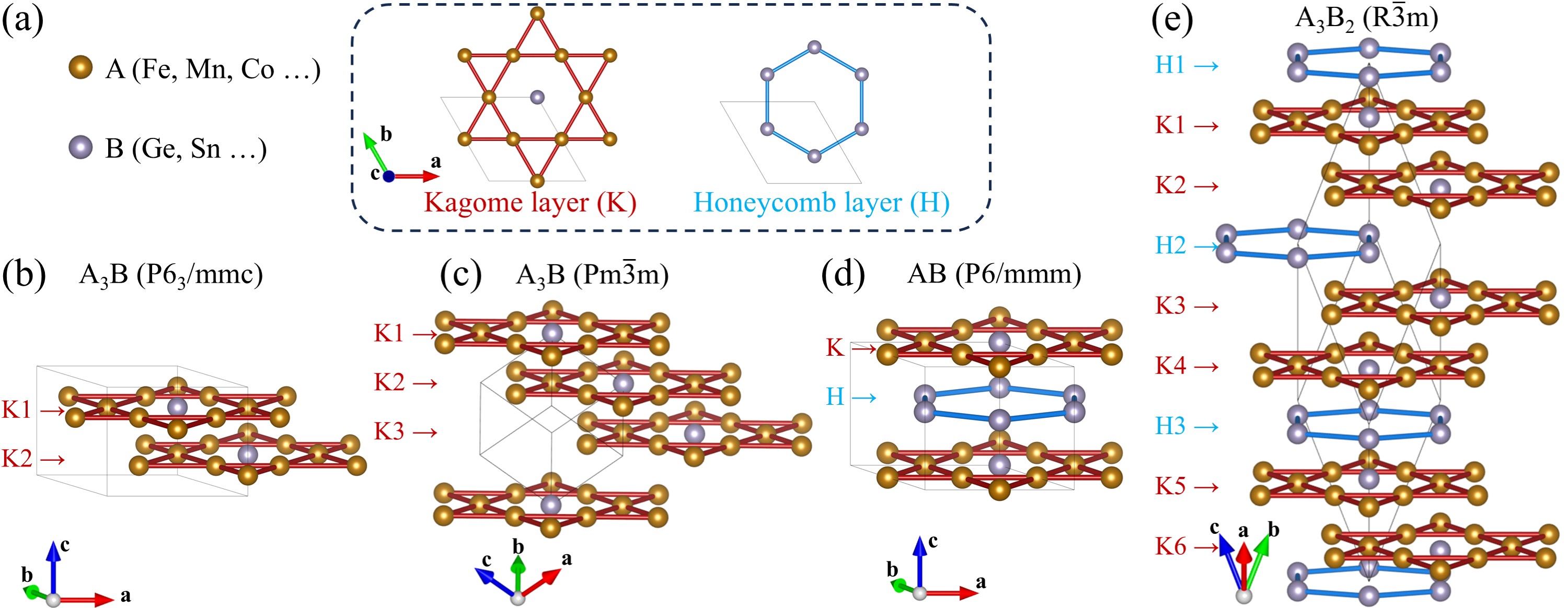}
		\caption{\textbf{Binary kagome structural families.} (a) Two common layer motifs used to construct the binary kagome prototypes. The A atoms (gold) form kagome layers (K), while the B atoms (grey) form honeycomb layers (H). (b--e) Crystal structures of four prototype families: (b) A$_3$B ($P6_3/mmc$), (c) A$_3$B ($Pm\bar{3}m$), (d) AB ($P6/mmm$), and (e) A$_3$B$_2$ ($R\bar{3}m$). These structures can be generated by stacking kagome and honeycomb layers in different sequences and registries. The drawings highlight the in-plane shifts between adjacent layers.}
		\label{fig:structural-families}
	\end{figure}

\begin{table}[!t]
	\caption{\textbf{Representative kagome superconducting candidates.} Columns list the estimated $T_\text{c}$ (calculated with a Coulomb pseudopotential $\mu^*=0.10$), logarithmic phonon frequency \(\omega_{\log}\), EPC constant \(\lambda\), density of states at \(\ef\) [$N(E_F)$], energy above the convex hull \(E_{\mathrm{hull}}\), and space group. The McMillan formula was used for \(\lambda<1.1\), and the Allen--Dynes correction was used for stronger-coupling entries (see Sec.~S1 of the Supplemental Material (SM)~\cite{supp}).}
	\label{tab:main-sc-candidates}
	\centering
	\begingroup
	\setlength{\tabcolsep}{1.5pt}
\renewcommand{\arraystretch}{0.85}
				\begin{tabular}{l>{\centering\arraybackslash}p{0.48cm}>{\centering\arraybackslash}p{0.95cm}>{\centering\arraybackslash}p{0.55cm}>{\centering\arraybackslash}p{1.85cm}>{\centering\arraybackslash}p{1.55cm}>{\centering\arraybackslash}p{1.3cm}}				
		\toprule
		Compound & \textcolor{black}{$T_\text{c}$ (K)} & $\omega_{\log}$ (K) & $\lambda$ & \textcolor{black}{\scriptsize $N(E_F)$ (eV$^{-1}$ atom$^{-1}$spin$^{-1}$)} & $E_{\mathrm{hull}}$ (meV/atom) & Space group \\
		\midrule
		Ti$_{3}$Si & \textcolor{black}{15.2} & 162.7 & 1.16 & \textcolor{black}{8.68} & 66.7 & $Pm\bar{3}m$ \\
		Ti$_{3}$Ge & \textcolor{black}{14.0} & 177.5 & 1.09 & \textcolor{black}{8.60} & 98.0 & $Pm\bar{3}m$ \\
		Tc$_{3}$Ge$_{2}$ & \textcolor{black}{12.1} & 119.8 & 1.33 & \textcolor{black}{6.48} & 51.7 & $R\bar{3}m$ \\
		Cr$_{3}$As & \textcolor{black}{11.2} & 211.9 & 0.85 & \textcolor{black}{8.07} & 190.0 & $P6_3/mmc$ \\
		Ti$_{3}$Sn & \textcolor{black}{10.5} & 147.9 & 1.01 & \textcolor{black}{10.05} & 39.0 & $Pm\bar{3}m$ \\
		Zr$_{3}$Pb & \textcolor{black}{9.7} & 120.9 & 1.10 & \textcolor{black}{8.48} & 0.0 & $Pm\bar{3}m$ \\
		Cr$_{3}$P & \textcolor{black}{9.1} & 265.2 & 0.70 & \textcolor{black}{6.94} & 199.4 & $P6_3/mmc$ \\
		Cd$_{3}$Ge & \textcolor{black}{8.7} & 85.3 & 1.35 & \textcolor{black}{2.72} & 147.4 & $Pm\bar{3}m$ \\
		Tc$_{3}$Sn$_{2}$ & \textcolor{black}{8.5} & 132.4 & 0.95 & \textcolor{black}{6.06} & 76.9 & $R\bar{3}m$ \\
		V$_{3}$P & \textcolor{black}{8.0} & 229.2 & 0.70 & \textcolor{black}{6.50} & 172.6 & $Pm\bar{3}m$ \\
		Ti$_{3}$Tl & \textcolor{black}{8.0} & 162.2 & 0.82 & \textcolor{black}{9.85} & 18.6 & $P6_3/mmc$ \\
		Zr$_{3}$Tl$_{2}$ & \textcolor{black}{7.9} & 80.8 & 1.20 & \textcolor{black}{7.54} & 195.4 & $R\bar{3}m$ \\
		Cr$_{3}$Ge & \textcolor{black}{7.0} & 237.1 & 0.66 & \textcolor{black}{6.45} & 171.7 & $P6_3/mmc$ \\
		Zr$_{3}$Sn & \textcolor{black}{6.9} & 114.5 & 0.91 & \textcolor{black}{7.20} & 11.5 & $Pm\bar{3}m$ \\
		Zr$_{3}$Ga & \textcolor{black}{6.6} & 152.1 & 0.77 & \textcolor{black}{6.70} & 39.1 & $P6_3/mmc$ \\
		Mo$_{3}$Ge & \textcolor{black}{6.6} & 206.0 & 0.68 & \textcolor{black}{4.66} & 200.0 & $P6_3/mmc$ \\
		Ti$_{3}$Sb & \textcolor{black}{6.6} & 125.4 & 0.85 & \textcolor{black}{8.78} & 87.7 & $Pm\bar{3}m$ \\
		Zr$_{3}$Tl & \textcolor{black}{6.2} & 139.6 & 0.78 & \textcolor{black}{7.74} & 24.3 & $P6_3/mmc$ \\
		V$_{3}$As & \textcolor{black}{6.2} & 207.8 & 0.66 & \textcolor{black}{6.78} & 126.1 & $Pm\bar{3}m$ \\
		Zn$_{3}$Sn$_{2}$ & \textcolor{black}{6.1} & 82.4 & 1.04 & \textcolor{black}{2.72} & 89.2 & $R\bar{3}m$ \\
		Zr$_{3}$Ga & \textcolor{black}{5.9} & 165.9 & 0.71 & \textcolor{black}{6.80} & 22.7 & $Pm\bar{3}m$ \\
		Zn$_{3}$Al & \textcolor{black}{5.9} & 103.9 & 0.89 & \textcolor{black}{2.20} & 0.0 & $P6_3/mmc$ \\
		VAs & \textcolor{black}{5.8} & 233.8 & 0.62 & \textcolor{black}{5.40} & 142.5 & $P6/mmm$ \\
		Nb$_{3}$Bi & \textcolor{black}{5.7} & 134.1 & 0.77 & \textcolor{black}{6.05} & 145.2 & $Pm\bar{3}m$ \\
		Nb$_{3}$Bi$_{2}$ & \textcolor{black}{5.7} & 156.6 & 0.71 & \textcolor{black}{4.76} & 68.0 & $R\bar{3}m$ \\
		Mo$_{3}$Sn & \textcolor{black}{5.5} & 184.6 & 0.66 & \textcolor{black}{4.90} & 174.4 & $P6_3/mmc$ \\
		Ir$_{3}$Al & \textcolor{black}{5.5} & 140.9 & 0.73 & \textcolor{black}{6.48} & 120.5 & $Pm\bar{3}m$ \\
		Hf$_{3}$Bi & \textcolor{black}{5.3} & 103.8 & 0.84 & \textcolor{black}{8.43} & 73.0 & $Pm\bar{3}m$ \\
		Zr$_{3}$Bi$_{2}$ & \textcolor{black}{5.3} & 114.4 & 0.80 & \textcolor{black}{6.01} & 69.4 & $R\bar{3}m$ \\
		Cd$_{3}$Sn & \textcolor{black}{5.3} & 79.5 & 0.97 & \textcolor{black}{2.92} & 46.5 & $Pm\bar{3}m$ \\
		Ti$_{3}$In & \textcolor{black}{5.2} & 195.9 & 0.63 & \textcolor{black}{9.31} & 0.0 & $P6_3/mmc$ \\
		Zr$_{3}$Al & \textcolor{black}{5.1} & 170.3 & 0.66 & \textcolor{black}{6.71} & 10.0 & $P6_3/mmc$ \\
		W$_{3}$Al & \textcolor{black}{5.1} & 165.1 & 0.67 & \textcolor{black}{4.11} & 157.9 & $P6_3/mmc$ \\
		Zr$_{3}$In & \textcolor{black}{5.0} & 155.1 & 0.68 & \textcolor{black}{7.46} & 21.6 & $P6_3/mmc$ \\
		Ti$_{3}$Ga & \textcolor{black}{4.9} & 229.4 & 0.59 & \textcolor{black}{7.90} & 9.7 & $Pm\bar{3}m$ \\
		Ti$_{3}$Ga & \textcolor{black}{4.9} & 231.7 & 0.58 & \textcolor{black}{8.28} & 2.2 & $P6_3/mmc$ \\
		Zr$_{3}$Al & \textcolor{black}{4.8} & 179.2 & 0.64 & \textcolor{black}{6.88} & 6.7 & $Pm\bar{3}m$ \\
		Zn$_{3}$Ga & \textcolor{black}{4.5} & 115.7 & 0.74 & \textcolor{black}{2.10} & 27.9 & $Pm\bar{3}m$ \\
		Hf$_{3}$Pb & \textcolor{black}{4.5} & 110.7 & 0.75 & \textcolor{black}{6.79} & 0.0 & $P6_3/mmc$ \\
		Ti$_{3}$Al & \textcolor{black}{4.4} & 243.4 & 0.56 & \textcolor{black}{8.40} & 20.7 & $Pm\bar{3}m$ \\
		Zr$_{3}$Tl & \textcolor{black}{4.4} & 149.2 & 0.65 & \textcolor{black}{7.55} & 9.6 & $Pm\bar{3}m$ \\
		Cd$_{3}$Pb$_{2}$ & \textcolor{black}{4.2} & 74.6 & 0.88 & \textcolor{black}{2.91} & 35.6 & $R\bar{3}m$ \\
		Hf$_{3}$Sn & \textcolor{black}{4.1} & 118.9 & 0.70 & \textcolor{black}{6.35} & 0.0 & $P6_3/mmc$ \\
		Ti$_{3}$Pb$_{2}$ & \textcolor{black}{4.1} & 164.9 & 0.62 & \textcolor{black}{7.23} & 33.2 & $R\bar{3}m$ \\
		NbBi      & \textcolor{black}{4.0} & 157.8 & 0.62 & \textcolor{black}{4.88} & 169.7 & $P6/mmm$ \\
		Ti$_{3}$Bi$_{2}$ & \textcolor{black}{3.9} & 182.9 & 0.59 & \textcolor{black}{6.75} & 0.0 & $R\bar{3}m$ \\
		Zr$_{3}$In & \textcolor{black}{3.7} & 163.3 & 0.60 & \textcolor{black}{7.35} & 10.2 & $Pm\bar{3}m$ \\
		Cd$_{3}$Tl$_{2}$ & \textcolor{black}{3.6} & 63.4 & 0.88 & \textcolor{black}{3.15} & 30.7 & $R\bar{3}m$ \\
		Cd$_{3}$In & \textcolor{black}{3.4} & 82.5 & 0.76 & \textcolor{black}{2.60} & 42.1 & $Pm\bar{3}m$ \\
		Pt$_{3}$Pb & \textcolor{black}{3.3} & 96.8 & 0.70 & \textcolor{black}{4.62} & 34.7 & $Pm\bar{3}m$ \\
		Hf$_{3}$Pb & \textcolor{black}{3.3} & 112.7 & 0.65 & \textcolor{black}{7.12} & 11.2 & $Pm\bar{3}m$ \\
		Hf$_{3}$Sn & \textcolor{black}{3.3} & 113.9 & 0.65 & \textcolor{black}{7.00} & 12.1 & $Pm\bar{3}m$ \\
		Ti$_{3}$Al & \textcolor{black}{3.1} & 258.6 & 0.50 & \textcolor{black}{8.09} & 1.5 & $P6_3/mmc$ \\
		\bottomrule
	\end{tabular}
	\endgroup
	
\end{table}

	\begin{figure*}[!t]
		\centering
		\includegraphics[width=\textwidth]{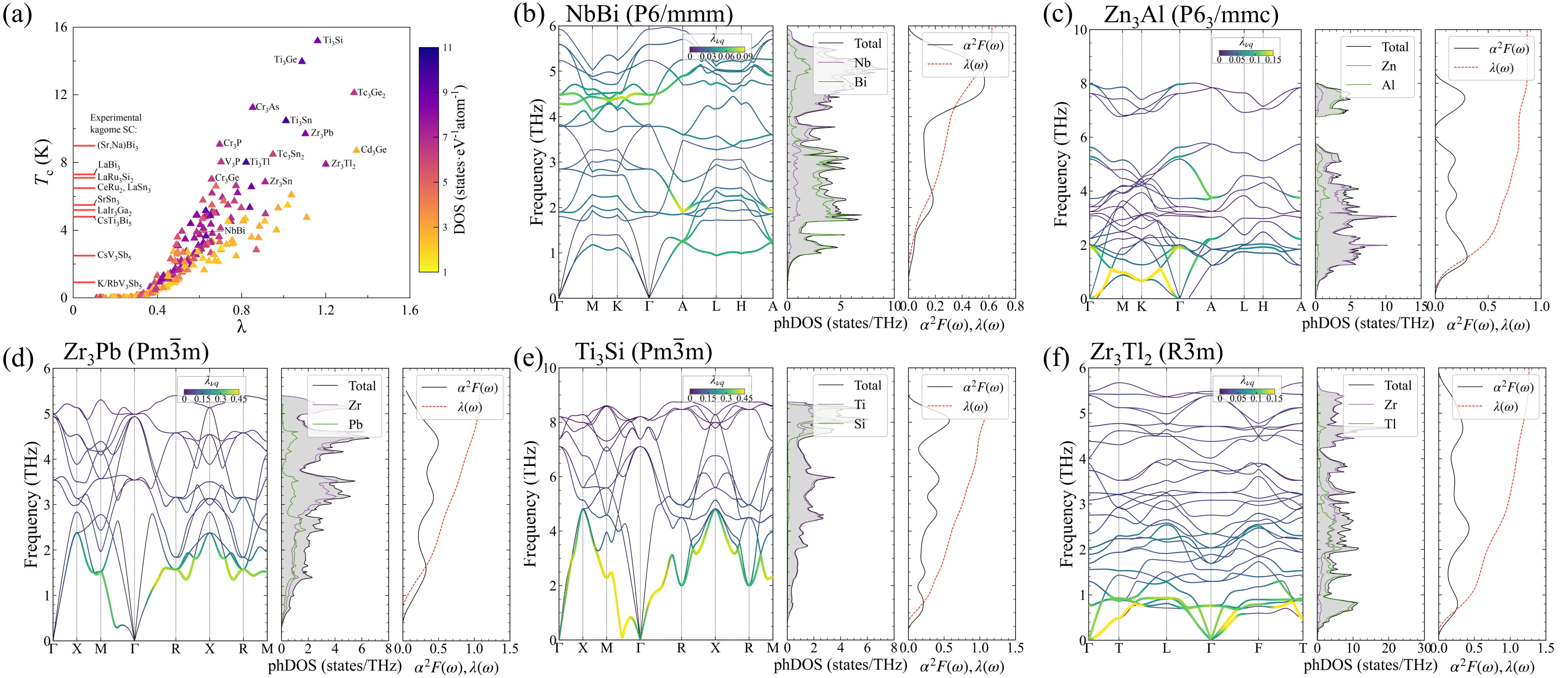}
			\caption{\textbf{Predicted kagome superconducting candidates.} (a) Estimated $T_c$ versus electron-phonon-coupling constant $\lambda$, with the $N(\ef)$ used as the color scale; representative ambient-pressure kagome materials are shown as references~\cite{RN967,RN1376,RN8,RN64,RN11,RN485,RN107,RN2165,RN3069, RN3070, RN3071}. (b--f) Phonon dispersions projected by the mode-resolved EPC $\lambda_{\nu\mathbf{q}}$, together with phonon density of states, Eliashberg spectral function $\alpha^2F(\omega)$, and accumulated $\lambda(\omega)$ for representative systems: (b) NbBi ($P6/mmm$), (c) Zn$_3$Al ($P6_3/mmc$), (d) Zr$_3$Pb ($Pm\bar{3}m$), (e) Ti$_3$Si ($Pm\bar{3}m$), and (f) Zr$_3$Tl$_2$ ($R\bar{3}m$). \(\lambda_{\nu\mathbf{q}}\) denotes the contribution to EPC \(\lambda\) of phonon branch \(\nu\) at wave vector \(\mathbf{q}\).}
		\label{fig:sc-landscape}
	\end{figure*}

	We consider binary intermetallic kagome compounds A$_m$B$_n$ with representative stoichiometries $m$:$n$=3:1, 1:1, and 3:2, constructed from different stacking sequences of kagome and honeycomb layers (Fig.~\ref{fig:structural-families}). These compounds are organized into four prototype families: A$_3$B ($P6_3/mmc$), A$_3$B ($Pm\bar{3}m$), AB ($P6/mmm$), and A$_3$B$_2$ ($R\bar{3}m$). Except for the AB prototype, adjacent layers are laterally shifted and change the interlayer coordination. \textcolor{black}{Although the cubic A$_3$B ($Pm\bar{3}m$) prototype is not conventionally layered, the face-centered A atoms form kagome nets within the (111) planes.} The kagome and honeycomb networks generate abundant flat bands, Dirac crossings, and vHs, while the distinct stacking patterns tune the three-dimensional connectivity, orbital content near $\ef$, Fermi-surface dimensionality, symmetry constraints, and phonon modes relevant to EPC. The A$_3$B ($Pm\bar{3}m$) prototype is also known as the Cu$_3$Au-type structure and includes several superconductors, such as LaSn$_3$ ($T_\text{c}=6.4$~K), LaBi$_3$ ($T_\text{c}=7.3$~K), and (Sr,Na)Bi$_3$ ($T_\text{c}\approx 9$~K)~\cite{RN3069,RN3070,RN3071}. Related compounds in these prototype families display diverse electronic and magnetic phenomena, including topological flat-band structures in CoSn and Fe$_3$Sn$_2$~\cite{RN246,RN253,RN93,RN285}, magnetic charge-density-wave order in FeGe~\cite{RN609}, and large anomalous Hall and Nernst responses in noncollinear antiferromagnets such as Mn$_3$Ge, Mn$_3$Sn, and Mn$_3$Ga~\cite{RN289,RN264,RN421,RN2219}.

	Starting from these prototype families, we constructed more than 2000 binary systems and screened them using a high-throughput first-principles workflow (Fig.~S1 of the SM~\cite{supp}). \textcolor{black}{Detailed computational procedures for the first-principles calculations, electron--phonon-coupling calculations, and topological analyses are provided in Sec.~S1 of the SM~\cite{supp}.} Formation-energy calculations and phonon spectra were used to exclude high-energy metastable entries with \(E_{\mathrm{hull}}>200\)~meV/atom and dynamically unstable entries with appreciable imaginary phonon modes. This procedure yielded more than 380 dynamically stable low-energy candidates within the adopted metastability window, including 31 entries on the convex hull with \(E_{\mathrm{hull}}=0\) that are especially promising for synthesis (Tables~S3--S7 in the SM~\cite{supp}). Most relaxed structures retained the symmetry of the parent prototype, whereas seven compounds, including Cu$_3$Bi and Ni$_3$B, lowered their symmetry after structural optimization. Literature and database checks further show that more than 20 screened compounds, including Ni$_3$Sn, Sc$_3$In, PtTl, and V$_3$Sb$_2$, have already been synthesized (see Tables~S3--S6 for details). We then performed magnetic-ground-state searches for the screened low-energy candidates, including ferromagnetic and representative antiferromagnetic configurations when applicable (Figs.~S2 and S3). Approximately 75\% of the materials were classified as nonmagnetic, while the remaining entries include 70 ferromagnets and 28 collinear or noncollinear antiferromagnets.
	
	\subsection{Kagome superconductors}
	
	We next focus on the nonmagnetic candidates, for which EPC provides a direct route to superconductivity. Within the isotropic EPC approximation, density functional perturbation theory (DFPT) calculations were performed for 286 nonmagnetic dynamically stable systems. The McMillan formula was used for \(\lambda<1.1\), and the Allen--Dynes correction was used for stronger-coupling entries (Sec.~S1 of the SM)~\cite{RN128, RN129,supp}.

	Fig.~\ref{fig:sc-landscape}(a) summarizes the superconducting landscape by plotting the estimated $T_\text{c}$ against \(\lambda\), with \(N(\ef)\) used as the color scale. The data show the expected broad tendency for $T_\text{c}$ to increase with both EPC strength and the Fermi-level density of states, but the spread at fixed \(\lambda\) indicates that phonon-energy scales and mode selectivity are also important. Representative experimental ambient-pressure kagome superconductors are included in Fig.~\ref{fig:sc-landscape}(a) for comparison. \textcolor{black}{The layered reference compounds span different superconducting scales: $A$V$_3$Sb$_5$ has $T_\text{c}\approx$0.9-2.5~K, CsTi$_3$Bi$_5$ has a reported onset near 4.8~K, and LaRu$_3$Si$_2$ has $T_\text{c}\approx$7~K. The cubic ($A_3B$)-type compound (Sr,Na)Bi$_3$, whose face-centered Bi sublattice forms kagome nets within the (111) planes, sets the upper reference $T_\text{c}$ scale at approximately 9~K.} Seven compounds have estimated transition temperatures above this value: Ti$_3$Si and Ti$_3$Ge give the largest values, 15.2 and 14.0~K, followed by Tc$_3$Ge$_2$ (12.1~K), Cr$_3$As (11.2~K), Ti$_3$Sn (10.5~K), Zr$_3$Pb (9.7~K), and Cr$_3$P (9.1~K). In total, 84 candidates exceed 3~K. Table~\ref{tab:main-sc-candidates} lists the leading superconducting candidates together with their \(N(\ef)\), \(\omega_{\log}\), \(\lambda\), estimated $T_\text{c}$, \(E_{\mathrm{hull}}\), and space group. The complete information for all calculated superconducting candidates is provided in Table~S8 of the SM~\cite{supp}. Notably, the higher-$T_\text{c}$ candidates are distributed across several prototype families introduced in Fig.~\ref{fig:structural-families}, rather than being confined to a single stacking motif.
	
	The mode-resolved spectra in Figs.~\ref{fig:sc-landscape}(b--f) show how the total EPC is assembled from different phonon windows. \textcolor{black}{The Eliashberg spectral function \(\alpha^2F(\omega)\) resolves the EPC by phonon frequency, and frequency ranges with large \(\alpha^2F(\omega)/\omega\) correspond to the most rapid increase in \(\lambda(\omega)\), thereby identifying the phonons that dominate the total EPC (Sec.~S1 of the SM~\cite{supp}). In Ti$_3$Si, Zr$_3$Tl$_2$, and Zn$_3$Al, the dominant low-frequency spectral weight, arising mainly from acoustic branches, produces a steep initial increase in \(\lambda(\omega)\). By contrast, the broader spectral weights of Zr$_3$Pb and NbBi distribute the EPC mainly over intermediate- and higher-frequency phonons, respectively.} In general, lighter atoms extend the phonon spectrum to higher frequencies, whereas the largest EPC weight often follows softened low- and intermediate-frequency branches with strong kagome-atom character. This trend is clearest in Ti$_3$Si, where softened Ti-dominated modes below about 4~THz carry large \(\lambda_{\nu\mathbf{q}}\) and produce the largest $T_\text{c}$ in the survey. \textcolor{black}{The corner-sharing Ti triangles in the (111) kagome net can accommodate collective shear- and bond-angle-changing distortions with partially cancelling restoring forces, providing a structural route to these low-frequency modes. The M- and R-point eigenvectors are predominantly Ti-like, with negligible Si displacement; because the states at \(E_{\mathrm F}\) are also dominated by Ti \(d\) orbitals, these vibrations efficiently modulate the local Ti--Ti and Ti--Si bonding environment, while their low frequencies further enhance their contribution to \(\lambda\) (Fig.~S775 of the SM~\cite{supp}).} Zr$_3$Pb, which lies on the convex hull and has $T_\text{c}=9.7$~K, shows a more distributed accumulation of \(\lambda(\omega)\): Zr and Pb both contribute at low frequency, while higher-frequency Zr modes supply additional EPC. Zn$_3$Al is also on the convex hull; its low-frequency acoustic branches, dominated by the kagome Zn sublattice, drive a rapid increase of \(\lambda(\omega)\) below about 2.5~THz and yield $T_\text{c}=5.9$~K. These cases with \(E_{\mathrm{hull}}=0\) are therefore especially attractive as experimentally accessible superconducting candidates.
	
	The remaining representative systems illustrate another two prototypes: AB ($P6/mmm$) and A$_3$B$_2$ ($R\bar{3}m$). In NbBi with $T_\text{c}=4.0$~K, Nb-dominated modes above roughly 3~THz coincide with the main peaks in \(\alpha^2F(\omega)\) and provide most of the EPC. Zr$_3$Tl$_2$ is the same structural family as Ti$_3$Bi$_2$-based topological superconducting candidates~\cite{RN1006}. In this compound, softened Tl-dominated low-frequency modes give large \(\lambda_{\nu\mathbf{q}}\), while Zr-derived modes at higher frequency contribute a substantial remaining fraction of \(\lambda\). Together, these examples indicate that binary kagome superconductivity is not controlled by a single chemical recipe, but by the combined tuning of \(N(\ef)\), softened phonons, and kagome-sublattice EPC across the available structural families.

	\subsection{\texorpdfstring{Flat bands and $\mathbb{Z}_2$ topology in the superconducting NbBi}{Flat bands and Z2 topology in superconducting NbBi}}
	
\begin{figure}[t]
	\centering
	\includegraphics[width=0.48\textwidth]{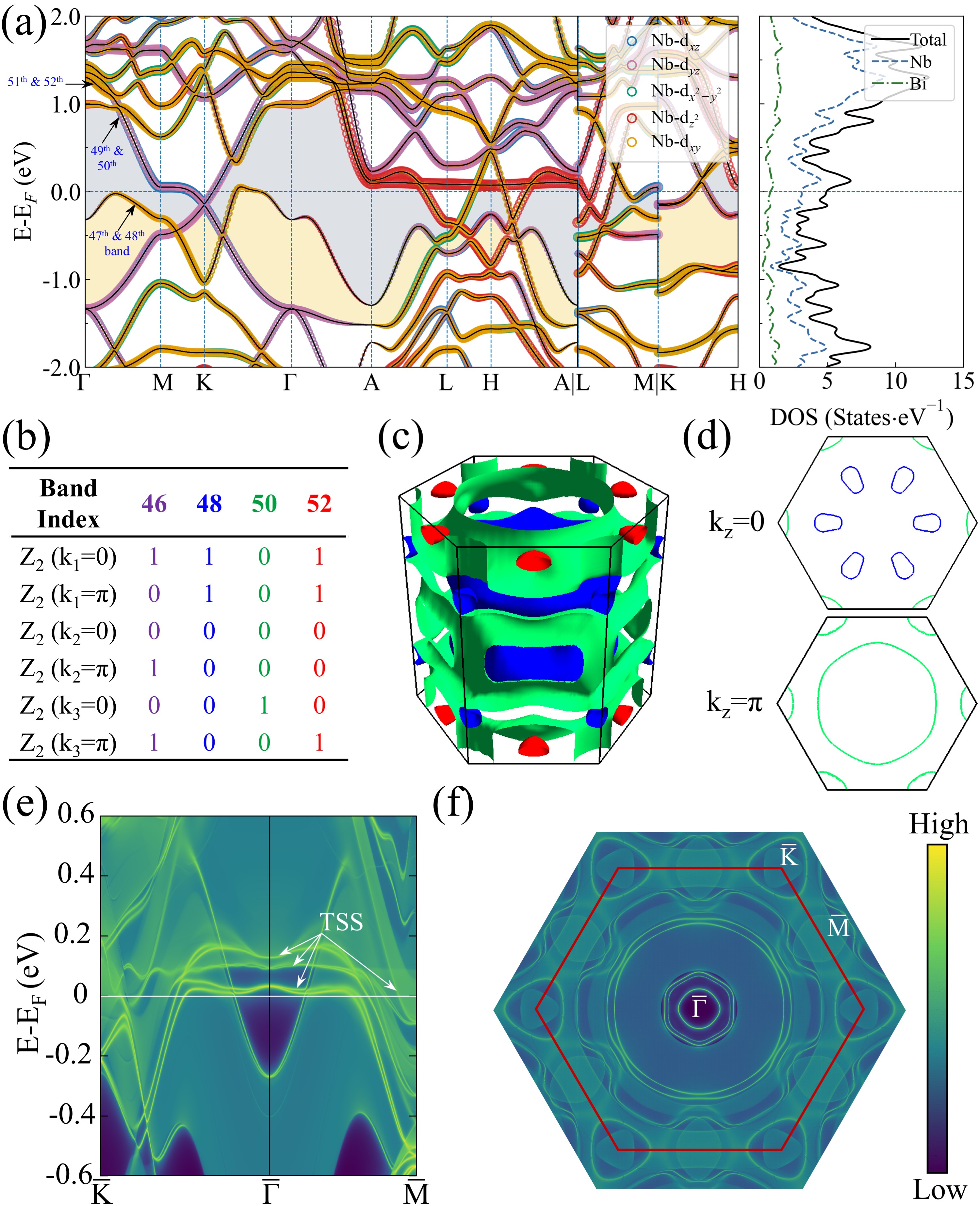}
	\caption{\textbf{Electronic topology of NbBi.} (a) Spin-orbit-coupled bulk band structure with Nb \(d\)-orbital projections and electronic density of states (DOS). \textcolor{black}{The continuous direct gaps between the 45/46 and 47/48 Kramers pairs and between the 47/48 and 49/50 Kramers pairs are highlighted by light-yellow and light-gray shading, respectively.} (b) Plane-resolved \(\mathbb{Z}_2\) indices for the direct gaps labeled by the upper band index of each Kramers pair, where $k_1$, $k_2$, and $k_3$ denote reciprocal primitive coordinates. (c) Three-dimensional Fermi surface and (d) two-dimensional Fermi-surface cuts at \(k_z=0\) and \(k_z=\pi\). (e) Surface spectral function along the \(\bar{K}\)--\(\bar{\Gamma}\)--\(\bar{M}\) path on the (001) surface. Topological surface states (TSS) are indicated by white arrows. (f) Constant-energy surface spectral function projected onto the (001) surface at \(E-\ef=0.015\)~eV.}
	\label{fig:nbbi-topology}
\end{figure}
	
	Although the estimated transition temperatures in the AB ($P6/mmm$) prototype are moderate compared with the leading superconductors in Table~\ref{tab:main-sc-candidates}, many members of this family provide a clean kagome-derived electronic structure (Sec.~S6 of the SM~\cite{supp}).
	
	We take NbBi as a representative AB-type candidate because it combines a clean kagome-derived electronic structure, weakly dispersive bands near \(\ef\), and nontrivial \(\mathbb{Z}_2\) topology. Fig.~\ref{fig:nbbi-topology}(a) shows its electronic band structure with spin-orbit coupling (SOC). The states near \(\ef\) are dominated by Nb \(d\) orbitals, consistent with a kagome-metal electronic manifold. In particular, an \(M\)-point vHs with dominant Nb-\(d_{xz}/d_{yz}\) character lies about \(0.05\)~eV above \(\ef\). A weakly dispersive band with mainly Nb-\(d_{z^2}\) character appears in the same energy window on the \(k_z=\pi\) plane, producing a peak in the density of states. The Fermi surface in Figs.~\ref{fig:nbbi-topology}(c) and (d) combines three-dimensional pockets with nearly hexagonal in-plane contours, reflecting finite interlayer coupling in the $P6/mmm$ stacking. The calculated phonon spectrum of NbBi shows no unstable branch [Fig.~\ref{fig:sc-landscape}(b)], despite the proximity of these kagome electronic features to \(\ef\). This contrasts with the lattice instabilities associated with charge-density-wave formation in \(A\)V$_3$Sb$_5$ kagome metals~\cite{RN2}. Electron doping, chemical substitution, alloying, or thin-film gating could therefore shift the flat band and vHs toward \(\ef\), increasing \(N(\ef)\) within a rigid-band picture and possibly leading to an enhanced \(T_\text{c}\).
	
	The bands near the Fermi level in NbBi carry nontrivial \(\mathbb{Z}_2\) topology. Because both inversion and time-reversal symmetries are present, the spin-orbit-coupled bands are Kramers degenerate. A direct gap between the 45/46 and 47/48 Kramers pairs persists throughout the Brillouin zone \textcolor{black}{[light-yellow shading in Fig.~\ref{fig:nbbi-topology}(a)]}. The isolated manifold comprising the lowest 46 bands has a nontrivial strong index, \(\nu_0=1\) [Fig.~\ref{fig:nbbi-topology}(b)]. The higher gap between the 47/48 and 49/50 Kramers pairs is also continuous \textcolor{black}{[light-grey shading in Fig.~\ref{fig:nbbi-topology}(a)]}, but its strong index is trivial. However, the two \(k_2\)-\(k_3\) time-reversal-invariant planes at \(k_1=0\) and \(k_1=\pi\) are nontrivial \textcolor{black}{[Fig.~\ref{fig:nbbi-topology}(b)]}, indicating a weak topological component in a nearby energy window.
	
	Higher bands add a crystalline-topological component. Along \(\Gamma\)-A and K-H, the little groups contain sixfold and threefold rotations, respectively. Crossings between Kramers doublets belonging to distinct double-group irreducible representations can therefore remain gapless with SOC. Several such crossings persist between 50 and 51, as well as 52 and 53 Kramers pairs in Fig.~\ref{fig:nbbi-topology}(a). For the 50- and 52-band fillings, bulk degeneracies prevent the definition of a globally isolated three-dimensional band manifold. We therefore report two-dimensional \(\mathbb{Z}_2\) indices only for individual time-reversal-invariant planes that remain directly gapped at the relevant filling [Fig.~\ref{fig:nbbi-topology}(b)]. Such nontrivial weak indices support topological surface states on suitable surfaces, depending on the surface orientation and the projection of the bulk bands.
	
	The (001) surface spectral function is consistent with this topology. Along \(\bar{K}\)--\(\bar{\Gamma}\)--\(\bar{M}\), Fig.~\ref{fig:nbbi-topology}(e) shows surface-localized branches near \(\ef\), and Fig.~\ref{fig:nbbi-topology}(f) shows circular and hexagonal surface contours around time-reversal-invariant momentum \(\bar{\Gamma}\) and the projected zone boundary. \textcolor{black}{Additional calculations for the (100) termination reveal prominent surface-state branches near the surface time-reversal-invariant momenta \(\bar{\Gamma}\) and \(\bar{M}\) (Fig.~S776 of the SM~\cite{supp}), providing an independent surface-orientation check consistent with the bulk--boundary correspondence.}
	
	\textcolor{black}{In a time-reversal-invariant metal, a lower band manifold separated from higher bands by a finite direct gap at every k can carry well-defined \(\mathbb{Z}_2\) indices despite an indirect band overlap. We use the term ``\(\mathbb{Z}_2\) topological metal" when this isolated manifold has nontrivial \(\mathbb{Z}_2\) topology ~\cite{RN372,RN146}.} Combining the nontrivial \(\mathbb{Z}_2\) index with the derived surface states identifies NbBi as a \(\mathbb{Z}_2\) topological metal. The coexistence of these spin-momentum-locked TSS near \(\ef\) with superconductivity in NbBi makes it a promising platform for exploring proximity-induced topological superconductivity.

	\subsection{\texorpdfstring{Nodal features in Ti$_3$Si}{Nodal features in Ti3Si}}
		
	\begin{figure}[t]
		\centering
		\includegraphics[width=0.48\textwidth]{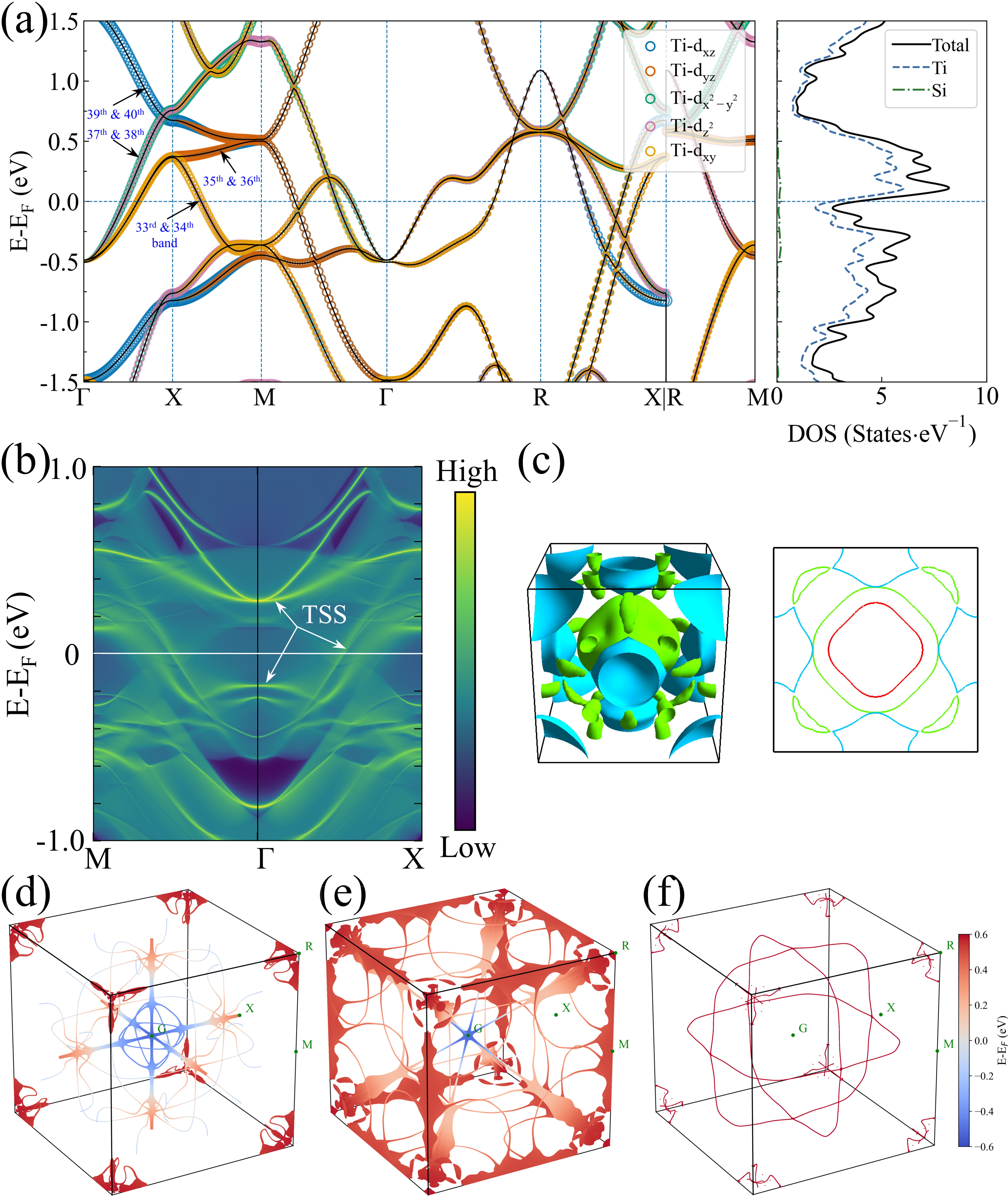}
		\caption{\textbf{Electronic topology of Ti$_3$Si.} (a) Spin-orbit-coupled bulk band structure with Ti \(d\)-orbital projections and electronic DOS. \textcolor{black}{(b) Surface spectral function along the \(\bar{M}\)--\(\bar{\Gamma}\)--\(\bar{X}\) path on the (001) surface with SOC. (c) Three-dimensional Fermi surface with SOC and its \(k_z=0\) two-dimensional cut. Panels (d--f) show representative nodal manifolds obtained without SOC, arising from crossings between the spin-degenerate band pairs 33/34 and 35/36 in (d), 35/36 and 37/38 in (e), and 37/38 and 39/40 in (f). The common color scale denotes the crossing energy relative to $\ef$.}}
		\label{fig:ti3si-topology}
	\end{figure}
	
	Ti$_3$Si is a representative centrosymmetric kagome metal with the largest estimated $T_\text{c}$ (15.2~K) in this work. Its electronic topology is shown in Fig.~\ref{fig:ti3si-topology}(a). The states near \(\ef\) are dominated by kagome Ti \(d\) orbitals, with little Si contribution. Its Fermi surface is also strongly three-dimensional [Fig.~\ref{fig:ti3si-topology}(c)], with nearly cubic symmetry rather than quasi-two-dimensional kagome cylinders. A supercell calculation with the kagome layer in the \(xy\) plane shows a weakly dispersive kagome-derived band about \(0.1\)~eV above \(\ef\) (Fig.~S774). In a rigid-band estimate, electron doping would shift \(\ef\) toward this band, more than double \(N(\ef)\), and may also strongly enhance superconductivity.
	
	Before SOC is included, the band topology of Ti$_3$Si is dominated by mirror-enforced nodal manifolds. Adjacent bands near \(\ef\) form nodal lines and sheet-like nodal structures on mirror-invariant planes, including \(k_{x/y/z}=0/\pi\) and diagonal planes such as \(k_x=k_y\) [Figs.~\ref{fig:ti3si-topology}(d--f)]. These are extended Brillouin-zone structures rather than isolated path crossings. Small closed loops linking representative nodal lines give a Berry phase of \(\pi\) for most tested loops, identifying them as spinless \(PT\)-symmetric nodal lines. The non-SOC surface spectral functions in Fig.~S777 of the SM~\cite{supp} show drumhead surface states associated with these nodal-line manifolds.

	SOC gaps most of the spinless nodal lines but leaves selected fourfold degeneracies along high-symmetry lines, where crossing Kramers doublets belong to distinct double-group irreducible representations (Figs.~S778 \textcolor{black}{and S779}). Because Ti$_3$Si is nonmagnetic and centrosymmetric, ordinary Weyl points are not expected. \textcolor{black}{The residual fourfold crossings occur along $\Gamma$-X, $\Gamma$-R, and R-M, where distinct double-group representations prevent hybridization under C$_4y$, C$_3[111]$, and C$_4z$, respectively (Sec.~S7 and Fig. S779 of the SM).} The (001) surface spectral function in Fig.~\ref{fig:ti3si-topology}(b) shows intense surface-localized branches along \(\bar{M}\)--\(\bar{\Gamma}\)--\(\bar{X}\). These branches track the non-SOC drumhead spectrum (Fig.~S777) and split into pairs after SOC, connecting the bulk continua on the two sides of the opened nodal-line gaps. We therefore interpret the dominant surface spectral weight as drumhead-derived surface resonances, rather than Weyl Fermi arcs~\cite{RN1206}. Plane-resolved \(\mathbb{Z}_2\) indices of gapped SOC slices may contribute to parts of the spectrum, while the similarity between the non-SOC and SOC surface spectra is consistent with a drumhead-derived interpretation.

	This mechanism differs from NbBi, where the surface states are tied mainly to nontrivial \(\mathbb{Z}_2\) indices of SOC direct gaps. Ti$_3$Si instead hosts drumhead-derived surface resonances inherited from the non-SOC mirror-protected nodal line phase. Beyond these two examples, Fig.~S780 shows surface spectral functions for several additional superconducting systems, many of which exhibit abundant surface-localized spectral weight near \(\ef\). These results indicate that many of the predicted superconducting candidates may serve as platforms where phonon-mediated superconductivity coexists with nontrivial or nodal-line-derived surface electronic states.

	\begingroup
	\color{black}
	\section{Discussion}
	\endgroup
	
	\textcolor{black}{Beyond the weakly dispersive band and nearby van Hove singularity in NbBi [Fig.~\ref{fig:nbbi-topology}(a)], many of the screened compounds exhibit characteristic kagome-derived electronic features. The phonon spectra and electronic band structures of the screened dynamically stable compounds are provided in Sec.~S6 of the SM~\cite{supp}. For the A$_3$B$_2$ (\(R\bar{3}m\)) and A$_3$B (\(Pm\bar{3}m\)) systems, supercells with the kagome planes parallel to the \(ab\) plane were employed to better resolve these features. Ti$_3$Si provides a representative example, in which the Ti \(d\) states on the kagome sublattice dominate the electronic states near \(\ef\), and a weakly dispersive kagome-derived flat band lies approximately 0.1~eV above \(\ef\) [Fig.~S774 of the SM~\cite{supp}]. The correspondingly large DOS at \(\ef\) is one of the key factors underlying its strong EPC and relatively high \(T_\text{c}\). More generally, kagome-derived flat bands and van Hove singularities can enhance the DOS near \(\ef\), thereby providing favorable electronic conditions for superconductivity and potentially contributing to an enhanced \(T_\text{c}\). These candidates therefore constitute a broad materials platform for exploring kagome superconductivity and related quantum phenomena.}
	
	\textcolor{black}{In the prototypical kagome-superconductor family \(A\mathrm{V}_3\mathrm{Sb}_5\), superconductivity coexists and competes with charge-density-wave order and cannot be fully accounted for by EPC alone~\cite{RN2}. By contrast, the present work focuses on EPC superconductivity in dynamically stable pristine phases, for which the estimated \(T_\mathrm{c}\) primarily reflects the balance between \(\lambda\) and the characteristic phonon energy scale. Moreover, the more three-dimensional Fermi surfaces of leading candidates such as Ti$_3$Si may weaken nesting-driven charge-ordering tendencies.}
	
	\textcolor{black}{The candidate landscape established here broadens our earlier structure-specific searches of the $A\mathrm{V}_3\mathrm{Sb}_5$-type and Ti$_6X_4$ families~\cite{RN985,RN1006} to four binary prototypes spanning the $A_3B$, $AB$, and $A_3B_2$ families. Within this materials space, 84 superconducting candidates are identified through stability and EPC screening, with selected representatives further examined through bulk topological indices, nodal structures, and surface spectral functions. By jointly examining phonon-mediated superconducting propensity and topological electronic states within related kagome structural families, this family-focused strategy complements broader high-throughput approaches, such as the symmetry-indicator-based classification of possible topological and nodal superconducting states~\cite{RN478}.}
	
	\textcolor{black}{Although the main focus of this work is binary kagome superconducting candidates, the screening also identifies Co$_3$Sb$_2$ as a rare semiconducting outlier with an indirect gap of 0.12~eV. Its Co-derived flat band edges and surface-localized resonances illustrate the broader electronic diversity of these structural families. Because it lies outside the metallic EPC-superconducting subset, the detailed analysis is presented in Sec.~S8 of the SM~\cite{supp}.}
	
	\section{Conclusion}
	
	In conclusion, we have carried out a high-throughput first-principles search for binary kagome superconducting candidates with topological electronic states. The survey covers \(A_3B\), \(AB\), and \(A_3B_2\)-type structural families, where kagome and honeycomb layers generate a broad range of electronic structures and phonon spectra. The resulting dataset shows that binary kagome materials provide a sizable search space for phonon-mediated superconducting candidates. EPC calculations identify Ti$_3$Si and Ti$_3$Ge as having the highest estimated $T_\text{c}$ values, about 15.2 and 14~K, in the present survey, while Zn$_3$Al and Zr$_3$Pb combine finite estimated $T_\text{c}$ values with $E_\text{hull}$=0. The topological analysis further shows that different candidates can host distinct surface-state mechanisms. In NbBi, the relevant spin-orbit-coupled direct gap carries a nontrivial \(\mathbb{Z}_2\) index and gives rise to \(\mathbb{Z}_2\)-derived surface states close to $\ef$ on the (001) surface. In Ti$_3$Si, the dominant surface spectral weight is instead connected to mirror-protected nodal-line structures in the non-SOC limit and their SOC-gapped descendants. Taken together, these results identify binary kagome compounds as promising topological superconducting material candidates, while pairing symmetry, possible surface superconductivity, and Majorana signatures remain to be established.

	\section{Acknowledgments}

	This work is supported in part by the CAS Superconducting Research Project (Grant No. SCZX-0101), the NSFC (Grant Nos. 12547131, 12534009, and 12447101), the Strategic Priority Research Program of the Chinese Academy of Sciences (Grant No. XDB1270000), and the Quantum Science and Technology–National Science and Technology Major Project (Grant No. 2024ZD0300500). We acknowledge the computational resources provided by HPC Cluster of ITP-CAS.
	
	\section{Data availability}
	
Detailed data on the crystal structures, superconducting properties, electronic structures, and related quantities are provided in the Supplemental Material. Additional data not shown in the manuscript or Supplemental Material are available from the authors upon reasonable request.
	

\begin{thebibliography}{108}%
		\makeatletter
		\providecommand \@ifxundefined [1]{%
			\@ifx{#1\undefined}
		}%
		\providecommand \@ifnum [1]{%
			\ifnum #1\expandafter \@firstoftwo
			\else \expandafter \@secondoftwo
			\fi
		}%
		\providecommand \@ifx [1]{%
			\ifx #1\expandafter \@firstoftwo
			\else \expandafter \@secondoftwo
			\fi
		}%
		\providecommand \natexlab [1]{#1}%
		\providecommand \enquote  [1]{``#1''}%
		\providecommand \bibnamefont  [1]{#1}%
		\providecommand \bibfnamefont [1]{#1}%
		\providecommand \citenamefont [1]{#1}%
		\providecommand \href@noop [0]{\@secondoftwo}%
		\providecommand \href [0]{\begingroup \@sanitize@url \@href}%
		\providecommand \@href[1]{\@@startlink{#1}\@@href}%
		\providecommand \@@href[1]{\endgroup#1\@@endlink}%
		\providecommand \@sanitize@url [0]{\catcode `\\12\catcode `\$12\catcode
			`\&12\catcode `\#12\catcode `\^12\catcode `\_12\catcode `\%12\relax}%
		\providecommand \@@startlink[1]{}%
		\providecommand \@@endlink[0]{}%
		\providecommand \url  [0]{\begingroup\@sanitize@url \@url }%
		\providecommand \@url [1]{\endgroup\@href {#1}{\urlprefix }}%
		\providecommand \urlprefix  [0]{URL }%
		\providecommand \Eprint [0]{\href }%
		\providecommand \doibase [0]{https://doi.org/}%
		\providecommand \selectlanguage [0]{\@gobble}%
		\providecommand \bibinfo  [0]{\@secondoftwo}%
		\providecommand \bibfield  [0]{\@secondoftwo}%
		\providecommand \translation [1]{[#1]}%
		\providecommand \BibitemOpen [0]{}%
		\providecommand \bibitemStop [0]{}%
		\providecommand \bibitemNoStop [0]{.\EOS\space}%
		\providecommand \EOS [0]{\spacefactor3000\relax}%
		\providecommand \BibitemShut  [1]{\csname bibitem#1\endcsname}%
		\let\auto@bib@innerbib\@empty
		\bibitem [{\citenamefont {Qi}\ and\ \citenamefont {Zhang}(2011)}]{RN529}%
		\BibitemOpen
		\bibfield  {author} {\bibinfo {author} {\bibfnamefont {X.-L.}\ \bibnamefont
				{Qi}}\ and\ \bibinfo {author} {\bibfnamefont {S.-C.}\ \bibnamefont {Zhang}},\
		}\bibfield  {title} {\bibinfo {title} {{Topological insulators and
					superconductors}},\ }\href {https://doi.org/10.1103/RevModPhys.83.1057}
		{\bibfield  {journal} {\bibinfo  {journal} {Rev. Mod. Phys.}\ }\textbf
			{\bibinfo {volume} {83}},\ \bibinfo {pages} {1057} (\bibinfo {year}
			{2011})}\BibitemShut {NoStop}%
		\bibitem [{\citenamefont {Alicea}(2012)}]{RN599}%
		\BibitemOpen
		\bibfield  {author} {\bibinfo {author} {\bibfnamefont {J.}~\bibnamefont
				{Alicea}},\ }\bibfield  {title} {\bibinfo {title} {{New directions in the
					pursuit of {Majorana} fermions in solid state systems}},\ }\href
		{https://doi.org/10.1088/0034-4885/75/7/076501} {\bibfield  {journal}
			{\bibinfo  {journal} {Rep. Prog. Phys.}\ }\textbf {\bibinfo {volume} {75}},\
			\bibinfo {pages} {076501} (\bibinfo {year} {2012})}\BibitemShut {NoStop}%
		\bibitem [{\citenamefont {Kitaev}(2003)}]{RN603}%
		\BibitemOpen
		\bibfield  {author} {\bibinfo {author} {\bibfnamefont {A.~Y.}\ \bibnamefont
				{Kitaev}},\ }\bibfield  {title} {\bibinfo {title} {{Fault-tolerant quantum
					computation by anyons}},\ }\href
		{https://doi.org/10.1016/S0003-4916(02)00018-0} {\bibfield  {journal}
			{\bibinfo  {journal} {Ann. Phys.}\ }\textbf {\bibinfo {volume} {303}},\
			\bibinfo {pages} {2} (\bibinfo {year} {2003})}\BibitemShut {NoStop}%
		\bibitem [{\citenamefont {Nayak}\ \emph {et~al.}(2008)\citenamefont {Nayak},
			\citenamefont {Simon}, \citenamefont {Stern}, \citenamefont {Freedman},\ and\
			\citenamefont {Das~Sarma}}]{RN516}%
		\BibitemOpen
		\bibfield  {author} {\bibinfo {author} {\bibfnamefont {C.}~\bibnamefont
				{Nayak}}, \bibinfo {author} {\bibfnamefont {S.~H.}\ \bibnamefont {Simon}},
			\bibinfo {author} {\bibfnamefont {A.}~\bibnamefont {Stern}}, \bibinfo
			{author} {\bibfnamefont {M.}~\bibnamefont {Freedman}},\ and\ \bibinfo
			{author} {\bibfnamefont {S.}~\bibnamefont {Das~Sarma}},\ }\bibfield  {title}
		{\bibinfo {title} {{Non-Abelian anyons and topological quantum
					computation}},\ }\href {https://doi.org/10.1103/RevModPhys.80.1083}
		{\bibfield  {journal} {\bibinfo  {journal} {Rev. Mod. Phys.}\ }\textbf
			{\bibinfo {volume} {80}},\ \bibinfo {pages} {1083} (\bibinfo {year}
			{2008})}\BibitemShut {NoStop}%
		\bibitem [{\citenamefont {Sato}\ and\ \citenamefont {Ando}(2017)}]{RN370}%
		\BibitemOpen
		\bibfield  {author} {\bibinfo {author} {\bibfnamefont {M.}~\bibnamefont
				{Sato}}\ and\ \bibinfo {author} {\bibfnamefont {Y.}~\bibnamefont {Ando}},\
		}\bibfield  {title} {\bibinfo {title} {{Topological superconductors: a
					review}},\ }\href {https://doi.org/10.1088/1361-6633/aa6ac7} {\bibfield
			{journal} {\bibinfo  {journal} {Rep. Prog. Phys.}\ }\textbf {\bibinfo
				{volume} {80}},\ \bibinfo {pages} {076501} (\bibinfo {year}
			{2017})}\BibitemShut {NoStop}%
		\bibitem [{\citenamefont {Sato}(2009)}]{RN538}%
		\BibitemOpen
		\bibfield  {author} {\bibinfo {author} {\bibfnamefont {M.}~\bibnamefont
				{Sato}},\ }\bibfield  {title} {\bibinfo {title} {{Topological properties of
					spin-triplet superconductors and {Fermi} surface topology in the normal
					state}},\ }\href {https://doi.org/10.1103/PhysRevB.79.214526} {\bibfield
			{journal} {\bibinfo  {journal} {Phys. Rev. B}\ }\textbf {\bibinfo {volume}
				{79}},\ \bibinfo {pages} {214526} (\bibinfo {year} {2009})}\BibitemShut
		{NoStop}%
		\bibitem [{\citenamefont {Sato}(2010)}]{RN539}%
		\BibitemOpen
		\bibfield  {author} {\bibinfo {author} {\bibfnamefont {M.}~\bibnamefont
				{Sato}},\ }\bibfield  {title} {\bibinfo {title} {{Topological odd-parity
					superconductors}},\ }\href {https://doi.org/10.1103/PhysRevB.81.220504}
		{\bibfield  {journal} {\bibinfo  {journal} {Phys. Rev. B}\ }\textbf {\bibinfo
				{volume} {81}},\ \bibinfo {pages} {220504} (\bibinfo {year}
			{2010})}\BibitemShut {NoStop}%
		\bibitem [{\citenamefont {Fu}\ and\ \citenamefont {Kane}(2008)}]{RN440}%
		\BibitemOpen
		\bibfield  {author} {\bibinfo {author} {\bibfnamefont {L.}~\bibnamefont
				{Fu}}\ and\ \bibinfo {author} {\bibfnamefont {C.~L.}\ \bibnamefont {Kane}},\
		}\bibfield  {title} {\bibinfo {title} {{Superconducting proximity effect and
					{Majorana} fermions at the surface of a topological insulator}},\ }\href
		{https://doi.org/10.1103/PhysRevLett.100.096407} {\bibfield  {journal}
			{\bibinfo  {journal} {Phys. Rev. Lett.}\ }\textbf {\bibinfo {volume} {100}},\
			\bibinfo {pages} {096407} (\bibinfo {year} {2008})}\BibitemShut {NoStop}%
		\bibitem [{\citenamefont {Sasaki}\ \emph {et~al.}(2011)\citenamefont {Sasaki},
			\citenamefont {Kriener}, \citenamefont {Segawa}, \citenamefont {Yada},
			\citenamefont {Tanaka}, \citenamefont {Sato},\ and\ \citenamefont
			{Ando}}]{RN532}%
		\BibitemOpen
		\bibfield  {author} {\bibinfo {author} {\bibfnamefont {S.}~\bibnamefont
				{Sasaki}}, \bibinfo {author} {\bibfnamefont {M.}~\bibnamefont {Kriener}},
			\bibinfo {author} {\bibfnamefont {K.}~\bibnamefont {Segawa}}, \bibinfo
			{author} {\bibfnamefont {K.}~\bibnamefont {Yada}}, \bibinfo {author}
			{\bibfnamefont {Y.}~\bibnamefont {Tanaka}}, \bibinfo {author} {\bibfnamefont
				{M.}~\bibnamefont {Sato}},\ and\ \bibinfo {author} {\bibfnamefont
				{Y.}~\bibnamefont {Ando}},\ }\bibfield  {title} {\bibinfo {title}
			{{Topological Superconductivity in {Cu$_x$Bi$_2$Se$_3$}}},\ }\href
		{https://doi.org/10.1103/PhysRevLett.107.217001} {\bibfield  {journal}
			{\bibinfo  {journal} {Phys. Rev. Lett.}\ }\textbf {\bibinfo {volume} {107}},\
			\bibinfo {pages} {217001} (\bibinfo {year} {2011})}\BibitemShut {NoStop}%
		\bibitem [{\citenamefont {Fu}\ and\ \citenamefont {Berg}(2010)}]{RN441}%
		\BibitemOpen
		\bibfield  {author} {\bibinfo {author} {\bibfnamefont {L.}~\bibnamefont
				{Fu}}\ and\ \bibinfo {author} {\bibfnamefont {E.}~\bibnamefont {Berg}},\
		}\bibfield  {title} {\bibinfo {title} {{Odd-parity topological
					superconductors: theory and application to {Cu$_x$Bi$_2$Se$_3$}}},\ }\href
		{https://doi.org/10.1103/PhysRevLett.105.097001} {\bibfield  {journal}
			{\bibinfo  {journal} {Phys. Rev. Lett.}\ }\textbf {\bibinfo {volume} {105}},\
			\bibinfo {pages} {097001} (\bibinfo {year} {2010})}\BibitemShut {NoStop}%
		\bibitem [{\citenamefont {Hsieh}\ and\ \citenamefont {Fu}(2012)}]{RN535}%
		\BibitemOpen
		\bibfield  {author} {\bibinfo {author} {\bibfnamefont {T.~H.}\ \bibnamefont
				{Hsieh}}\ and\ \bibinfo {author} {\bibfnamefont {L.}~\bibnamefont {Fu}},\
		}\bibfield  {title} {\bibinfo {title} {{{Majorana} fermions and exotic
					surface Andreev bound states in topological superconductors: application to
					{Cu$_x$Bi$_2$Se$_3$}}},\ }\href
		{https://doi.org/10.1103/PhysRevLett.108.107005} {\bibfield  {journal}
			{\bibinfo  {journal} {Phys. Rev. Lett.}\ }\textbf {\bibinfo {volume} {108}},\
			\bibinfo {pages} {107005} (\bibinfo {year} {2012})}\BibitemShut {NoStop}%
		\bibitem [{\citenamefont {Sasaki}\ \emph {et~al.}(2012)\citenamefont {Sasaki},
			\citenamefont {Ren}, \citenamefont {Taskin}, \citenamefont {Segawa},
			\citenamefont {Fu},\ and\ \citenamefont {Ando}}]{RN584}%
		\BibitemOpen
		\bibfield  {author} {\bibinfo {author} {\bibfnamefont {S.}~\bibnamefont
				{Sasaki}}, \bibinfo {author} {\bibfnamefont {Z.}~\bibnamefont {Ren}},
			\bibinfo {author} {\bibfnamefont {A.~A.}\ \bibnamefont {Taskin}}, \bibinfo
			{author} {\bibfnamefont {K.}~\bibnamefont {Segawa}}, \bibinfo {author}
			{\bibfnamefont {L.}~\bibnamefont {Fu}},\ and\ \bibinfo {author}
			{\bibfnamefont {Y.}~\bibnamefont {Ando}},\ }\bibfield  {title} {\bibinfo
			{title} {{Odd-parity pairing and topological superconductivity in a strongly
					spin-orbit coupled semiconductor}},\ }\href
		{https://doi.org/10.1103/PhysRevLett.109.217004} {\bibfield  {journal}
			{\bibinfo  {journal} {Phys. Rev. Lett.}\ }\textbf {\bibinfo {volume} {109}},\
			\bibinfo {pages} {217004} (\bibinfo {year} {2012})}\BibitemShut {NoStop}%
		\bibitem [{\citenamefont {Sato}\ \emph {et~al.}(2013)\citenamefont {Sato},
			\citenamefont {Tanaka}, \citenamefont {Nakayama}, \citenamefont {Souma},
			\citenamefont {Takahashi}, \citenamefont {Sasaki}, \citenamefont {Ren},
			\citenamefont {Taskin}, \citenamefont {Segawa},\ and\ \citenamefont
			{Ando}}]{RN515}%
		\BibitemOpen
		\bibfield  {author} {\bibinfo {author} {\bibfnamefont {T.}~\bibnamefont
				{Sato}}, \bibinfo {author} {\bibfnamefont {Y.}~\bibnamefont {Tanaka}},
			\bibinfo {author} {\bibfnamefont {K.}~\bibnamefont {Nakayama}}, \bibinfo
			{author} {\bibfnamefont {S.}~\bibnamefont {Souma}}, \bibinfo {author}
			{\bibfnamefont {T.}~\bibnamefont {Takahashi}}, \bibinfo {author}
			{\bibfnamefont {S.}~\bibnamefont {Sasaki}}, \bibinfo {author} {\bibfnamefont
				{Z.}~\bibnamefont {Ren}}, \bibinfo {author} {\bibfnamefont {A.~A.}\
				\bibnamefont {Taskin}}, \bibinfo {author} {\bibfnamefont {K.}~\bibnamefont
				{Segawa}},\ and\ \bibinfo {author} {\bibfnamefont {Y.}~\bibnamefont {Ando}},\
		}\bibfield  {title} {\bibinfo {title} {{Fermiology of the strongly spin-orbit
					coupled superconductor {Sn$_{1-x}$In$_x$Te}: implications for topological
					superconductivity}},\ }\href {https://doi.org/10.1103/PhysRevLett.110.206804}
		{\bibfield  {journal} {\bibinfo  {journal} {Phys. Rev. Lett.}\ }\textbf
			{\bibinfo {volume} {110}},\ \bibinfo {pages} {206804} (\bibinfo {year}
			{2013})}\BibitemShut {NoStop}%
		\bibitem [{\citenamefont {Xu}\ \emph {et~al.}(2014)\citenamefont {Xu},
			\citenamefont {Liu}, \citenamefont {Wang}, \citenamefont {Ge}, \citenamefont
			{Liu}, \citenamefont {Yang}, \citenamefont {Chen}, \citenamefont {Liu},
			\citenamefont {Xu}, \citenamefont {Gao}, \citenamefont {Qian}, \citenamefont
			{Zhang},\ and\ \citenamefont {Jia}}]{RN572}%
		\BibitemOpen
		\bibfield  {author} {\bibinfo {author} {\bibfnamefont {J.-P.}\ \bibnamefont
				{Xu}}, \bibinfo {author} {\bibfnamefont {C.}~\bibnamefont {Liu}}, \bibinfo
			{author} {\bibfnamefont {M.-X.}\ \bibnamefont {Wang}}, \bibinfo {author}
			{\bibfnamefont {J.}~\bibnamefont {Ge}}, \bibinfo {author} {\bibfnamefont
				{Z.-L.}\ \bibnamefont {Liu}}, \bibinfo {author} {\bibfnamefont
				{X.}~\bibnamefont {Yang}}, \bibinfo {author} {\bibfnamefont {Y.}~\bibnamefont
				{Chen}}, \bibinfo {author} {\bibfnamefont {Y.}~\bibnamefont {Liu}}, \bibinfo
			{author} {\bibfnamefont {Z.-A.}\ \bibnamefont {Xu}}, \bibinfo {author}
			{\bibfnamefont {C.-L.}\ \bibnamefont {Gao}}, \bibinfo {author} {\bibfnamefont
				{D.}~\bibnamefont {Qian}}, \bibinfo {author} {\bibfnamefont {F.-C.}\
				\bibnamefont {Zhang}},\ and\ \bibinfo {author} {\bibfnamefont {J.-F.}\
				\bibnamefont {Jia}},\ }\bibfield  {title} {\bibinfo {title} {{Artificial
					Topological Superconductor by the Proximity Effect}},\ }\href
		{https://doi.org/10.1103/PhysRevLett.112.217001} {\bibfield  {journal}
			{\bibinfo  {journal} {Phys. Rev. Lett.}\ }\textbf {\bibinfo {volume} {112}},\
			\bibinfo {pages} {217001} (\bibinfo {year} {2014})}\BibitemShut {NoStop}%
		\bibitem [{\citenamefont {Xu}\ \emph {et~al.}(2015)\citenamefont {Xu},
			\citenamefont {Wang}, \citenamefont {Liu}, \citenamefont {Ge}, \citenamefont
			{Yang}, \citenamefont {Liu}, \citenamefont {Xu}, \citenamefont {Guan},
			\citenamefont {Gao}, \citenamefont {Qian}, \citenamefont {Liu}, \citenamefont
			{Wang}, \citenamefont {Zhang}, \citenamefont {Xue},\ and\ \citenamefont
			{Jia}}]{RN606}%
		\BibitemOpen
		\bibfield  {author} {\bibinfo {author} {\bibfnamefont {J.~P.}\ \bibnamefont
				{Xu}}, \bibinfo {author} {\bibfnamefont {M.~X.}\ \bibnamefont {Wang}},
			\bibinfo {author} {\bibfnamefont {Z.~L.}\ \bibnamefont {Liu}}, \bibinfo
			{author} {\bibfnamefont {J.~F.}\ \bibnamefont {Ge}}, \bibinfo {author}
			{\bibfnamefont {X.}~\bibnamefont {Yang}}, \bibinfo {author} {\bibfnamefont
				{C.}~\bibnamefont {Liu}}, \bibinfo {author} {\bibfnamefont {Z.~A.}\
				\bibnamefont {Xu}}, \bibinfo {author} {\bibfnamefont {D.}~\bibnamefont
				{Guan}}, \bibinfo {author} {\bibfnamefont {C.~L.}\ \bibnamefont {Gao}},
			\bibinfo {author} {\bibfnamefont {D.}~\bibnamefont {Qian}}, \bibinfo {author}
			{\bibfnamefont {Y.}~\bibnamefont {Liu}}, \bibinfo {author} {\bibfnamefont
				{Q.~H.}\ \bibnamefont {Wang}}, \bibinfo {author} {\bibfnamefont {F.~C.}\
				\bibnamefont {Zhang}}, \bibinfo {author} {\bibfnamefont {Q.~K.}\ \bibnamefont
				{Xue}},\ and\ \bibinfo {author} {\bibfnamefont {J.~F.}\ \bibnamefont {Jia}},\
		}\bibfield  {title} {\bibinfo {title} {{Experimental detection of a
					{Majorana} mode in the core of a magnetic vortex inside a topological
					insulator-superconductor {Bi$_2$Te$_3$/NbSe$_2$} heterostructure}},\ }\href
		{https://doi.org/10.1103/PhysRevLett.114.017001} {\bibfield  {journal}
			{\bibinfo  {journal} {Phys. Rev. Lett.}\ }\textbf {\bibinfo {volume} {114}},\
			\bibinfo {pages} {017001} (\bibinfo {year} {2015})}\BibitemShut {NoStop}%
		\bibitem [{\citenamefont {Lutchyn}\ \emph {et~al.}(2010)\citenamefont
			{Lutchyn}, \citenamefont {Sau},\ and\ \citenamefont {Das~Sarma}}]{RN574}%
		\BibitemOpen
		\bibfield  {author} {\bibinfo {author} {\bibfnamefont {R.~M.}\ \bibnamefont
				{Lutchyn}}, \bibinfo {author} {\bibfnamefont {J.~D.}\ \bibnamefont {Sau}},\
			and\ \bibinfo {author} {\bibfnamefont {S.}~\bibnamefont {Das~Sarma}},\
		}\bibfield  {title} {\bibinfo {title} {{{Majorana} fermions and a topological
					phase transition in semiconductor-superconductor heterostructures}},\ }\href
		{https://doi.org/10.1103/PhysRevLett.105.077001} {\bibfield  {journal}
			{\bibinfo  {journal} {Phys. Rev. Lett.}\ }\textbf {\bibinfo {volume} {105}},\
			\bibinfo {pages} {077001} (\bibinfo {year} {2010})}\BibitemShut {NoStop}%
		\bibitem [{\citenamefont {Oreg}\ \emph {et~al.}(2010)\citenamefont {Oreg},
			\citenamefont {Refael},\ and\ \citenamefont {von Oppen}}]{RN575}%
		\BibitemOpen
		\bibfield  {author} {\bibinfo {author} {\bibfnamefont {Y.}~\bibnamefont
				{Oreg}}, \bibinfo {author} {\bibfnamefont {G.}~\bibnamefont {Refael}},\ and\
			\bibinfo {author} {\bibfnamefont {F.}~\bibnamefont {von Oppen}},\ }\bibfield
		{title} {\bibinfo {title} {{Helical liquids and {Majorana} bound states in
					quantum wires}},\ }\href {https://doi.org/10.1103/PhysRevLett.105.177002}
		{\bibfield  {journal} {\bibinfo  {journal} {Phys. Rev. Lett.}\ }\textbf
			{\bibinfo {volume} {105}},\ \bibinfo {pages} {177002} (\bibinfo {year}
			{2010})}\BibitemShut {NoStop}%
		\bibitem [{\citenamefont {Mourik}\ \emph {et~al.}(2012)\citenamefont {Mourik},
			\citenamefont {Zuo}, \citenamefont {Frolov}, \citenamefont {Plissard},
			\citenamefont {Bakkers},\ and\ \citenamefont {Kouwenhoven}}]{RN577}%
		\BibitemOpen
		\bibfield  {author} {\bibinfo {author} {\bibfnamefont {V.}~\bibnamefont
				{Mourik}}, \bibinfo {author} {\bibfnamefont {K.}~\bibnamefont {Zuo}},
			\bibinfo {author} {\bibfnamefont {S.~M.}\ \bibnamefont {Frolov}}, \bibinfo
			{author} {\bibfnamefont {S.~R.}\ \bibnamefont {Plissard}}, \bibinfo {author}
			{\bibfnamefont {E.~P.}\ \bibnamefont {Bakkers}},\ and\ \bibinfo {author}
			{\bibfnamefont {L.~P.}\ \bibnamefont {Kouwenhoven}},\ }\bibfield  {title}
		{\bibinfo {title} {{Signatures of {Majorana} fermions in hybrid
					superconductor-semiconductor nanowire devices}},\ }\href
		{https://doi.org/10.1126/science.1222360} {\bibfield  {journal} {\bibinfo
				{journal} {Science}\ }\textbf {\bibinfo {volume} {336}},\ \bibinfo {pages}
			{1003} (\bibinfo {year} {2012})}\BibitemShut {NoStop}%
		\bibitem [{\citenamefont {Nadj-Perge}\ \emph {et~al.}(2014)\citenamefont
			{Nadj-Perge}, \citenamefont {Drozdov}, \citenamefont {Li}, \citenamefont
			{Chen}, \citenamefont {Jeon}, \citenamefont {Seo}, \citenamefont {MacDonald},
			\citenamefont {Bernevig},\ and\ \citenamefont {Yazdani}}]{RN578}%
		\BibitemOpen
		\bibfield  {author} {\bibinfo {author} {\bibfnamefont {S.}~\bibnamefont
				{Nadj-Perge}}, \bibinfo {author} {\bibfnamefont {I.~K.}\ \bibnamefont
				{Drozdov}}, \bibinfo {author} {\bibfnamefont {J.}~\bibnamefont {Li}},
			\bibinfo {author} {\bibfnamefont {H.}~\bibnamefont {Chen}}, \bibinfo {author}
			{\bibfnamefont {S.}~\bibnamefont {Jeon}}, \bibinfo {author} {\bibfnamefont
				{J.}~\bibnamefont {Seo}}, \bibinfo {author} {\bibfnamefont {A.~H.}\
				\bibnamefont {MacDonald}}, \bibinfo {author} {\bibfnamefont {B.~A.}\
				\bibnamefont {Bernevig}},\ and\ \bibinfo {author} {\bibfnamefont
				{A.}~\bibnamefont {Yazdani}},\ }\bibfield  {title} {\bibinfo {title}
			{{Observation of {Majorana} fermions in ferromagnetic atomic chains on a
					superconductor}},\ }\href {https://doi.org/10.1126/science.1259327}
		{\bibfield  {journal} {\bibinfo  {journal} {Science}\ }\textbf {\bibinfo
				{volume} {346}},\ \bibinfo {pages} {602} (\bibinfo {year}
			{2014})}\BibitemShut {NoStop}%
		\bibitem [{\citenamefont {Xu}\ \emph {et~al.}(2016)\citenamefont {Xu},
			\citenamefont {Lian}, \citenamefont {Tang}, \citenamefont {Qi},\ and\
			\citenamefont {Zhang}}]{RN533}%
		\BibitemOpen
		\bibfield  {author} {\bibinfo {author} {\bibfnamefont {G.}~\bibnamefont
				{Xu}}, \bibinfo {author} {\bibfnamefont {B.}~\bibnamefont {Lian}}, \bibinfo
			{author} {\bibfnamefont {P.}~\bibnamefont {Tang}}, \bibinfo {author}
			{\bibfnamefont {X.~L.}\ \bibnamefont {Qi}},\ and\ \bibinfo {author}
			{\bibfnamefont {S.~C.}\ \bibnamefont {Zhang}},\ }\bibfield  {title} {\bibinfo
			{title} {{Topological Superconductivity on the Surface of Fe-Based
					Superconductors}},\ }\href {https://doi.org/10.1103/PhysRevLett.117.047001}
		{\bibfield  {journal} {\bibinfo  {journal} {Phys. Rev. Lett.}\ }\textbf
			{\bibinfo {volume} {117}},\ \bibinfo {pages} {047001} (\bibinfo {year}
			{2016})}\BibitemShut {NoStop}%
		\bibitem [{\citenamefont {Zhu}\ \emph {et~al.}(2020)\citenamefont {Zhu},
			\citenamefont {Kong}, \citenamefont {Cao}, \citenamefont {Chen},
			\citenamefont {Papaj}, \citenamefont {Du}, \citenamefont {Xing},
			\citenamefont {Liu}, \citenamefont {Wang}, \citenamefont {Shen},
			\citenamefont {Yang}, \citenamefont {Schneeloch}, \citenamefont {Zhong},
			\citenamefont {Gu}, \citenamefont {Fu}, \citenamefont {Zhang}, \citenamefont
			{Ding},\ and\ \citenamefont {Gao}}]{RN592}%
		\BibitemOpen
		\bibfield  {author} {\bibinfo {author} {\bibfnamefont {S.~Y.}\ \bibnamefont
				{Zhu}}, \bibinfo {author} {\bibfnamefont {L.~Y.}\ \bibnamefont {Kong}},
			\bibinfo {author} {\bibfnamefont {L.}~\bibnamefont {Cao}}, \bibinfo {author}
			{\bibfnamefont {H.}~\bibnamefont {Chen}}, \bibinfo {author} {\bibfnamefont
				{M.}~\bibnamefont {Papaj}}, \bibinfo {author} {\bibfnamefont {S.~X.}\
				\bibnamefont {Du}}, \bibinfo {author} {\bibfnamefont {Y.~Q.}\ \bibnamefont
				{Xing}}, \bibinfo {author} {\bibfnamefont {W.~Y.}\ \bibnamefont {Liu}},
			\bibinfo {author} {\bibfnamefont {D.~F.}\ \bibnamefont {Wang}}, \bibinfo
			{author} {\bibfnamefont {C.~M.}\ \bibnamefont {Shen}}, \bibinfo {author}
			{\bibfnamefont {F.~Z.}\ \bibnamefont {Yang}}, \bibinfo {author}
			{\bibfnamefont {J.}~\bibnamefont {Schneeloch}}, \bibinfo {author}
			{\bibfnamefont {R.~D.}\ \bibnamefont {Zhong}}, \bibinfo {author}
			{\bibfnamefont {G.~D.}\ \bibnamefont {Gu}}, \bibinfo {author} {\bibfnamefont
				{L.}~\bibnamefont {Fu}}, \bibinfo {author} {\bibfnamefont {Y.~Y.}\
				\bibnamefont {Zhang}}, \bibinfo {author} {\bibfnamefont {H.}~\bibnamefont
				{Ding}},\ and\ \bibinfo {author} {\bibfnamefont {H.~J.}\ \bibnamefont
				{Gao}},\ }\bibfield  {title} {\bibinfo {title} {{Nearly quantized conductance
					plateau of vortex zero mode in an iron-based superconductor}},\ }\href
		{https://doi.org/10.1126/science.aax0274} {\bibfield  {journal} {\bibinfo
				{journal} {Science}\ }\textbf {\bibinfo {volume} {367}},\ \bibinfo {pages}
			{189} (\bibinfo {year} {2020})}\BibitemShut {NoStop}%
		\bibitem [{\citenamefont {Wang}\ \emph {et~al.}(2018)\citenamefont {Wang},
			\citenamefont {Kong}, \citenamefont {Fan}, \citenamefont {Chen},
			\citenamefont {Zhu}, \citenamefont {Liu}, \citenamefont {Cao}, \citenamefont
			{Sun}, \citenamefont {Du}, \citenamefont {Schneeloch}, \citenamefont {Zhong},
			\citenamefont {Gu}, \citenamefont {Fu}, \citenamefont {Ding},\ and\
			\citenamefont {Gao}}]{RN594}%
		\BibitemOpen
		\bibfield  {author} {\bibinfo {author} {\bibfnamefont {D.}~\bibnamefont
				{Wang}}, \bibinfo {author} {\bibfnamefont {L.}~\bibnamefont {Kong}}, \bibinfo
			{author} {\bibfnamefont {P.}~\bibnamefont {Fan}}, \bibinfo {author}
			{\bibfnamefont {H.}~\bibnamefont {Chen}}, \bibinfo {author} {\bibfnamefont
				{S.}~\bibnamefont {Zhu}}, \bibinfo {author} {\bibfnamefont {W.}~\bibnamefont
				{Liu}}, \bibinfo {author} {\bibfnamefont {L.}~\bibnamefont {Cao}}, \bibinfo
			{author} {\bibfnamefont {Y.}~\bibnamefont {Sun}}, \bibinfo {author}
			{\bibfnamefont {S.}~\bibnamefont {Du}}, \bibinfo {author} {\bibfnamefont
				{J.}~\bibnamefont {Schneeloch}}, \bibinfo {author} {\bibfnamefont
				{R.}~\bibnamefont {Zhong}}, \bibinfo {author} {\bibfnamefont
				{G.}~\bibnamefont {Gu}}, \bibinfo {author} {\bibfnamefont {L.}~\bibnamefont
				{Fu}}, \bibinfo {author} {\bibfnamefont {H.}~\bibnamefont {Ding}},\ and\
			\bibinfo {author} {\bibfnamefont {H.~J.}\ \bibnamefont {Gao}},\ }\bibfield
		{title} {\bibinfo {title} {{Evidence for {Majorana} bound states in an
					iron-based superconductor}},\ }\href
		{https://doi.org/10.1126/science.aao1797} {\bibfield  {journal} {\bibinfo
				{journal} {Science}\ }\textbf {\bibinfo {volume} {362}},\ \bibinfo {pages}
			{333} (\bibinfo {year} {2018})}\BibitemShut {NoStop}%
		\bibitem [{\citenamefont {Zhang}\ \emph {et~al.}(2018)\citenamefont {Zhang},
			\citenamefont {Yaji}, \citenamefont {Hashimoto}, \citenamefont {Ota},
			\citenamefont {Kondo}, \citenamefont {Okazaki}, \citenamefont {Wang},
			\citenamefont {Wen}, \citenamefont {Gu}, \citenamefont {Ding},\ and\
			\citenamefont {Shin}}]{RN536}%
		\BibitemOpen
		\bibfield  {author} {\bibinfo {author} {\bibfnamefont {P.}~\bibnamefont
				{Zhang}}, \bibinfo {author} {\bibfnamefont {K.}~\bibnamefont {Yaji}},
			\bibinfo {author} {\bibfnamefont {T.}~\bibnamefont {Hashimoto}}, \bibinfo
			{author} {\bibfnamefont {Y.}~\bibnamefont {Ota}}, \bibinfo {author}
			{\bibfnamefont {T.}~\bibnamefont {Kondo}}, \bibinfo {author} {\bibfnamefont
				{K.}~\bibnamefont {Okazaki}}, \bibinfo {author} {\bibfnamefont
				{Z.}~\bibnamefont {Wang}}, \bibinfo {author} {\bibfnamefont {J.}~\bibnamefont
				{Wen}}, \bibinfo {author} {\bibfnamefont {G.~D.}\ \bibnamefont {Gu}},
			\bibinfo {author} {\bibfnamefont {H.}~\bibnamefont {Ding}},\ and\ \bibinfo
			{author} {\bibfnamefont {S.}~\bibnamefont {Shin}},\ }\bibfield  {title}
		{\bibinfo {title} {{Observation of topological superconductivity on the
					surface of an iron-based superconductor}},\ }\href
		{https://doi.org/10.1126/science.aan4596} {\bibfield  {journal} {\bibinfo
				{journal} {Science}\ }\textbf {\bibinfo {volume} {360}},\ \bibinfo {pages}
			{182} (\bibinfo {year} {2018})}\BibitemShut {NoStop}%
		\bibitem [{\citenamefont {Liu}\ \emph {et~al.}(2020)\citenamefont {Liu},
			\citenamefont {Cao}, \citenamefont {Zhu}, \citenamefont {Kong}, \citenamefont
			{Wang}, \citenamefont {Papaj}, \citenamefont {Zhang}, \citenamefont {Liu},
			\citenamefont {Chen}, \citenamefont {Li}, \citenamefont {Yang}, \citenamefont
			{Kondo}, \citenamefont {Du}, \citenamefont {Cao}, \citenamefont {Shin},
			\citenamefont {Fu}, \citenamefont {Yin}, \citenamefont {Gao},\ and\
			\citenamefont {Ding}}]{RN548}%
		\BibitemOpen
		\bibfield  {author} {\bibinfo {author} {\bibfnamefont {W.}~\bibnamefont
				{Liu}}, \bibinfo {author} {\bibfnamefont {L.}~\bibnamefont {Cao}}, \bibinfo
			{author} {\bibfnamefont {S.}~\bibnamefont {Zhu}}, \bibinfo {author}
			{\bibfnamefont {L.}~\bibnamefont {Kong}}, \bibinfo {author} {\bibfnamefont
				{G.}~\bibnamefont {Wang}}, \bibinfo {author} {\bibfnamefont {M.}~\bibnamefont
				{Papaj}}, \bibinfo {author} {\bibfnamefont {P.}~\bibnamefont {Zhang}},
			\bibinfo {author} {\bibfnamefont {Y.~B.}\ \bibnamefont {Liu}}, \bibinfo
			{author} {\bibfnamefont {H.}~\bibnamefont {Chen}}, \bibinfo {author}
			{\bibfnamefont {G.}~\bibnamefont {Li}}, \bibinfo {author} {\bibfnamefont
				{F.}~\bibnamefont {Yang}}, \bibinfo {author} {\bibfnamefont {T.}~\bibnamefont
				{Kondo}}, \bibinfo {author} {\bibfnamefont {S.}~\bibnamefont {Du}}, \bibinfo
			{author} {\bibfnamefont {G.~H.}\ \bibnamefont {Cao}}, \bibinfo {author}
			{\bibfnamefont {S.}~\bibnamefont {Shin}}, \bibinfo {author} {\bibfnamefont
				{L.}~\bibnamefont {Fu}}, \bibinfo {author} {\bibfnamefont {Z.}~\bibnamefont
				{Yin}}, \bibinfo {author} {\bibfnamefont {H.~J.}\ \bibnamefont {Gao}},\ and\
			\bibinfo {author} {\bibfnamefont {H.}~\bibnamefont {Ding}},\ }\bibfield
		{title} {\bibinfo {title} {{{A new {Majorana} platform in an {Fe--As} bilayer
						superconductor}}},\ }\href {https://doi.org/10.1038/s41467-020-19487-1}
		{\bibfield  {journal} {\bibinfo  {journal} {Nat. Commun.}\ }\textbf {\bibinfo
				{volume} {11}},\ \bibinfo {pages} {5688} (\bibinfo {year}
			{2020})}\BibitemShut {NoStop}%
		\bibitem [{\citenamefont {Kong}\ \emph {et~al.}(2021)\citenamefont {Kong},
			\citenamefont {Cao}, \citenamefont {Zhu}, \citenamefont {Papaj},
			\citenamefont {Dai}, \citenamefont {Li}, \citenamefont {Fan}, \citenamefont
			{Liu}, \citenamefont {Yang}, \citenamefont {Wang}, \citenamefont {Du},
			\citenamefont {Jin}, \citenamefont {Fu}, \citenamefont {Gao},\ and\
			\citenamefont {Ding}}]{RN549}%
		\BibitemOpen
		\bibfield  {author} {\bibinfo {author} {\bibfnamefont {L.}~\bibnamefont
				{Kong}}, \bibinfo {author} {\bibfnamefont {L.}~\bibnamefont {Cao}}, \bibinfo
			{author} {\bibfnamefont {S.}~\bibnamefont {Zhu}}, \bibinfo {author}
			{\bibfnamefont {M.}~\bibnamefont {Papaj}}, \bibinfo {author} {\bibfnamefont
				{G.}~\bibnamefont {Dai}}, \bibinfo {author} {\bibfnamefont {G.}~\bibnamefont
				{Li}}, \bibinfo {author} {\bibfnamefont {P.}~\bibnamefont {Fan}}, \bibinfo
			{author} {\bibfnamefont {W.}~\bibnamefont {Liu}}, \bibinfo {author}
			{\bibfnamefont {F.}~\bibnamefont {Yang}}, \bibinfo {author} {\bibfnamefont
				{X.}~\bibnamefont {Wang}}, \bibinfo {author} {\bibfnamefont {S.}~\bibnamefont
				{Du}}, \bibinfo {author} {\bibfnamefont {C.}~\bibnamefont {Jin}}, \bibinfo
			{author} {\bibfnamefont {L.}~\bibnamefont {Fu}}, \bibinfo {author}
			{\bibfnamefont {H.~J.}\ \bibnamefont {Gao}},\ and\ \bibinfo {author}
			{\bibfnamefont {H.}~\bibnamefont {Ding}},\ }\bibfield  {title} {\bibinfo
			{title} {{{{Majorana} zero modes in impurity-assisted vortex of {LiFeAs}
						superconductor}}},\ }\href {https://doi.org/10.1038/s41467-021-24372-6}
		{\bibfield  {journal} {\bibinfo  {journal} {Nat. Commun.}\ }\textbf {\bibinfo
				{volume} {12}},\ \bibinfo {pages} {4146} (\bibinfo {year}
			{2021})}\BibitemShut {NoStop}%
		\bibitem [{\citenamefont {Li}\ \emph {et~al.}(2022)\citenamefont {Li},
			\citenamefont {Li}, \citenamefont {Cao}, \citenamefont {Zhou}, \citenamefont
			{Wang}, \citenamefont {Jin}, \citenamefont {Chiu}, \citenamefont {Pennycook},
			\citenamefont {Wang},\ and\ \citenamefont {Gao}}]{RN604}%
		\BibitemOpen
		\bibfield  {author} {\bibinfo {author} {\bibfnamefont {M.}~\bibnamefont
				{Li}}, \bibinfo {author} {\bibfnamefont {G.}~\bibnamefont {Li}}, \bibinfo
			{author} {\bibfnamefont {L.}~\bibnamefont {Cao}}, \bibinfo {author}
			{\bibfnamefont {X.}~\bibnamefont {Zhou}}, \bibinfo {author} {\bibfnamefont
				{X.}~\bibnamefont {Wang}}, \bibinfo {author} {\bibfnamefont {C.}~\bibnamefont
				{Jin}}, \bibinfo {author} {\bibfnamefont {C.~K.}\ \bibnamefont {Chiu}},
			\bibinfo {author} {\bibfnamefont {S.~J.}\ \bibnamefont {Pennycook}}, \bibinfo
			{author} {\bibfnamefont {Z.}~\bibnamefont {Wang}},\ and\ \bibinfo {author}
			{\bibfnamefont {H.~J.}\ \bibnamefont {Gao}},\ }\bibfield  {title} {\bibinfo
			{title} {{Ordered and tunable {Majorana}-zero-mode lattice in naturally
					strained {LiFeAs}}},\ }\href {https://doi.org/10.1038/s41586-022-04744-8}
		{\bibfield  {journal} {\bibinfo  {journal} {Nature}\ }\textbf {\bibinfo
				{volume} {606}},\ \bibinfo {pages} {890} (\bibinfo {year}
			{2022})}\BibitemShut {NoStop}%
		\bibitem [{\citenamefont {Tang}\ \emph {et~al.}(2022)\citenamefont {Tang},
			\citenamefont {Ono}, \citenamefont {Wan},\ and\ \citenamefont
			{Watanabe}}]{RN478}%
		\BibitemOpen
		\bibfield  {author} {\bibinfo {author} {\bibfnamefont {F.}~\bibnamefont
				{Tang}}, \bibinfo {author} {\bibfnamefont {S.}~\bibnamefont {Ono}}, \bibinfo
			{author} {\bibfnamefont {X.}~\bibnamefont {Wan}},\ and\ \bibinfo {author}
			{\bibfnamefont {H.}~\bibnamefont {Watanabe}},\ }\bibfield  {title} {\bibinfo
			{title} {{High-Throughput Investigations of Topological and Nodal
					Superconductors}},\ }\href {https://doi.org/10.1103/PhysRevLett.129.027001}
		{\bibfield  {journal} {\bibinfo  {journal} {Phys. Rev. Lett.}\ }\textbf
			{\bibinfo {volume} {129}},\ \bibinfo {pages} {027001} (\bibinfo {year}
			{2022})}\BibitemShut {NoStop}%
		\bibitem [{\citenamefont {Yi}\ \emph {et~al.}(2023{\natexlab{a}})\citenamefont
			{Yi}, \citenamefont {Liao}, \citenamefont {You}, \citenamefont {Gu},\ and\
			\citenamefont {Su}}]{RN1006}%
		\BibitemOpen
		\bibfield  {author} {\bibinfo {author} {\bibfnamefont {X.~W.}\ \bibnamefont
				{Yi}}, \bibinfo {author} {\bibfnamefont {Z.~W.}\ \bibnamefont {Liao}},
			\bibinfo {author} {\bibfnamefont {J.~Y.}\ \bibnamefont {You}}, \bibinfo
			{author} {\bibfnamefont {B.}~\bibnamefont {Gu}},\ and\ \bibinfo {author}
			{\bibfnamefont {G.}~\bibnamefont {Su}},\ }\bibfield  {title} {\bibinfo
			{title} {{Topological superconductivity and large spin {Hall} effect in the
					kagome family {Ti$_6$X$_4$ (X = Bi, Sb, Pb, Tl, and In)}}},\ }\href
		{https://doi.org/10.1016/j.isci.2022.105813} {\bibfield  {journal} {\bibinfo
				{journal} {iScience}\ }\textbf {\bibinfo {volume} {26}},\ \bibinfo {pages}
			{105813} (\bibinfo {year} {2023}{\natexlab{a}})}\BibitemShut {NoStop}%
		\bibitem [{\citenamefont {You}\ \emph {et~al.}(2022)\citenamefont {You},
			\citenamefont {Gu}, \citenamefont {Su},\ and\ \citenamefont {Feng}}]{RN514}%
		\BibitemOpen
		\bibfield  {author} {\bibinfo {author} {\bibfnamefont {J.~Y.}\ \bibnamefont
				{You}}, \bibinfo {author} {\bibfnamefont {B.}~\bibnamefont {Gu}}, \bibinfo
			{author} {\bibfnamefont {G.}~\bibnamefont {Su}},\ and\ \bibinfo {author}
			{\bibfnamefont {Y.~P.}\ \bibnamefont {Feng}},\ }\bibfield  {title} {\bibinfo
			{title} {{Emergent Kagome Electrides}},\ }\href
		{https://doi.org/10.1021/jacs.2c00177} {\bibfield  {journal} {\bibinfo
				{journal} {J. Am. Chem. Soc.}\ }\textbf {\bibinfo {volume} {144}},\ \bibinfo
			{pages} {5527} (\bibinfo {year} {2022})}\BibitemShut {NoStop}%
		\bibitem [{\citenamefont {Syozi}(1951)}]{RN1235}%
		\BibitemOpen
		\bibfield  {author} {\bibinfo {author} {\bibfnamefont {I.}~\bibnamefont
				{Syozi}},\ }\bibfield  {title} {\bibinfo {title} {{Statistics of Kagome
					Lattice}},\ }\href {https://doi.org/10.1143/ptp/6.3.306} {\bibfield
			{journal} {\bibinfo  {journal} {Prog. Theor. Phys.}\ }\textbf {\bibinfo
				{volume} {6}},\ \bibinfo {pages} {306} (\bibinfo {year} {1951})}\BibitemShut
		{NoStop}%
		\bibitem [{\citenamefont {Jiang}\ \emph {et~al.}(2023)\citenamefont {Jiang},
			\citenamefont {Wu}, \citenamefont {Yin}, \citenamefont {Wang}, \citenamefont
			{Hasan}, \citenamefont {Wilson}, \citenamefont {Chen},\ and\ \citenamefont
			{Hu}}]{RN1043}%
		\BibitemOpen
		\bibfield  {author} {\bibinfo {author} {\bibfnamefont {K.}~\bibnamefont
				{Jiang}}, \bibinfo {author} {\bibfnamefont {T.}~\bibnamefont {Wu}}, \bibinfo
			{author} {\bibfnamefont {J.~X.}\ \bibnamefont {Yin}}, \bibinfo {author}
			{\bibfnamefont {Z.}~\bibnamefont {Wang}}, \bibinfo {author} {\bibfnamefont
				{M.~Z.}\ \bibnamefont {Hasan}}, \bibinfo {author} {\bibfnamefont {S.~D.}\
				\bibnamefont {Wilson}}, \bibinfo {author} {\bibfnamefont {X.}~\bibnamefont
				{Chen}},\ and\ \bibinfo {author} {\bibfnamefont {J.}~\bibnamefont {Hu}},\
		}\bibfield  {title} {\bibinfo {title} {{Kagome superconductors {A}V$_3$Sb$_5$
					({A} = K, Rb, Cs)}},\ }\href {https://doi.org/10.1093/nsr/nwac199} {\bibfield
			{journal} {\bibinfo  {journal} {Natl. Sci. Rev.}\ }\textbf {\bibinfo
				{volume} {10}},\ \bibinfo {pages} {nwac199} (\bibinfo {year}
			{2023})}\BibitemShut {NoStop}%
		\bibitem [{\citenamefont {Yin}\ \emph {et~al.}(2022)\citenamefont {Yin},
			\citenamefont {Lian},\ and\ \citenamefont {Hasan}}]{RN976}%
		\BibitemOpen
		\bibfield  {author} {\bibinfo {author} {\bibfnamefont {J.-X.}\ \bibnamefont
				{Yin}}, \bibinfo {author} {\bibfnamefont {B.}~\bibnamefont {Lian}},\ and\
			\bibinfo {author} {\bibfnamefont {M.~Z.}\ \bibnamefont {Hasan}},\ }\bibfield
		{title} {\bibinfo {title} {{Topological kagome magnets and
					superconductors}},\ }\href {https://doi.org/10.1038/s41586-022-05516-0}
		{\bibfield  {journal} {\bibinfo  {journal} {Nature}\ }\textbf {\bibinfo
				{volume} {612}},\ \bibinfo {pages} {647} (\bibinfo {year}
			{2022})}\BibitemShut {NoStop}%
		\bibitem [{\citenamefont {Baidya}\ \emph {et~al.}(2020)\citenamefont {Baidya},
			\citenamefont {Mallik}, \citenamefont {Bhattacharjee},\ and\ \citenamefont
			{Saha-Dasgupta}}]{RN253}%
		\BibitemOpen
		\bibfield  {author} {\bibinfo {author} {\bibfnamefont {S.}~\bibnamefont
				{Baidya}}, \bibinfo {author} {\bibfnamefont {A.~V.}\ \bibnamefont {Mallik}},
			\bibinfo {author} {\bibfnamefont {S.}~\bibnamefont {Bhattacharjee}},\ and\
			\bibinfo {author} {\bibfnamefont {T.}~\bibnamefont {Saha-Dasgupta}},\
		}\bibfield  {title} {\bibinfo {title} {{Interplay of Magnetism and
					Topological Superconductivity in Bilayer Kagome Metals}},\ }\href
		{https://doi.org/10.1103/PhysRevLett.125.026401} {\bibfield  {journal}
			{\bibinfo  {journal} {Phys. Rev. Lett.}\ }\textbf {\bibinfo {volume} {125}},\
			\bibinfo {pages} {026401} (\bibinfo {year} {2020})}\BibitemShut {NoStop}%
		\bibitem [{\citenamefont {Zeng}\ \emph {et~al.}(2023)\citenamefont {Zeng},
			\citenamefont {Li}, \citenamefont {Yang}, \citenamefont {Xu},\ and\
			\citenamefont {Wang}}]{RN1374}%
		\BibitemOpen
		\bibfield  {author} {\bibinfo {author} {\bibfnamefont {J.}~\bibnamefont
				{Zeng}}, \bibinfo {author} {\bibfnamefont {Q.}~\bibnamefont {Li}}, \bibinfo
			{author} {\bibfnamefont {X.}~\bibnamefont {Yang}}, \bibinfo {author}
			{\bibfnamefont {D.-H.}\ \bibnamefont {Xu}},\ and\ \bibinfo {author}
			{\bibfnamefont {R.}~\bibnamefont {Wang}},\ }\bibfield  {title} {\bibinfo
			{title} {{Chiral-Flux-Phase-Based Topological Superconductivity in Kagome
					Systems with Mixed Edge Chiralities}},\ }\href
		{https://doi.org/10.1103/PhysRevLett.131.086601} {\bibfield  {journal}
			{\bibinfo  {journal} {Phys. Rev. Lett.}\ }\textbf {\bibinfo {volume} {131}},\
			\bibinfo {pages} {086601} (\bibinfo {year} {2023})}\BibitemShut {NoStop}%
		\bibitem [{\citenamefont {Hutter}\ \emph {et~al.}(2015)\citenamefont {Hutter},
			\citenamefont {Wootton},\ and\ \citenamefont {Loss}}]{RN1220}%
		\BibitemOpen
		\bibfield  {author} {\bibinfo {author} {\bibfnamefont {A.}~\bibnamefont
				{Hutter}}, \bibinfo {author} {\bibfnamefont {J.~R.}\ \bibnamefont
				{Wootton}},\ and\ \bibinfo {author} {\bibfnamefont {D.}~\bibnamefont
				{Loss}},\ }\bibfield  {title} {\bibinfo {title} {{Parafermions in a Kagome
					Lattice of Qubits for Topological Quantum Computation}},\ }\href
		{https://doi.org/10.1103/PhysRevX.5.041040} {\bibfield  {journal} {\bibinfo
				{journal} {Phys. Rev. X}\ }\textbf {\bibinfo {volume} {5}},\ \bibinfo {pages}
			{041040} (\bibinfo {year} {2015})}\BibitemShut {NoStop}%
		\bibitem [{\citenamefont {Wilson}\ and\ \citenamefont {Ortiz}(2024)}]{RN1660}%
		\BibitemOpen
		\bibfield  {author} {\bibinfo {author} {\bibfnamefont {S.~D.}\ \bibnamefont
				{Wilson}}\ and\ \bibinfo {author} {\bibfnamefont {B.~R.}\ \bibnamefont
				{Ortiz}},\ }\bibfield  {title} {\bibinfo {title} {{{A}{V}$_{3}${Sb}$_{5}$
					kagome superconductors}},\ }\href
		{https://doi.org/10.1038/s41578-024-00677-y} {\bibfield  {journal} {\bibinfo
				{journal} {Nat. Rev. Mater.}\ }\textbf {\bibinfo {volume} {9}},\ \bibinfo
			{pages} {420} (\bibinfo {year} {2024})}\BibitemShut {NoStop}%
		\bibitem [{\citenamefont {Neupert}\ \emph {et~al.}(2022)\citenamefont
			{Neupert}, \citenamefont {Denner}, \citenamefont {Yin}, \citenamefont
			{Thomale},\ and\ \citenamefont {Hasan}}]{RN82}%
		\BibitemOpen
		\bibfield  {author} {\bibinfo {author} {\bibfnamefont {T.}~\bibnamefont
				{Neupert}}, \bibinfo {author} {\bibfnamefont {M.~M.}\ \bibnamefont {Denner}},
			\bibinfo {author} {\bibfnamefont {J.-X.}\ \bibnamefont {Yin}}, \bibinfo
			{author} {\bibfnamefont {R.}~\bibnamefont {Thomale}},\ and\ \bibinfo {author}
			{\bibfnamefont {M.~Z.}\ \bibnamefont {Hasan}},\ }\bibfield  {title} {\bibinfo
			{title} {{Charge order and superconductivity in kagome materials}},\ }\href
		{https://doi.org/10.1038/s41567-021-01404-y} {\bibfield  {journal} {\bibinfo
				{journal} {Nat. Phys.}\ }\textbf {\bibinfo {volume} {18}},\ \bibinfo {pages}
			{137} (\bibinfo {year} {2022})}\BibitemShut {NoStop}%
		\bibitem [{\citenamefont {Ortiz}\ \emph {et~al.}(2019)\citenamefont {Ortiz},
			\citenamefont {Gomes}, \citenamefont {Morey}, \citenamefont {Winiarski},
			\citenamefont {Bordelon}, \citenamefont {Mangum}, \citenamefont {Oswald},
			\citenamefont {Rodriguez-Rivera}, \citenamefont {Neilson}, \citenamefont
			{Wilson}, \citenamefont {Ertekin}, \citenamefont {McQueen},\ and\
			\citenamefont {Toberer}}]{RN5}%
		\BibitemOpen
		\bibfield  {author} {\bibinfo {author} {\bibfnamefont {B.~R.}\ \bibnamefont
				{Ortiz}}, \bibinfo {author} {\bibfnamefont {L.~C.}\ \bibnamefont {Gomes}},
			\bibinfo {author} {\bibfnamefont {J.~R.}\ \bibnamefont {Morey}}, \bibinfo
			{author} {\bibfnamefont {M.}~\bibnamefont {Winiarski}}, \bibinfo {author}
			{\bibfnamefont {M.}~\bibnamefont {Bordelon}}, \bibinfo {author}
			{\bibfnamefont {J.~S.}\ \bibnamefont {Mangum}}, \bibinfo {author}
			{\bibfnamefont {I.~W.~H.}\ \bibnamefont {Oswald}}, \bibinfo {author}
			{\bibfnamefont {J.~A.}\ \bibnamefont {Rodriguez-Rivera}}, \bibinfo {author}
			{\bibfnamefont {J.~R.}\ \bibnamefont {Neilson}}, \bibinfo {author}
			{\bibfnamefont {S.~D.}\ \bibnamefont {Wilson}}, \bibinfo {author}
			{\bibfnamefont {E.}~\bibnamefont {Ertekin}}, \bibinfo {author} {\bibfnamefont
				{T.~M.}\ \bibnamefont {McQueen}},\ and\ \bibinfo {author} {\bibfnamefont
				{E.~S.}\ \bibnamefont {Toberer}},\ }\bibfield  {title} {\bibinfo {title}
			{{New kagome prototype materials: discovery of {KV$_{3}$Sb$_{5}$},
					{RbV$_{3}$Sb$_{5}$}, and {{CsV$_{3}$Sb$_{5}$}}}},\ }\href
		{https://doi.org/10.1103/PhysRevMaterials.3.094407} {\bibfield  {journal}
			{\bibinfo  {journal} {Phys. Rev. Mater.}\ }\textbf {\bibinfo {volume} {3}},\
			\bibinfo {pages} {094407} (\bibinfo {year} {2019})}\BibitemShut {NoStop}%
		\bibitem [{\citenamefont {Hu}\ \emph {et~al.}(2022)\citenamefont {Hu},
			\citenamefont {Teicher}, \citenamefont {Ortiz}, \citenamefont {Luo},
			\citenamefont {Peng}, \citenamefont {Huai}, \citenamefont {Ma}, \citenamefont
			{Plumb}, \citenamefont {Wilson}, \citenamefont {He},\ and\ \citenamefont
			{Shi}}]{RN461}%
		\BibitemOpen
		\bibfield  {author} {\bibinfo {author} {\bibfnamefont {Y.}~\bibnamefont
				{Hu}}, \bibinfo {author} {\bibfnamefont {S.~M.~L.}\ \bibnamefont {Teicher}},
			\bibinfo {author} {\bibfnamefont {B.~R.}\ \bibnamefont {Ortiz}}, \bibinfo
			{author} {\bibfnamefont {Y.}~\bibnamefont {Luo}}, \bibinfo {author}
			{\bibfnamefont {S.}~\bibnamefont {Peng}}, \bibinfo {author} {\bibfnamefont
				{L.}~\bibnamefont {Huai}}, \bibinfo {author} {\bibfnamefont {J.}~\bibnamefont
				{Ma}}, \bibinfo {author} {\bibfnamefont {N.~C.}\ \bibnamefont {Plumb}},
			\bibinfo {author} {\bibfnamefont {S.~D.}\ \bibnamefont {Wilson}}, \bibinfo
			{author} {\bibfnamefont {J.}~\bibnamefont {He}},\ and\ \bibinfo {author}
			{\bibfnamefont {M.}~\bibnamefont {Shi}},\ }\bibfield  {title} {\bibinfo
			{title} {{Topological surface states and flat bands in the kagome
					superconductor {{CsV$_3$Sb$_5$}}}},\ }\href
		{https://doi.org/10.1016/j.scib.2021.11.026} {\bibfield  {journal} {\bibinfo
				{journal} {Sci. Bull.}\ }\textbf {\bibinfo {volume} {67}},\ \bibinfo {pages}
			{495} (\bibinfo {year} {2022})}\BibitemShut {NoStop}%
		\bibitem [{\citenamefont {Ortiz}\ \emph {et~al.}(2020)\citenamefont {Ortiz},
			\citenamefont {Teicher}, \citenamefont {Hu}, \citenamefont {Zuo},
			\citenamefont {Sarte}, \citenamefont {Schueller}, \citenamefont {Abeykoon},
			\citenamefont {Krogstad}, \citenamefont {Rosenkranz}, \citenamefont {Osborn},
			\citenamefont {Seshadri}, \citenamefont {Balents}, \citenamefont {He},\ and\
			\citenamefont {Wilson}}]{RN8}%
		\BibitemOpen
		\bibfield  {author} {\bibinfo {author} {\bibfnamefont {B.~R.}\ \bibnamefont
				{Ortiz}}, \bibinfo {author} {\bibfnamefont {S.~M.~L.}\ \bibnamefont
				{Teicher}}, \bibinfo {author} {\bibfnamefont {Y.}~\bibnamefont {Hu}},
			\bibinfo {author} {\bibfnamefont {J.~L.}\ \bibnamefont {Zuo}}, \bibinfo
			{author} {\bibfnamefont {P.~M.}\ \bibnamefont {Sarte}}, \bibinfo {author}
			{\bibfnamefont {E.~C.}\ \bibnamefont {Schueller}}, \bibinfo {author}
			{\bibfnamefont {A.~M.~M.}\ \bibnamefont {Abeykoon}}, \bibinfo {author}
			{\bibfnamefont {M.~J.}\ \bibnamefont {Krogstad}}, \bibinfo {author}
			{\bibfnamefont {S.}~\bibnamefont {Rosenkranz}}, \bibinfo {author}
			{\bibfnamefont {R.}~\bibnamefont {Osborn}}, \bibinfo {author} {\bibfnamefont
				{R.}~\bibnamefont {Seshadri}}, \bibinfo {author} {\bibfnamefont
				{L.}~\bibnamefont {Balents}}, \bibinfo {author} {\bibfnamefont
				{J.}~\bibnamefont {He}},\ and\ \bibinfo {author} {\bibfnamefont {S.~D.}\
				\bibnamefont {Wilson}},\ }\bibfield  {title} {\bibinfo {title}
			{{{{CsV$_{3}$Sb$_{5}$}}: A {Z$_{2}$} Topological Kagome Metal with a
					Superconducting Ground State}},\ }\href
		{https://doi.org/10.1103/PhysRevLett.125.247002} {\bibfield  {journal}
			{\bibinfo  {journal} {Phys. Rev. Lett.}\ }\textbf {\bibinfo {volume} {125}},\
			\bibinfo {pages} {247002} (\bibinfo {year} {2020})}\BibitemShut {NoStop}%
		\bibitem [{\citenamefont {Liang}\ \emph {et~al.}(2021)\citenamefont {Liang},
			\citenamefont {Hou}, \citenamefont {Zhang}, \citenamefont {Ma}, \citenamefont
			{Wu}, \citenamefont {Zhang}, \citenamefont {Yu}, \citenamefont {Ying},
			\citenamefont {Jiang}, \citenamefont {Shan}, \citenamefont {Wang},\ and\
			\citenamefont {Chen}}]{RN1}%
		\BibitemOpen
		\bibfield  {author} {\bibinfo {author} {\bibfnamefont {Z.}~\bibnamefont
				{Liang}}, \bibinfo {author} {\bibfnamefont {X.}~\bibnamefont {Hou}}, \bibinfo
			{author} {\bibfnamefont {F.}~\bibnamefont {Zhang}}, \bibinfo {author}
			{\bibfnamefont {W.}~\bibnamefont {Ma}}, \bibinfo {author} {\bibfnamefont
				{P.}~\bibnamefont {Wu}}, \bibinfo {author} {\bibfnamefont {Z.}~\bibnamefont
				{Zhang}}, \bibinfo {author} {\bibfnamefont {F.}~\bibnamefont {Yu}}, \bibinfo
			{author} {\bibfnamefont {J.~J.}\ \bibnamefont {Ying}}, \bibinfo {author}
			{\bibfnamefont {K.}~\bibnamefont {Jiang}}, \bibinfo {author} {\bibfnamefont
				{L.}~\bibnamefont {Shan}}, \bibinfo {author} {\bibfnamefont {Z.}~\bibnamefont
				{Wang}},\ and\ \bibinfo {author} {\bibfnamefont {X.~H.}\ \bibnamefont
				{Chen}},\ }\bibfield  {title} {\bibinfo {title} {{Three-Dimensional charge
					density wave and surface-dependent vortex-core states in a kagome
					superconductor {CsV$_3$Sb$_5$}}},\ }\href
		{https://doi.org/10.1103/PhysRevX.11.031026} {\bibfield  {journal} {\bibinfo
				{journal} {Phys. Rev. X}\ }\textbf {\bibinfo {volume} {11}},\ \bibinfo
			{pages} {031026} (\bibinfo {year} {2021})}\BibitemShut {NoStop}%
		\bibitem [{\citenamefont {Yi}\ \emph {et~al.}(2022)\citenamefont {Yi},
			\citenamefont {Ma}, \citenamefont {Zhang}, \citenamefont {Liao},
			\citenamefont {You},\ and\ \citenamefont {Su}}]{RN985}%
		\BibitemOpen
		\bibfield  {author} {\bibinfo {author} {\bibfnamefont {X.-W.}\ \bibnamefont
				{Yi}}, \bibinfo {author} {\bibfnamefont {X.-Y.}\ \bibnamefont {Ma}}, \bibinfo
			{author} {\bibfnamefont {Z.}~\bibnamefont {Zhang}}, \bibinfo {author}
			{\bibfnamefont {Z.-W.}\ \bibnamefont {Liao}}, \bibinfo {author}
			{\bibfnamefont {J.-Y.}\ \bibnamefont {You}},\ and\ \bibinfo {author}
			{\bibfnamefont {G.}~\bibnamefont {Su}},\ }\bibfield  {title} {\bibinfo
			{title} {{Large kagome family candidates with topological superconductivity
					and charge density waves}},\ }\href
		{https://doi.org/10.1103/PhysRevB.106.L220505} {\bibfield  {journal}
			{\bibinfo  {journal} {Phys. Rev. B}\ }\textbf {\bibinfo {volume} {106}},\
			\bibinfo {pages} {L220505} (\bibinfo {year} {2022})}\BibitemShut {NoStop}%
		\bibitem [{\citenamefont {Yang}\ \emph {et~al.}(2024)\citenamefont {Yang},
			\citenamefont {Ye}, \citenamefont {Zhao}, \citenamefont {Liu}, \citenamefont
			{Yi}, \citenamefont {Zhang}, \citenamefont {Xiao}, \citenamefont {Shi},
			\citenamefont {You}, \citenamefont {Huang}, \citenamefont {Wang},
			\citenamefont {Wang}, \citenamefont {Guo}, \citenamefont {Lin}, \citenamefont
			{Shen}, \citenamefont {Zhou}, \citenamefont {Chen}, \citenamefont {Dong},
			\citenamefont {Su}, \citenamefont {Wang},\ and\ \citenamefont {Gao}}]{RN967}%
		\BibitemOpen
		\bibfield  {author} {\bibinfo {author} {\bibfnamefont {H.}~\bibnamefont
				{Yang}}, \bibinfo {author} {\bibfnamefont {Y.}~\bibnamefont {Ye}}, \bibinfo
			{author} {\bibfnamefont {Z.}~\bibnamefont {Zhao}}, \bibinfo {author}
			{\bibfnamefont {J.}~\bibnamefont {Liu}}, \bibinfo {author} {\bibfnamefont
				{X.-W.}\ \bibnamefont {Yi}}, \bibinfo {author} {\bibfnamefont
				{Y.}~\bibnamefont {Zhang}}, \bibinfo {author} {\bibfnamefont
				{H.}~\bibnamefont {Xiao}}, \bibinfo {author} {\bibfnamefont {J.}~\bibnamefont
				{Shi}}, \bibinfo {author} {\bibfnamefont {J.-Y.}\ \bibnamefont {You}},
			\bibinfo {author} {\bibfnamefont {Z.}~\bibnamefont {Huang}}, \bibinfo
			{author} {\bibfnamefont {B.}~\bibnamefont {Wang}}, \bibinfo {author}
			{\bibfnamefont {J.}~\bibnamefont {Wang}}, \bibinfo {author} {\bibfnamefont
				{H.}~\bibnamefont {Guo}}, \bibinfo {author} {\bibfnamefont {X.}~\bibnamefont
				{Lin}}, \bibinfo {author} {\bibfnamefont {C.}~\bibnamefont {Shen}}, \bibinfo
			{author} {\bibfnamefont {W.}~\bibnamefont {Zhou}}, \bibinfo {author}
			{\bibfnamefont {H.}~\bibnamefont {Chen}}, \bibinfo {author} {\bibfnamefont
				{X.}~\bibnamefont {Dong}}, \bibinfo {author} {\bibfnamefont {G.}~\bibnamefont
				{Su}}, \bibinfo {author} {\bibfnamefont {Z.}~\bibnamefont {Wang}},\ and\
			\bibinfo {author} {\bibfnamefont {H.-J.}\ \bibnamefont {Gao}},\ }\bibfield
		{title} {\bibinfo {title} {{Superconductivity and nematic order in a new
					titanium-based kagome metal {Cs}{Ti}$_{3}${Bi}$_{5}$ without charge density
					wave order}},\ }\href {https://doi.org/10.1038/s41467-024-53870-6} {\bibfield
			{journal} {\bibinfo  {journal} {Nat. Commun.}\ }\textbf {\bibinfo {volume}
				{15}},\ \bibinfo {pages} {9626} (\bibinfo {year} {2024})}\BibitemShut
		{NoStop}%
		\bibitem [{\citenamefont {Werhahn}\ \emph {et~al.}(2022)\citenamefont
			{Werhahn}, \citenamefont {Ortiz}, \citenamefont {Hay}, \citenamefont
			{Wilson}, \citenamefont {Seshadri},\ and\ \citenamefont {Johrendt}}]{RN795}%
		\BibitemOpen
		\bibfield  {author} {\bibinfo {author} {\bibfnamefont {D.}~\bibnamefont
				{Werhahn}}, \bibinfo {author} {\bibfnamefont {B.~R.}\ \bibnamefont {Ortiz}},
			\bibinfo {author} {\bibfnamefont {A.~K.}\ \bibnamefont {Hay}}, \bibinfo
			{author} {\bibfnamefont {S.~D.}\ \bibnamefont {Wilson}}, \bibinfo {author}
			{\bibfnamefont {R.}~\bibnamefont {Seshadri}},\ and\ \bibinfo {author}
			{\bibfnamefont {D.}~\bibnamefont {Johrendt}},\ }\bibfield  {title} {\bibinfo
			{title} {{The kagom{\'e} metals {RbTi$_3$Bi$_5$} and {CsTi$_3$Bi$_5$}}},\
		}\href {https://doi.org/10.1515/znb-2022-0125} {\bibfield  {journal}
			{\bibinfo  {journal} {Z. Naturforsch. B}\ }\textbf {\bibinfo {volume} {77}},\
			\bibinfo {pages} {757} (\bibinfo {year} {2022})}\BibitemShut {NoStop}%
		\bibitem [{\citenamefont {Yang}\ \emph {et~al.}(2023)\citenamefont {Yang},
			\citenamefont {Yi}, \citenamefont {Zhao}, \citenamefont {Xie}, \citenamefont
			{Miao}, \citenamefont {Luo}, \citenamefont {Chen}, \citenamefont {Liang},
			\citenamefont {Zhu}, \citenamefont {Ye}, \citenamefont {You}, \citenamefont
			{Gu}, \citenamefont {Zhang}, \citenamefont {Zhang}, \citenamefont {Yang},
			\citenamefont {Wang}, \citenamefont {Peng}, \citenamefont {Mao},
			\citenamefont {Liu}, \citenamefont {Xu}, \citenamefont {Chen}, \citenamefont
			{Yang}, \citenamefont {Su}, \citenamefont {Gao}, \citenamefont {Zhao},\ and\
			\citenamefont {Zhou}}]{RN966}%
		\BibitemOpen
		\bibfield  {author} {\bibinfo {author} {\bibfnamefont {J.}~\bibnamefont
				{Yang}}, \bibinfo {author} {\bibfnamefont {X.}~\bibnamefont {Yi}}, \bibinfo
			{author} {\bibfnamefont {Z.}~\bibnamefont {Zhao}}, \bibinfo {author}
			{\bibfnamefont {Y.}~\bibnamefont {Xie}}, \bibinfo {author} {\bibfnamefont
				{T.}~\bibnamefont {Miao}}, \bibinfo {author} {\bibfnamefont {H.}~\bibnamefont
				{Luo}}, \bibinfo {author} {\bibfnamefont {H.}~\bibnamefont {Chen}}, \bibinfo
			{author} {\bibfnamefont {B.}~\bibnamefont {Liang}}, \bibinfo {author}
			{\bibfnamefont {W.}~\bibnamefont {Zhu}}, \bibinfo {author} {\bibfnamefont
				{Y.}~\bibnamefont {Ye}}, \bibinfo {author} {\bibfnamefont {J.~Y.}\
				\bibnamefont {You}}, \bibinfo {author} {\bibfnamefont {B.}~\bibnamefont
				{Gu}}, \bibinfo {author} {\bibfnamefont {S.}~\bibnamefont {Zhang}}, \bibinfo
			{author} {\bibfnamefont {F.}~\bibnamefont {Zhang}}, \bibinfo {author}
			{\bibfnamefont {F.}~\bibnamefont {Yang}}, \bibinfo {author} {\bibfnamefont
				{Z.}~\bibnamefont {Wang}}, \bibinfo {author} {\bibfnamefont {Q.}~\bibnamefont
				{Peng}}, \bibinfo {author} {\bibfnamefont {H.}~\bibnamefont {Mao}}, \bibinfo
			{author} {\bibfnamefont {G.}~\bibnamefont {Liu}}, \bibinfo {author}
			{\bibfnamefont {Z.}~\bibnamefont {Xu}}, \bibinfo {author} {\bibfnamefont
				{H.}~\bibnamefont {Chen}}, \bibinfo {author} {\bibfnamefont {H.}~\bibnamefont
				{Yang}}, \bibinfo {author} {\bibfnamefont {G.}~\bibnamefont {Su}}, \bibinfo
			{author} {\bibfnamefont {H.}~\bibnamefont {Gao}}, \bibinfo {author}
			{\bibfnamefont {L.}~\bibnamefont {Zhao}},\ and\ \bibinfo {author}
			{\bibfnamefont {X.~J.}\ \bibnamefont {Zhou}},\ }\bibfield  {title} {\bibinfo
			{title} {{Observation of flat band, Dirac nodal lines and topological surface
					states in Kagome superconductor {CsTi$_{3}$Bi$_{5}$}}},\ }\href
		{https://doi.org/10.1038/s41467-023-39620-0} {\bibfield  {journal} {\bibinfo
				{journal} {Nat. Commun.}\ }\textbf {\bibinfo {volume} {14}},\ \bibinfo
			{pages} {4089} (\bibinfo {year} {2023})}\BibitemShut {NoStop}%
		\bibitem [{\citenamefont {Yi}\ \emph {et~al.}(2023{\natexlab{b}})\citenamefont
			{Yi}, \citenamefont {Liao}, \citenamefont {You}, \citenamefont {Gu},\ and\
			\citenamefont {Su}}]{RN1284}%
		\BibitemOpen
		\bibfield  {author} {\bibinfo {author} {\bibfnamefont {X.~W.}\ \bibnamefont
				{Yi}}, \bibinfo {author} {\bibfnamefont {Z.~W.}\ \bibnamefont {Liao}},
			\bibinfo {author} {\bibfnamefont {J.~Y.}\ \bibnamefont {You}}, \bibinfo
			{author} {\bibfnamefont {B.}~\bibnamefont {Gu}},\ and\ \bibinfo {author}
			{\bibfnamefont {G.}~\bibnamefont {Su}},\ }\bibfield  {title} {\bibinfo
			{title} {{Superconducting, Topological, and Transport Properties of Kagome
					Metals {CsTi$_{3}$Bi$_{5}$} and {RbTi$_{3}$Bi$_{5}$}}},\ }\href
		{https://doi.org/10.34133/research.0238} {\bibfield  {journal} {\bibinfo
				{journal} {Research}\ }\textbf {\bibinfo {volume} {6}},\ \bibinfo {pages}
			{0238} (\bibinfo {year} {2023}{\natexlab{b}})}\BibitemShut {NoStop}%
		\bibitem [{sup()}]{supp}%
		\BibitemOpen
		\href@noop {} {}\bibinfo {note} {See Supplemental Material [url] for more
			calculated details and results (see also Refs.
			\cite{RN141,RN142,RN143,RN139,RN1116,RN138,RN139,RN128,MorelAnderson1962,RN129,RN144,RN140,RN1284,Uzunok2018LaSn3,RN3070,
				RN967,RN3148, RN890,RN437,
				Turbil1969Fe3Ge,Nial1943Fe3Sn,Nial1947FeSnAlloys,Singh1968Mn3Sn,Ye2024Ni3In,Lyubimtsev2002Ni3Sn,Palenzona1996ScIn,Raghavan2005AlTi,Du2019Ti3Sn,Fernandez1976Fe3Ge,Chen1998Fe3GeL12DO19,Ochiai1984Ni3X,Predel2016PdSn,Cui2021PtSn,Murray1992AlZr,Nial1938CoSn,Watanabe1966FeGe,Richardson1967FeGe,Ehret1933FeSn,Yamaguchi1967FeSn,Meier2020CoSnType,Malaman1976Fe3Sn2,Wang2022V3Sb2,RN327,RN1260}
			therein)}\BibitemShut {NoStop}%
		\bibitem [{\citenamefont {Matthias}\ \emph {et~al.}(1958)\citenamefont
			{Matthias}, \citenamefont {Suhl},\ and\ \citenamefont {Corenzwit}}]{RN1376}%
		\BibitemOpen
		\bibfield  {author} {\bibinfo {author} {\bibfnamefont {B.~T.}\ \bibnamefont
				{Matthias}}, \bibinfo {author} {\bibfnamefont {H.}~\bibnamefont {Suhl}},\
			and\ \bibinfo {author} {\bibfnamefont {E.}~\bibnamefont {Corenzwit}},\
		}\bibfield  {title} {\bibinfo {title} {{Ferromagnetic Superconductors}},\
		}\href {https://doi.org/10.1103/PhysRevLett.1.449} {\bibfield  {journal}
			{\bibinfo  {journal} {Phys. Rev. Lett.}\ }\textbf {\bibinfo {volume} {1}},\
			\bibinfo {pages} {449} (\bibinfo {year} {1958})}\BibitemShut {NoStop}%
		\bibitem [{\citenamefont {Yin}\ \emph {et~al.}(2021)\citenamefont {Yin},
			\citenamefont {Tu}, \citenamefont {Gong}, \citenamefont {Fu}, \citenamefont
			{Yan},\ and\ \citenamefont {Lei}}]{RN64}%
		\BibitemOpen
		\bibfield  {author} {\bibinfo {author} {\bibfnamefont {Q.}~\bibnamefont
				{Yin}}, \bibinfo {author} {\bibfnamefont {Z.}~\bibnamefont {Tu}}, \bibinfo
			{author} {\bibfnamefont {C.}~\bibnamefont {Gong}}, \bibinfo {author}
			{\bibfnamefont {Y.}~\bibnamefont {Fu}}, \bibinfo {author} {\bibfnamefont
				{S.}~\bibnamefont {Yan}},\ and\ \bibinfo {author} {\bibfnamefont
				{H.}~\bibnamefont {Lei}},\ }\bibfield  {title} {\bibinfo {title}
			{{Superconductivity and Normal-State Properties of Kagome Metal
					{RbV$_3$Sb$_5$} Single Crystals}},\ }\href
		{https://doi.org/10.1088/0256-307x/38/3/037403} {\bibfield  {journal}
			{\bibinfo  {journal} {Chin. Phys. Lett.}\ }\textbf {\bibinfo {volume} {38}},\
			\bibinfo {pages} {037403} (\bibinfo {year} {2021})}\BibitemShut {NoStop}%
		\bibitem [{\citenamefont {Ortiz}\ \emph {et~al.}(2021)\citenamefont {Ortiz},
			\citenamefont {Sarte}, \citenamefont {Kenney}, \citenamefont {Graf},
			\citenamefont {Teicher}, \citenamefont {Seshadri},\ and\ \citenamefont
			{Wilson}}]{RN11}%
		\BibitemOpen
		\bibfield  {author} {\bibinfo {author} {\bibfnamefont {B.~R.}\ \bibnamefont
				{Ortiz}}, \bibinfo {author} {\bibfnamefont {P.~M.}\ \bibnamefont {Sarte}},
			\bibinfo {author} {\bibfnamefont {E.~M.}\ \bibnamefont {Kenney}}, \bibinfo
			{author} {\bibfnamefont {M.~J.}\ \bibnamefont {Graf}}, \bibinfo {author}
			{\bibfnamefont {S.~M.~L.}\ \bibnamefont {Teicher}}, \bibinfo {author}
			{\bibfnamefont {R.}~\bibnamefont {Seshadri}},\ and\ \bibinfo {author}
			{\bibfnamefont {S.~D.}\ \bibnamefont {Wilson}},\ }\bibfield  {title}
		{\bibinfo {title} {{Superconductivity in the {Z$_2$} kagome metal
					{KV$_3$Sb$_5$}}},\ }\href {https://doi.org/10.1103/PhysRevMaterials.5.034801}
		{\bibfield  {journal} {\bibinfo  {journal} {Phys. Rev. Mater.}\ }\textbf
			{\bibinfo {volume} {5}},\ \bibinfo {pages} {034801} (\bibinfo {year}
			{2021})}\BibitemShut {NoStop}%
		\bibitem [{\citenamefont {Gui}\ and\ \citenamefont {Cava}(2022)}]{RN485}%
		\BibitemOpen
		\bibfield  {author} {\bibinfo {author} {\bibfnamefont {X.}~\bibnamefont
				{Gui}}\ and\ \bibinfo {author} {\bibfnamefont {R.~J.}\ \bibnamefont {Cava}},\
		}\bibfield  {title} {\bibinfo {title} {{{La}{Ir}$_{3}${Ga}$_{2}$: A
					Superconductor Based on a Kagome Lattice of Ir}},\ }\href
		{https://doi.org/10.1021/acs.chemmater.2c00280} {\bibfield  {journal}
			{\bibinfo  {journal} {Chem. Mater.}\ }\textbf {\bibinfo {volume} {34}},\
			\bibinfo {pages} {2824} (\bibinfo {year} {2022})}\BibitemShut {NoStop}%
		\bibitem [{\citenamefont {Mielke}\ \emph {et~al.}(2021)\citenamefont {Mielke},
			\citenamefont {Qin}, \citenamefont {Yin}, \citenamefont {Nakamura},
			\citenamefont {Das}, \citenamefont {Guo}, \citenamefont {Khasanov},
			\citenamefont {Chang}, \citenamefont {Wang}, \citenamefont {Jia},
			\citenamefont {Nakatsuji}, \citenamefont {Amato}, \citenamefont {Luetkens},
			\citenamefont {Xu}, \citenamefont {Hasan},\ and\ \citenamefont
			{Guguchia}}]{RN107}%
		\BibitemOpen
		\bibfield  {author} {\bibinfo {author} {\bibfnamefont {C.}~\bibnamefont
				{Mielke}}, \bibinfo {author} {\bibfnamefont {Y.}~\bibnamefont {Qin}},
			\bibinfo {author} {\bibfnamefont {J.~X.}\ \bibnamefont {Yin}}, \bibinfo
			{author} {\bibfnamefont {H.}~\bibnamefont {Nakamura}}, \bibinfo {author}
			{\bibfnamefont {D.}~\bibnamefont {Das}}, \bibinfo {author} {\bibfnamefont
				{K.}~\bibnamefont {Guo}}, \bibinfo {author} {\bibfnamefont {R.}~\bibnamefont
				{Khasanov}}, \bibinfo {author} {\bibfnamefont {J.}~\bibnamefont {Chang}},
			\bibinfo {author} {\bibfnamefont {Z.~Q.}\ \bibnamefont {Wang}}, \bibinfo
			{author} {\bibfnamefont {S.}~\bibnamefont {Jia}}, \bibinfo {author}
			{\bibfnamefont {S.}~\bibnamefont {Nakatsuji}}, \bibinfo {author}
			{\bibfnamefont {A.}~\bibnamefont {Amato}}, \bibinfo {author} {\bibfnamefont
				{H.}~\bibnamefont {Luetkens}}, \bibinfo {author} {\bibfnamefont
				{G.}~\bibnamefont {Xu}}, \bibinfo {author} {\bibfnamefont {M.~Z.}\
				\bibnamefont {Hasan}},\ and\ \bibinfo {author} {\bibfnamefont
				{Z.}~\bibnamefont {Guguchia}},\ }\bibfield  {title} {\bibinfo {title}
			{{Nodeless kagome superconductivity in {La}{Ru}$_{3}${Si}$_{2}$}},\ }\href
		{https://doi.org/10.1103/PhysRevMaterials.5.034803} {\bibfield  {journal}
			{\bibinfo  {journal} {Phys. Rev. Mater.}\ }\textbf {\bibinfo {volume} {5}},\
			\bibinfo {pages} {034803} (\bibinfo {year} {2021})}\BibitemShut {NoStop}%
		\bibitem [{\citenamefont {Yang}\ \emph {et~al.}(2025)\citenamefont {Yang},
			\citenamefont {Li}, \citenamefont {Rehman}, \citenamefont {Ning},
			\citenamefont {Hao}, \citenamefont {Zhu},\ and\ \citenamefont
			{Tian}}]{RN2165}%
		\BibitemOpen
		\bibfield  {author} {\bibinfo {author} {\bibfnamefont {Y.}~\bibnamefont
				{Yang}}, \bibinfo {author} {\bibfnamefont {J.}~\bibnamefont {Li}}, \bibinfo
			{author} {\bibfnamefont {M.~U.}\ \bibnamefont {Rehman}}, \bibinfo {author}
			{\bibfnamefont {W.}~\bibnamefont {Ning}}, \bibinfo {author} {\bibfnamefont
				{N.}~\bibnamefont {Hao}}, \bibinfo {author} {\bibfnamefont {X.}~\bibnamefont
				{Zhu}},\ and\ \bibinfo {author} {\bibfnamefont {M.}~\bibnamefont {Tian}},\
		}\bibfield  {title} {\bibinfo {title} {{Anisotropic superconductivity in the
					topological semimetal {Sr}{Sn}$_{3}$ with layered Sn kagome network}},\
		}\href {https://doi.org/10.1103/PhysRevB.111.184517} {\bibfield  {journal}
			{\bibinfo  {journal} {Phys. Rev. B}\ }\textbf {\bibinfo {volume} {111}},\
			\bibinfo {pages} {184517} (\bibinfo {year} {2025})}\BibitemShut {NoStop}%
		\bibitem [{\citenamefont {Iyo}\ \emph {et~al.}(2015)\citenamefont {Iyo},
			\citenamefont {Yanagi}, \citenamefont {Kinjo}, \citenamefont {Nishio},
			\citenamefont {Hase}, \citenamefont {Yanagisawa}, \citenamefont {Ishida},
			\citenamefont {Kito}, \citenamefont {Takeshita}, \citenamefont {Oka},
			\citenamefont {Yoshida},\ and\ \citenamefont {Eisaki}}]{RN3069}%
		\BibitemOpen
		\bibfield  {author} {\bibinfo {author} {\bibfnamefont {A.}~\bibnamefont
				{Iyo}}, \bibinfo {author} {\bibfnamefont {Y.}~\bibnamefont {Yanagi}},
			\bibinfo {author} {\bibfnamefont {T.}~\bibnamefont {Kinjo}}, \bibinfo
			{author} {\bibfnamefont {T.}~\bibnamefont {Nishio}}, \bibinfo {author}
			{\bibfnamefont {I.}~\bibnamefont {Hase}}, \bibinfo {author} {\bibfnamefont
				{T.}~\bibnamefont {Yanagisawa}}, \bibinfo {author} {\bibfnamefont
				{S.}~\bibnamefont {Ishida}}, \bibinfo {author} {\bibfnamefont
				{H.}~\bibnamefont {Kito}}, \bibinfo {author} {\bibfnamefont {N.}~\bibnamefont
				{Takeshita}}, \bibinfo {author} {\bibfnamefont {K.}~\bibnamefont {Oka}},
			\bibinfo {author} {\bibfnamefont {Y.}~\bibnamefont {Yoshida}},\ and\ \bibinfo
			{author} {\bibfnamefont {H.}~\bibnamefont {Eisaki}},\ }\bibfield  {title}
		{\bibinfo {title} {{Large enhancement of superconducting transition
					temperature of {SrBi$_{3}$} induced by Na substitution for Sr}},\ }\href
		{https://doi.org/10.1038/srep10089} {\bibfield  {journal} {\bibinfo
				{journal} {Sci. Rep.}\ }\textbf {\bibinfo {volume} {5}},\ \bibinfo {pages}
			{10089} (\bibinfo {year} {2015})}\BibitemShut {NoStop}%
		\bibitem [{\citenamefont {Gambino}\ \emph {et~al.}(1968)\citenamefont
			{Gambino}, \citenamefont {Stemple},\ and\ \citenamefont {Toxen}}]{RN3070}%
		\BibitemOpen
		\bibfield  {author} {\bibinfo {author} {\bibfnamefont {R.~J.}\ \bibnamefont
				{Gambino}}, \bibinfo {author} {\bibfnamefont {N.~R.}\ \bibnamefont
				{Stemple}},\ and\ \bibinfo {author} {\bibfnamefont {A.~M.}\ \bibnamefont
				{Toxen}},\ }\bibfield  {title} {\bibinfo {title} {{Superconductivity of
					lanthanum intermetallic compounds with the {Cu$_3$Au} structure}},\ }\href
		{https://doi.org/10.1016/0022-3697(68)90074-7} {\bibfield  {journal}
			{\bibinfo  {journal} {J. Phys. Chem. Solids}\ }\textbf {\bibinfo {volume}
				{29}},\ \bibinfo {pages} {295} (\bibinfo {year} {1968})}\BibitemShut
		{NoStop}%
		\bibitem [{\citenamefont {Kinjo}\ \emph {et~al.}(2016)\citenamefont {Kinjo},
			\citenamefont {Kajino}, \citenamefont {Nishio}, \citenamefont {Kawashima},
			\citenamefont {Yanagi}, \citenamefont {Hase}, \citenamefont {Yanagisawa},
			\citenamefont {Ishida}, \citenamefont {Kito}, \citenamefont {Takeshita},
			\citenamefont {Oka}, \citenamefont {Eisaki}, \citenamefont {Yoshida},\ and\
			\citenamefont {Iyo}}]{RN3071}%
		\BibitemOpen
		\bibfield  {author} {\bibinfo {author} {\bibfnamefont {T.}~\bibnamefont
				{Kinjo}}, \bibinfo {author} {\bibfnamefont {S.}~\bibnamefont {Kajino}},
			\bibinfo {author} {\bibfnamefont {T.}~\bibnamefont {Nishio}}, \bibinfo
			{author} {\bibfnamefont {K.}~\bibnamefont {Kawashima}}, \bibinfo {author}
			{\bibfnamefont {Y.}~\bibnamefont {Yanagi}}, \bibinfo {author} {\bibfnamefont
				{I.}~\bibnamefont {Hase}}, \bibinfo {author} {\bibfnamefont {T.}~\bibnamefont
				{Yanagisawa}}, \bibinfo {author} {\bibfnamefont {S.}~\bibnamefont {Ishida}},
			\bibinfo {author} {\bibfnamefont {H.}~\bibnamefont {Kito}}, \bibinfo {author}
			{\bibfnamefont {N.}~\bibnamefont {Takeshita}}, \bibinfo {author}
			{\bibfnamefont {K.}~\bibnamefont {Oka}}, \bibinfo {author} {\bibfnamefont
				{H.}~\bibnamefont {Eisaki}}, \bibinfo {author} {\bibfnamefont
				{Y.}~\bibnamefont {Yoshida}},\ and\ \bibinfo {author} {\bibfnamefont
				{A.}~\bibnamefont {Iyo}},\ }\bibfield  {title} {\bibinfo {title}
			{{Superconductivity in {LaBi$_{3}$} with {AuCu$_{3}$}-type structure}},\
		}\href {https://doi.org/10.1088/0953-2048/29/3/03lt02} {\bibfield  {journal}
			{\bibinfo  {journal} {Supercond. Sci. Technol.}\ }\textbf {\bibinfo {volume}
				{29}},\ \bibinfo {pages} {03LT02} (\bibinfo {year} {2016})}\BibitemShut
		{NoStop}%
		\bibitem [{\citenamefont {Kang}\ \emph {et~al.}(2020)\citenamefont {Kang},
			\citenamefont {Fang}, \citenamefont {Ye}, \citenamefont {Po}, \citenamefont
			{Denlinger}, \citenamefont {Jozwiak}, \citenamefont {Bostwick}, \citenamefont
			{Rotenberg}, \citenamefont {Kaxiras}, \citenamefont {Checkelsky},\ and\
			\citenamefont {Comin}}]{RN246}%
		\BibitemOpen
		\bibfield  {author} {\bibinfo {author} {\bibfnamefont {M.}~\bibnamefont
				{Kang}}, \bibinfo {author} {\bibfnamefont {S.}~\bibnamefont {Fang}}, \bibinfo
			{author} {\bibfnamefont {L.}~\bibnamefont {Ye}}, \bibinfo {author}
			{\bibfnamefont {H.~C.}\ \bibnamefont {Po}}, \bibinfo {author} {\bibfnamefont
				{J.}~\bibnamefont {Denlinger}}, \bibinfo {author} {\bibfnamefont
				{C.}~\bibnamefont {Jozwiak}}, \bibinfo {author} {\bibfnamefont
				{A.}~\bibnamefont {Bostwick}}, \bibinfo {author} {\bibfnamefont
				{E.}~\bibnamefont {Rotenberg}}, \bibinfo {author} {\bibfnamefont
				{E.}~\bibnamefont {Kaxiras}}, \bibinfo {author} {\bibfnamefont {J.~G.}\
				\bibnamefont {Checkelsky}},\ and\ \bibinfo {author} {\bibfnamefont
				{R.}~\bibnamefont {Comin}},\ }\bibfield  {title} {\bibinfo {title}
			{{Topological flat bands in frustrated kagome lattice CoSn}},\ }\href
		{https://doi.org/10.1038/s41467-020-17465-1} {\bibfield  {journal} {\bibinfo
				{journal} {Nat. Commun.}\ }\textbf {\bibinfo {volume} {11}},\ \bibinfo
			{pages} {4004} (\bibinfo {year} {2020})}\BibitemShut {NoStop}%
		\bibitem [{\citenamefont {Ye}\ \emph {et~al.}(2018)\citenamefont {Ye},
			\citenamefont {Kang}, \citenamefont {Liu}, \citenamefont {von Cube},
			\citenamefont {Wicker}, \citenamefont {Suzuki}, \citenamefont {Jozwiak},
			\citenamefont {Bostwick}, \citenamefont {Rotenberg}, \citenamefont {Bell},
			\citenamefont {Fu}, \citenamefont {Comin},\ and\ \citenamefont
			{Checkelsky}}]{RN93}%
		\BibitemOpen
		\bibfield  {author} {\bibinfo {author} {\bibfnamefont {L.}~\bibnamefont
				{Ye}}, \bibinfo {author} {\bibfnamefont {M.}~\bibnamefont {Kang}}, \bibinfo
			{author} {\bibfnamefont {J.}~\bibnamefont {Liu}}, \bibinfo {author}
			{\bibfnamefont {F.}~\bibnamefont {von Cube}}, \bibinfo {author}
			{\bibfnamefont {C.~R.}\ \bibnamefont {Wicker}}, \bibinfo {author}
			{\bibfnamefont {T.}~\bibnamefont {Suzuki}}, \bibinfo {author} {\bibfnamefont
				{C.}~\bibnamefont {Jozwiak}}, \bibinfo {author} {\bibfnamefont
				{A.}~\bibnamefont {Bostwick}}, \bibinfo {author} {\bibfnamefont
				{E.}~\bibnamefont {Rotenberg}}, \bibinfo {author} {\bibfnamefont {D.~C.}\
				\bibnamefont {Bell}}, \bibinfo {author} {\bibfnamefont {L.}~\bibnamefont
				{Fu}}, \bibinfo {author} {\bibfnamefont {R.}~\bibnamefont {Comin}},\ and\
			\bibinfo {author} {\bibfnamefont {J.~G.}\ \bibnamefont {Checkelsky}},\
		}\bibfield  {title} {\bibinfo {title} {{Massive {Dirac} fermions in a
					ferromagnetic kagome metal}},\ }\href {https://doi.org/10.1038/nature25987}
		{\bibfield  {journal} {\bibinfo  {journal} {Nature}\ }\textbf {\bibinfo
				{volume} {555}},\ \bibinfo {pages} {638} (\bibinfo {year}
			{2018})}\BibitemShut {NoStop}%
		\bibitem [{\citenamefont {Yin}\ \emph {et~al.}(2018)\citenamefont {Yin},
			\citenamefont {Zhang}, \citenamefont {Li}, \citenamefont {Jiang},
			\citenamefont {Chang}, \citenamefont {Zhang}, \citenamefont {Lian},
			\citenamefont {Xiang}, \citenamefont {Belopolski}, \citenamefont {Zheng},
			\citenamefont {Cochran}, \citenamefont {Xu}, \citenamefont {Bian},
			\citenamefont {Liu}, \citenamefont {Chang}, \citenamefont {Lin},
			\citenamefont {Lu}, \citenamefont {Wang}, \citenamefont {Jia}, \citenamefont
			{Wang},\ and\ \citenamefont {Hasan}}]{RN285}%
		\BibitemOpen
		\bibfield  {author} {\bibinfo {author} {\bibfnamefont {J.~X.}\ \bibnamefont
				{Yin}}, \bibinfo {author} {\bibfnamefont {S.~S.}\ \bibnamefont {Zhang}},
			\bibinfo {author} {\bibfnamefont {H.}~\bibnamefont {Li}}, \bibinfo {author}
			{\bibfnamefont {K.}~\bibnamefont {Jiang}}, \bibinfo {author} {\bibfnamefont
				{G.}~\bibnamefont {Chang}}, \bibinfo {author} {\bibfnamefont
				{B.}~\bibnamefont {Zhang}}, \bibinfo {author} {\bibfnamefont
				{B.}~\bibnamefont {Lian}}, \bibinfo {author} {\bibfnamefont {C.}~\bibnamefont
				{Xiang}}, \bibinfo {author} {\bibfnamefont {I.}~\bibnamefont {Belopolski}},
			\bibinfo {author} {\bibfnamefont {H.}~\bibnamefont {Zheng}}, \bibinfo
			{author} {\bibfnamefont {T.~A.}\ \bibnamefont {Cochran}}, \bibinfo {author}
			{\bibfnamefont {S.~Y.}\ \bibnamefont {Xu}}, \bibinfo {author} {\bibfnamefont
				{G.}~\bibnamefont {Bian}}, \bibinfo {author} {\bibfnamefont {K.}~\bibnamefont
				{Liu}}, \bibinfo {author} {\bibfnamefont {T.~R.}\ \bibnamefont {Chang}},
			\bibinfo {author} {\bibfnamefont {H.}~\bibnamefont {Lin}}, \bibinfo {author}
			{\bibfnamefont {Z.~Y.}\ \bibnamefont {Lu}}, \bibinfo {author} {\bibfnamefont
				{Z.}~\bibnamefont {Wang}}, \bibinfo {author} {\bibfnamefont {S.}~\bibnamefont
				{Jia}}, \bibinfo {author} {\bibfnamefont {W.}~\bibnamefont {Wang}},\ and\
			\bibinfo {author} {\bibfnamefont {M.~Z.}\ \bibnamefont {Hasan}},\ }\bibfield
		{title} {\bibinfo {title} {{Giant and anisotropic many-body spin-orbit
					tunability in a strongly correlated kagome magnet}},\ }\href
		{https://doi.org/10.1038/s41586-018-0502-7} {\bibfield  {journal} {\bibinfo
				{journal} {Nature}\ }\textbf {\bibinfo {volume} {562}},\ \bibinfo {pages}
			{91} (\bibinfo {year} {2018})}\BibitemShut {NoStop}%
		\bibitem [{\citenamefont {Teng}\ \emph {et~al.}(2022)\citenamefont {Teng},
			\citenamefont {Chen}, \citenamefont {Ye}, \citenamefont {Rosenberg},
			\citenamefont {Liu}, \citenamefont {Yin}, \citenamefont {Jiang},
			\citenamefont {Oh}, \citenamefont {Hasan}, \citenamefont {Neubauer},
			\citenamefont {Gao}, \citenamefont {Xie}, \citenamefont {Hashimoto},
			\citenamefont {Lu}, \citenamefont {Jozwiak}, \citenamefont {Bostwick},
			\citenamefont {Rotenberg}, \citenamefont {Birgeneau}, \citenamefont {Chu},
			\citenamefont {Yi},\ and\ \citenamefont {Dai}}]{RN609}%
		\BibitemOpen
		\bibfield  {author} {\bibinfo {author} {\bibfnamefont {X.}~\bibnamefont
				{Teng}}, \bibinfo {author} {\bibfnamefont {L.}~\bibnamefont {Chen}}, \bibinfo
			{author} {\bibfnamefont {F.}~\bibnamefont {Ye}}, \bibinfo {author}
			{\bibfnamefont {E.}~\bibnamefont {Rosenberg}}, \bibinfo {author}
			{\bibfnamefont {Z.}~\bibnamefont {Liu}}, \bibinfo {author} {\bibfnamefont
				{J.~X.}\ \bibnamefont {Yin}}, \bibinfo {author} {\bibfnamefont {Y.~X.}\
				\bibnamefont {Jiang}}, \bibinfo {author} {\bibfnamefont {J.~S.}\ \bibnamefont
				{Oh}}, \bibinfo {author} {\bibfnamefont {M.~Z.}\ \bibnamefont {Hasan}},
			\bibinfo {author} {\bibfnamefont {K.~J.}\ \bibnamefont {Neubauer}}, \bibinfo
			{author} {\bibfnamefont {B.}~\bibnamefont {Gao}}, \bibinfo {author}
			{\bibfnamefont {Y.}~\bibnamefont {Xie}}, \bibinfo {author} {\bibfnamefont
				{M.}~\bibnamefont {Hashimoto}}, \bibinfo {author} {\bibfnamefont
				{D.}~\bibnamefont {Lu}}, \bibinfo {author} {\bibfnamefont {C.}~\bibnamefont
				{Jozwiak}}, \bibinfo {author} {\bibfnamefont {A.}~\bibnamefont {Bostwick}},
			\bibinfo {author} {\bibfnamefont {E.}~\bibnamefont {Rotenberg}}, \bibinfo
			{author} {\bibfnamefont {R.~J.}\ \bibnamefont {Birgeneau}}, \bibinfo {author}
			{\bibfnamefont {J.~H.}\ \bibnamefont {Chu}}, \bibinfo {author} {\bibfnamefont
				{M.}~\bibnamefont {Yi}},\ and\ \bibinfo {author} {\bibfnamefont
				{P.}~\bibnamefont {Dai}},\ }\bibfield  {title} {\bibinfo {title} {{Discovery
					of charge density wave in a kagome lattice antiferromagnet}},\ }\href
		{https://doi.org/10.1038/s41586-022-05034-z} {\bibfield  {journal} {\bibinfo
				{journal} {Nature}\ }\textbf {\bibinfo {volume} {609}},\ \bibinfo {pages}
			{490} (\bibinfo {year} {2022})}\BibitemShut {NoStop}%
		\bibitem [{\citenamefont {Chen}\ \emph {et~al.}(2014)\citenamefont {Chen},
			\citenamefont {Niu},\ and\ \citenamefont {MacDonald}}]{RN289}%
		\BibitemOpen
		\bibfield  {author} {\bibinfo {author} {\bibfnamefont {H.}~\bibnamefont
				{Chen}}, \bibinfo {author} {\bibfnamefont {Q.}~\bibnamefont {Niu}},\ and\
			\bibinfo {author} {\bibfnamefont {A.~H.}\ \bibnamefont {MacDonald}},\
		}\bibfield  {title} {\bibinfo {title} {{Anomalous {Hall} effect arising from
					noncollinear antiferromagnetism}},\ }\href
		{https://doi.org/10.1103/PhysRevLett.112.017205} {\bibfield  {journal}
			{\bibinfo  {journal} {Phys. Rev. Lett.}\ }\textbf {\bibinfo {volume} {112}},\
			\bibinfo {pages} {017205} (\bibinfo {year} {2014})}\BibitemShut {NoStop}%
		\bibitem [{\citenamefont {Nayak}\ \emph {et~al.}(2016)\citenamefont {Nayak},
			\citenamefont {Fischer}, \citenamefont {Sun}, \citenamefont {Yan},
			\citenamefont {Karel}, \citenamefont {Komarek}, \citenamefont {Shekhar},
			\citenamefont {Kumar}, \citenamefont {Schnelle}, \citenamefont {Kubler},
			\citenamefont {Felser},\ and\ \citenamefont {Parkin}}]{RN264}%
		\BibitemOpen
		\bibfield  {author} {\bibinfo {author} {\bibfnamefont {A.~K.}\ \bibnamefont
				{Nayak}}, \bibinfo {author} {\bibfnamefont {J.~E.}\ \bibnamefont {Fischer}},
			\bibinfo {author} {\bibfnamefont {Y.}~\bibnamefont {Sun}}, \bibinfo {author}
			{\bibfnamefont {B.}~\bibnamefont {Yan}}, \bibinfo {author} {\bibfnamefont
				{J.}~\bibnamefont {Karel}}, \bibinfo {author} {\bibfnamefont {A.~C.}\
				\bibnamefont {Komarek}}, \bibinfo {author} {\bibfnamefont {C.}~\bibnamefont
				{Shekhar}}, \bibinfo {author} {\bibfnamefont {N.}~\bibnamefont {Kumar}},
			\bibinfo {author} {\bibfnamefont {W.}~\bibnamefont {Schnelle}}, \bibinfo
			{author} {\bibfnamefont {J.}~\bibnamefont {Kubler}}, \bibinfo {author}
			{\bibfnamefont {C.}~\bibnamefont {Felser}},\ and\ \bibinfo {author}
			{\bibfnamefont {S.~S.}\ \bibnamefont {Parkin}},\ }\bibfield  {title}
		{\bibinfo {title} {{Large anomalous {Hall} effect driven by a nonvanishing
					Berry curvature in the noncolinear antiferromagnet {Mn$_{3}$Ge}}},\ }\href
		{https://doi.org/10.1126/sciadv.1501870} {\bibfield  {journal} {\bibinfo
				{journal} {Sci. Adv.}\ }\textbf {\bibinfo {volume} {2}},\ \bibinfo {pages}
			{e1501870} (\bibinfo {year} {2016})}\BibitemShut {NoStop}%
		\bibitem [{\citenamefont {Guo}\ and\ \citenamefont {Wang}(2017)}]{RN421}%
		\BibitemOpen
		\bibfield  {author} {\bibinfo {author} {\bibfnamefont {G.-Y.}\ \bibnamefont
				{Guo}}\ and\ \bibinfo {author} {\bibfnamefont {T.-C.}\ \bibnamefont {Wang}},\
		}\bibfield  {title} {\bibinfo {title} {{Large anomalous Nernst and spin
					Nernst effects in the noncollinear antiferromagnets {Mn}$_{3}${X} ({X} = Sn,
					Ge, Ga)}},\ }\href {https://doi.org/10.1103/PhysRevB.96.224415} {\bibfield
			{journal} {\bibinfo  {journal} {Phys. Rev. B}\ }\textbf {\bibinfo {volume}
				{96}},\ \bibinfo {pages} {224415} (\bibinfo {year} {2017})}\BibitemShut
		{NoStop}%
		\bibitem [{\citenamefont {{\v{S}}mejkal}\ \emph {et~al.}(2022)\citenamefont
			{{\v{S}}mejkal}, \citenamefont {MacDonald}, \citenamefont {Sinova},
			\citenamefont {Nakatsuji},\ and\ \citenamefont {Jungwirth}}]{RN2219}%
		\BibitemOpen
		\bibfield  {author} {\bibinfo {author} {\bibfnamefont {L.}~\bibnamefont
				{{\v{S}}mejkal}}, \bibinfo {author} {\bibfnamefont {A.~H.}\ \bibnamefont
				{MacDonald}}, \bibinfo {author} {\bibfnamefont {J.}~\bibnamefont {Sinova}},
			\bibinfo {author} {\bibfnamefont {S.}~\bibnamefont {Nakatsuji}},\ and\
			\bibinfo {author} {\bibfnamefont {T.}~\bibnamefont {Jungwirth}},\ }\bibfield
		{title} {\bibinfo {title} {{Anomalous Hall antiferromagnets}},\ }\href
		{https://doi.org/10.1038/s41578-022-00430-3} {\bibfield  {journal} {\bibinfo
				{journal} {Nat. Rev. Mater.}\ }\textbf {\bibinfo {volume} {7}},\ \bibinfo
			{pages} {482} (\bibinfo {year} {2022})}\BibitemShut {NoStop}%
		\bibitem [{\citenamefont {McMillan}(1968)}]{RN128}%
		\BibitemOpen
		\bibfield  {author} {\bibinfo {author} {\bibfnamefont {W.~L.}\ \bibnamefont
				{McMillan}},\ }\bibfield  {title} {\bibinfo {title} {{Transition Temperature
					of Strong-Coupled Superconductors}},\ }\href
		{https://doi.org/10.1103/PhysRev.167.331} {\bibfield  {journal} {\bibinfo
				{journal} {Phys. Rev.}\ }\textbf {\bibinfo {volume} {167}},\ \bibinfo {pages}
			{331} (\bibinfo {year} {1968})}\BibitemShut {NoStop}%
		\bibitem [{\citenamefont {Allen}\ and\ \citenamefont {Dynes}(1975)}]{RN129}%
		\BibitemOpen
		\bibfield  {author} {\bibinfo {author} {\bibfnamefont {P.~B.}\ \bibnamefont
				{Allen}}\ and\ \bibinfo {author} {\bibfnamefont {R.~C.}\ \bibnamefont
				{Dynes}},\ }\bibfield  {title} {\bibinfo {title} {{Transition temperature of
					strong-coupled superconductors reanalyzed}},\ }\href
		{https://doi.org/10.1103/PhysRevB.12.905} {\bibfield  {journal} {\bibinfo
				{journal} {Phys. Rev. B}\ }\textbf {\bibinfo {volume} {12}},\ \bibinfo
			{pages} {905} (\bibinfo {year} {1975})}\BibitemShut {NoStop}%
		\bibitem [{\citenamefont {Tan}\ \emph {et~al.}(2021)\citenamefont {Tan},
			\citenamefont {Liu}, \citenamefont {Wang},\ and\ \citenamefont {Yan}}]{RN2}%
		\BibitemOpen
		\bibfield  {author} {\bibinfo {author} {\bibfnamefont {H.}~\bibnamefont
				{Tan}}, \bibinfo {author} {\bibfnamefont {Y.}~\bibnamefont {Liu}}, \bibinfo
			{author} {\bibfnamefont {Z.}~\bibnamefont {Wang}},\ and\ \bibinfo {author}
			{\bibfnamefont {B.}~\bibnamefont {Yan}},\ }\bibfield  {title} {\bibinfo
			{title} {{Charge Density Waves and Electronic Properties of Superconducting
					Kagome Metals}},\ }\href {https://doi.org/10.1103/PhysRevLett.127.046401}
		{\bibfield  {journal} {\bibinfo  {journal} {Phys. Rev. Lett.}\ }\textbf
			{\bibinfo {volume} {127}},\ \bibinfo {pages} {046401} (\bibinfo {year}
			{2021})}\BibitemShut {NoStop}%
		\bibitem [{\citenamefont {Fu}\ and\ \citenamefont {Kane}(2007)}]{RN372}%
		\BibitemOpen
		\bibfield  {author} {\bibinfo {author} {\bibfnamefont {L.}~\bibnamefont
				{Fu}}\ and\ \bibinfo {author} {\bibfnamefont {C.~L.}\ \bibnamefont {Kane}},\
		}\bibfield  {title} {\bibinfo {title} {{Topological insulators with inversion
					symmetry}},\ }\href {https://doi.org/10.1103/PhysRevB.76.045302} {\bibfield
			{journal} {\bibinfo  {journal} {Phys. Rev. B}\ }\textbf {\bibinfo {volume}
				{76}},\ \bibinfo {pages} {045302} (\bibinfo {year} {2007})}\BibitemShut
		{NoStop}%
		\bibitem [{\citenamefont {Nayak}\ \emph {et~al.}(2017)\citenamefont {Nayak},
			\citenamefont {Wu}, \citenamefont {Kumar}, \citenamefont {Shekhar},
			\citenamefont {Singh}, \citenamefont {Fink}, \citenamefont {Rienks},
			\citenamefont {Fecher}, \citenamefont {Parkin}, \citenamefont {Yan},\ and\
			\citenamefont {Felser}}]{RN146}%
		\BibitemOpen
		\bibfield  {author} {\bibinfo {author} {\bibfnamefont {J.}~\bibnamefont
				{Nayak}}, \bibinfo {author} {\bibfnamefont {S.~C.}\ \bibnamefont {Wu}},
			\bibinfo {author} {\bibfnamefont {N.}~\bibnamefont {Kumar}}, \bibinfo
			{author} {\bibfnamefont {C.}~\bibnamefont {Shekhar}}, \bibinfo {author}
			{\bibfnamefont {S.}~\bibnamefont {Singh}}, \bibinfo {author} {\bibfnamefont
				{J.}~\bibnamefont {Fink}}, \bibinfo {author} {\bibfnamefont {E.~E.}\
				\bibnamefont {Rienks}}, \bibinfo {author} {\bibfnamefont {G.~H.}\
				\bibnamefont {Fecher}}, \bibinfo {author} {\bibfnamefont {S.~S.}\
				\bibnamefont {Parkin}}, \bibinfo {author} {\bibfnamefont {B.}~\bibnamefont
				{Yan}},\ and\ \bibinfo {author} {\bibfnamefont {C.}~\bibnamefont {Felser}},\
		}\bibfield  {title} {\bibinfo {title} {{Multiple {Dirac} cones at the surface
					of the topological metal LaBi}},\ }\href
		{https://doi.org/10.1038/ncomms13942} {\bibfield  {journal} {\bibinfo
				{journal} {Nat. Commun.}\ }\textbf {\bibinfo {volume} {8}},\ \bibinfo {pages}
			{13942} (\bibinfo {year} {2017})}\BibitemShut {NoStop}%
		\bibitem [{\citenamefont {You}\ \emph {et~al.}(2024)\citenamefont {You},
			\citenamefont {Su},\ and\ \citenamefont {Feng}}]{RN1206}%
		\BibitemOpen
		\bibfield  {author} {\bibinfo {author} {\bibfnamefont {J.-Y.}\ \bibnamefont
				{You}}, \bibinfo {author} {\bibfnamefont {G.}~\bibnamefont {Su}},\ and\
			\bibinfo {author} {\bibfnamefont {Y.~P.}\ \bibnamefont {Feng}},\ }\bibfield
		{title} {\bibinfo {title} {{A versatile model with three-dimensional
					triangular lattice for unconventional transport and various topological
					effects}},\ }\href {https://doi.org/10.1093/nsr/nwad114} {\bibfield
			{journal} {\bibinfo  {journal} {Natl. Sci. Rev.}\ }\textbf {\bibinfo {volume}
				{11}},\ \bibinfo {pages} {nwad114} (\bibinfo {year} {2024})}\BibitemShut
		{NoStop}%
		\bibitem [{\citenamefont {Blochl}(1994)}]{RN141}%
		\BibitemOpen
		\bibfield  {author} {\bibinfo {author} {\bibfnamefont {P.~E.}\ \bibnamefont
				{Blochl}},\ }\bibfield  {title} {\bibinfo {title} {{Projector augmented-wave
					method}},\ }\href {https://doi.org/10.1103/physrevb.50.17953} {\bibfield
			{journal} {\bibinfo  {journal} {Phys. Rev. B}\ }\textbf {\bibinfo {volume}
				{50}},\ \bibinfo {pages} {17953} (\bibinfo {year} {1994})}\BibitemShut
		{NoStop}%
		\bibitem [{\citenamefont {Kresse}\ and\ \citenamefont
			{Furthmuller}(1996)}]{RN142}%
		\BibitemOpen
		\bibfield  {author} {\bibinfo {author} {\bibfnamefont {G.}~\bibnamefont
				{Kresse}}\ and\ \bibinfo {author} {\bibfnamefont {J.}~\bibnamefont
				{Furthmuller}},\ }\bibfield  {title} {\bibinfo {title} {{Efficient iterative
					schemes for ab initio total-energy calculations using a plane-wave basis
					set}},\ }\href {https://doi.org/10.1103/physrevb.54.11169} {\bibfield
			{journal} {\bibinfo  {journal} {Phys. Rev. B}\ }\textbf {\bibinfo {volume}
				{54}},\ \bibinfo {pages} {11169} (\bibinfo {year} {1996})}\BibitemShut
		{NoStop}%
		\bibitem [{\citenamefont {Perdew}\ \emph {et~al.}(1996)\citenamefont {Perdew},
			\citenamefont {Burke},\ and\ \citenamefont {Ernzerhof}}]{RN143}%
		\BibitemOpen
		\bibfield  {author} {\bibinfo {author} {\bibfnamefont {J.~P.}\ \bibnamefont
				{Perdew}}, \bibinfo {author} {\bibfnamefont {K.}~\bibnamefont {Burke}},\ and\
			\bibinfo {author} {\bibfnamefont {M.}~\bibnamefont {Ernzerhof}},\ }\bibfield
		{title} {\bibinfo {title} {{Generalized Gradient Approximation Made
					Simple}},\ }\href {https://doi.org/10.1103/PhysRevLett.77.3865} {\bibfield
			{journal} {\bibinfo  {journal} {Phys. Rev. Lett.}\ }\textbf {\bibinfo
				{volume} {77}},\ \bibinfo {pages} {3865} (\bibinfo {year}
			{1996})}\BibitemShut {NoStop}%
		\bibitem [{\citenamefont {Giannozzi}\ \emph {et~al.}(2009)\citenamefont
			{Giannozzi}, \citenamefont {Baroni}, \citenamefont {Bonini}, \citenamefont
			{Calandra}, \citenamefont {Car}, \citenamefont {Cavazzoni}, \citenamefont
			{Ceresoli}, \citenamefont {Chiarotti}, \citenamefont {Cococcioni},
			\citenamefont {Dabo}, \citenamefont {Dal~Corso}, \citenamefont
			{de~Gironcoli}, \citenamefont {Fabris}, \citenamefont {Fratesi},
			\citenamefont {Gebauer}, \citenamefont {Gerstmann}, \citenamefont
			{Gougoussis}, \citenamefont {Kokalj}, \citenamefont {Lazzeri}, \citenamefont
			{Martin-Samos}, \citenamefont {Marzari}, \citenamefont {Mauri}, \citenamefont
			{Mazzarello}, \citenamefont {Paolini}, \citenamefont {Pasquarello},
			\citenamefont {Paulatto}, \citenamefont {Sbraccia}, \citenamefont {Scandolo},
			\citenamefont {Sclauzero}, \citenamefont {Seitsonen}, \citenamefont
			{Smogunov}, \citenamefont {Umari},\ and\ \citenamefont
			{Wentzcovitch}}]{RN139}%
		\BibitemOpen
		\bibfield  {author} {\bibinfo {author} {\bibfnamefont {P.}~\bibnamefont
				{Giannozzi}}, \bibinfo {author} {\bibfnamefont {S.}~\bibnamefont {Baroni}},
			\bibinfo {author} {\bibfnamefont {N.}~\bibnamefont {Bonini}}, \bibinfo
			{author} {\bibfnamefont {M.}~\bibnamefont {Calandra}}, \bibinfo {author}
			{\bibfnamefont {R.}~\bibnamefont {Car}}, \bibinfo {author} {\bibfnamefont
				{C.}~\bibnamefont {Cavazzoni}}, \bibinfo {author} {\bibfnamefont
				{D.}~\bibnamefont {Ceresoli}}, \bibinfo {author} {\bibfnamefont {G.~L.}\
				\bibnamefont {Chiarotti}}, \bibinfo {author} {\bibfnamefont {M.}~\bibnamefont
				{Cococcioni}}, \bibinfo {author} {\bibfnamefont {I.}~\bibnamefont {Dabo}},
			\bibinfo {author} {\bibfnamefont {A.}~\bibnamefont {Dal~Corso}}, \bibinfo
			{author} {\bibfnamefont {S.}~\bibnamefont {de~Gironcoli}}, \bibinfo {author}
			{\bibfnamefont {S.}~\bibnamefont {Fabris}}, \bibinfo {author} {\bibfnamefont
				{G.}~\bibnamefont {Fratesi}}, \bibinfo {author} {\bibfnamefont
				{R.}~\bibnamefont {Gebauer}}, \bibinfo {author} {\bibfnamefont
				{U.}~\bibnamefont {Gerstmann}}, \bibinfo {author} {\bibfnamefont
				{C.}~\bibnamefont {Gougoussis}}, \bibinfo {author} {\bibfnamefont
				{A.}~\bibnamefont {Kokalj}}, \bibinfo {author} {\bibfnamefont
				{M.}~\bibnamefont {Lazzeri}}, \bibinfo {author} {\bibfnamefont
				{L.}~\bibnamefont {Martin-Samos}}, \bibinfo {author} {\bibfnamefont
				{N.}~\bibnamefont {Marzari}}, \bibinfo {author} {\bibfnamefont
				{F.}~\bibnamefont {Mauri}}, \bibinfo {author} {\bibfnamefont
				{R.}~\bibnamefont {Mazzarello}}, \bibinfo {author} {\bibfnamefont
				{S.}~\bibnamefont {Paolini}}, \bibinfo {author} {\bibfnamefont
				{A.}~\bibnamefont {Pasquarello}}, \bibinfo {author} {\bibfnamefont
				{L.}~\bibnamefont {Paulatto}}, \bibinfo {author} {\bibfnamefont
				{C.}~\bibnamefont {Sbraccia}}, \bibinfo {author} {\bibfnamefont
				{S.}~\bibnamefont {Scandolo}}, \bibinfo {author} {\bibfnamefont
				{G.}~\bibnamefont {Sclauzero}}, \bibinfo {author} {\bibfnamefont {A.~P.}\
				\bibnamefont {Seitsonen}}, \bibinfo {author} {\bibfnamefont {A.}~\bibnamefont
				{Smogunov}}, \bibinfo {author} {\bibfnamefont {P.}~\bibnamefont {Umari}},\
			and\ \bibinfo {author} {\bibfnamefont {R.~M.}\ \bibnamefont {Wentzcovitch}},\
		}\bibfield  {title} {\bibinfo {title} {{{QUANTUM ESPRESSO}: a modular and
					open-source software project for quantum simulations of materials}},\ }\href
		{https://doi.org/10.1088/0953-8984/21/39/395502} {\bibfield  {journal}
			{\bibinfo  {journal} {J. Phys.: Condens. Matter}\ }\textbf {\bibinfo {volume}
				{21}},\ \bibinfo {pages} {395502} (\bibinfo {year} {2009})}\BibitemShut
		{NoStop}%
		\bibitem [{\citenamefont {Jain}\ \emph {et~al.}(2011)\citenamefont {Jain},
			\citenamefont {Hautier}, \citenamefont {Moore}, \citenamefont {Ping~Ong},
			\citenamefont {Fischer}, \citenamefont {Mueller}, \citenamefont {Persson},\
			and\ \citenamefont {Ceder}}]{RN1116}%
		\BibitemOpen
		\bibfield  {author} {\bibinfo {author} {\bibfnamefont {A.}~\bibnamefont
				{Jain}}, \bibinfo {author} {\bibfnamefont {G.}~\bibnamefont {Hautier}},
			\bibinfo {author} {\bibfnamefont {C.~J.}\ \bibnamefont {Moore}}, \bibinfo
			{author} {\bibfnamefont {S.}~\bibnamefont {Ping~Ong}}, \bibinfo {author}
			{\bibfnamefont {C.~C.}\ \bibnamefont {Fischer}}, \bibinfo {author}
			{\bibfnamefont {T.}~\bibnamefont {Mueller}}, \bibinfo {author} {\bibfnamefont
				{K.~A.}\ \bibnamefont {Persson}},\ and\ \bibinfo {author} {\bibfnamefont
				{G.}~\bibnamefont {Ceder}},\ }\bibfield  {title} {\bibinfo {title} {{A
					high-throughput infrastructure for density functional theory calculations}},\
		}\href {https://doi.org/10.1016/j.commatsci.2011.02.023} {\bibfield
			{journal} {\bibinfo  {journal} {Comput. Mater. Sci.}\ }\textbf {\bibinfo
				{volume} {50}},\ \bibinfo {pages} {2295} (\bibinfo {year}
			{2011})}\BibitemShut {NoStop}%
		\bibitem [{\citenamefont {Togo}\ and\ \citenamefont {Tanaka}(2015)}]{RN138}%
		\BibitemOpen
		\bibfield  {author} {\bibinfo {author} {\bibfnamefont {A.}~\bibnamefont
				{Togo}}\ and\ \bibinfo {author} {\bibfnamefont {I.}~\bibnamefont {Tanaka}},\
		}\bibfield  {title} {\bibinfo {title} {{First principles phonon calculations
					in materials science}},\ }\href
		{https://doi.org/10.1016/j.scriptamat.2015.07.021} {\bibfield  {journal}
			{\bibinfo  {journal} {Scr. Mater.}\ }\textbf {\bibinfo {volume} {108}},\
			\bibinfo {pages} {1} (\bibinfo {year} {2015})}\BibitemShut {NoStop}%
		\bibitem [{\citenamefont {Morel}\ and\ \citenamefont
			{Anderson}(1962)}]{MorelAnderson1962}%
		\BibitemOpen
		\bibfield  {author} {\bibinfo {author} {\bibfnamefont {P.}~\bibnamefont
				{Morel}}\ and\ \bibinfo {author} {\bibfnamefont {P.~W.}\ \bibnamefont
				{Anderson}},\ }\bibfield  {title} {\bibinfo {title} {Calculation of the
				superconducting state parameters with retarded electron-phonon interaction},\
		}\href {https://doi.org/10.1103/PhysRev.125.1263} {\bibfield  {journal}
			{\bibinfo  {journal} {Physical Review}\ }\textbf {\bibinfo {volume} {125}},\
			\bibinfo {pages} {1263} (\bibinfo {year} {1962})}\BibitemShut {NoStop}%
		\bibitem [{\citenamefont {Mostofi}\ \emph {et~al.}(2014)\citenamefont
			{Mostofi}, \citenamefont {Yates}, \citenamefont {Pizzi}, \citenamefont {Lee},
			\citenamefont {Souza}, \citenamefont {Vanderbilt},\ and\ \citenamefont
			{Marzari}}]{RN144}%
		\BibitemOpen
		\bibfield  {author} {\bibinfo {author} {\bibfnamefont {A.~A.}\ \bibnamefont
				{Mostofi}}, \bibinfo {author} {\bibfnamefont {J.~R.}\ \bibnamefont {Yates}},
			\bibinfo {author} {\bibfnamefont {G.}~\bibnamefont {Pizzi}}, \bibinfo
			{author} {\bibfnamefont {Y.-S.}\ \bibnamefont {Lee}}, \bibinfo {author}
			{\bibfnamefont {I.}~\bibnamefont {Souza}}, \bibinfo {author} {\bibfnamefont
				{D.}~\bibnamefont {Vanderbilt}},\ and\ \bibinfo {author} {\bibfnamefont
				{N.}~\bibnamefont {Marzari}},\ }\bibfield  {title} {\bibinfo {title} {{{An
						updated version of {wannier90}: A tool for obtaining maximally-localised
						Wannier functions}}},\ }\href {https://doi.org/10.1016/j.cpc.2014.05.003}
		{\bibfield  {journal} {\bibinfo  {journal} {Comput. Phys. Commun.}\ }\textbf
			{\bibinfo {volume} {185}},\ \bibinfo {pages} {2309} (\bibinfo {year}
			{2014})}\BibitemShut {NoStop}%
		\bibitem [{\citenamefont {Sancho}\ \emph {et~al.}(1985)\citenamefont {Sancho},
			\citenamefont {Sancho}, \citenamefont {Sancho},\ and\ \citenamefont
			{Rubio}}]{RN140}%
		\BibitemOpen
		\bibfield  {author} {\bibinfo {author} {\bibfnamefont {M.~P.~L.}\
				\bibnamefont {Sancho}}, \bibinfo {author} {\bibfnamefont {J.~M.~L.}\
				\bibnamefont {Sancho}}, \bibinfo {author} {\bibfnamefont {J.~M.~L.}\
				\bibnamefont {Sancho}},\ and\ \bibinfo {author} {\bibfnamefont
				{J.}~\bibnamefont {Rubio}},\ }\bibfield  {title} {\bibinfo {title} {{Highly
					convergent schemes for the calculation of bulk and surface Green
					functions}},\ }\href {https://doi.org/10.1088/0305-4608/15/4/009} {\bibfield
			{journal} {\bibinfo  {journal} {J. Phys. F: Met. Phys.}\ }\textbf {\bibinfo
				{volume} {15}},\ \bibinfo {pages} {851} (\bibinfo {year} {1985})}\BibitemShut
		{NoStop}%
		\bibitem [{\citenamefont {Uzunok}\ \emph {et~al.}(2018)\citenamefont {Uzunok},
			\citenamefont {T\"ut\"unc\"u}, \citenamefont {Karaca}, \citenamefont
			{Ba\c{s}o\u{g}lu},\ and\ \citenamefont {Srivastava}}]{Uzunok2018LaSn3}%
		\BibitemOpen
		\bibfield  {author} {\bibinfo {author} {\bibfnamefont {H.~Y.}\ \bibnamefont
				{Uzunok}}, \bibinfo {author} {\bibfnamefont {H.~M.}\ \bibnamefont
				{T\"ut\"unc\"u}}, \bibinfo {author} {\bibfnamefont {E.}~\bibnamefont
				{Karaca}}, \bibinfo {author} {\bibfnamefont {A.}~\bibnamefont
				{Ba\c{s}o\u{g}lu}},\ and\ \bibinfo {author} {\bibfnamefont {G.~P.}\
				\bibnamefont {Srivastava}},\ }\bibfield  {title} {\bibinfo {title} {Ab initio
				investigation of electron--phonon interaction in {LaSn$_3$} and {CaSn$_3$}},\
		}\href {https://doi.org/10.1080/09500839.2018.1554273} {\bibfield  {journal}
			{\bibinfo  {journal} {Philosophical Magazine Letters}\ }\textbf {\bibinfo
				{volume} {98}},\ \bibinfo {pages} {375} (\bibinfo {year} {2018})}\BibitemShut
		{NoStop}%
		\bibitem [{\citenamefont {You}\ \emph {et~al.}(2025)\citenamefont {You},
			\citenamefont {Hsu}, \citenamefont {Del~Ben},\ and\ \citenamefont
			{Li}}]{RN3148}%
		\BibitemOpen
		\bibfield  {author} {\bibinfo {author} {\bibfnamefont {J.~Y.}\ \bibnamefont
				{You}}, \bibinfo {author} {\bibfnamefont {C.~E.}\ \bibnamefont {Hsu}},
			\bibinfo {author} {\bibfnamefont {M.}~\bibnamefont {Del~Ben}},\ and\ \bibinfo
			{author} {\bibfnamefont {Z.}~\bibnamefont {Li}},\ }\bibfield  {title}
		{\bibinfo {title} {Diverse manifestations of electron-phonon coupling in a
				kagome superconductor},\ }\href
		{https://doi.org/10.1103/PhysRevLett.134.106401} {\bibfield  {journal}
			{\bibinfo  {journal} {Phys Rev Lett}\ }\textbf {\bibinfo {volume} {134}},\
			\bibinfo {pages} {106401} (\bibinfo {year} {2025})}\BibitemShut {NoStop}%
		\bibitem [{\citenamefont {Wu}\ \emph {et~al.}(2018)\citenamefont {Wu},
			\citenamefont {Zhang}, \citenamefont {Song}, \citenamefont {Troyer},\ and\
			\citenamefont {Soluyanov}}]{RN890}%
		\BibitemOpen
		\bibfield  {author} {\bibinfo {author} {\bibfnamefont {Q.}~\bibnamefont
				{Wu}}, \bibinfo {author} {\bibfnamefont {S.}~\bibnamefont {Zhang}}, \bibinfo
			{author} {\bibfnamefont {H.-F.}\ \bibnamefont {Song}}, \bibinfo {author}
			{\bibfnamefont {M.}~\bibnamefont {Troyer}},\ and\ \bibinfo {author}
			{\bibfnamefont {A.~A.}\ \bibnamefont {Soluyanov}},\ }\bibfield  {title}
		{\bibinfo {title} {{{{WannierTools}: An open-source software package for
						novel topological materials}}},\ }\href
		{https://doi.org/10.1016/j.cpc.2017.09.033} {\bibfield  {journal} {\bibinfo
				{journal} {Comput. Phys. Commun.}\ }\textbf {\bibinfo {volume} {224}},\
			\bibinfo {pages} {405} (\bibinfo {year} {2018})}\BibitemShut {NoStop}%
		\bibitem [{\citenamefont {Gao}\ \emph {et~al.}(2021)\citenamefont {Gao},
			\citenamefont {Wu}, \citenamefont {Persson},\ and\ \citenamefont
			{Wang}}]{RN437}%
		\BibitemOpen
		\bibfield  {author} {\bibinfo {author} {\bibfnamefont {J.}~\bibnamefont
				{Gao}}, \bibinfo {author} {\bibfnamefont {Q.}~\bibnamefont {Wu}}, \bibinfo
			{author} {\bibfnamefont {C.}~\bibnamefont {Persson}},\ and\ \bibinfo {author}
			{\bibfnamefont {Z.}~\bibnamefont {Wang}},\ }\bibfield  {title} {\bibinfo
			{title} {{Irvsp: To obtain irreducible representations of electronic states
					in the VASP}},\ }\href {https://doi.org/10.1016/j.cpc.2020.107760} {\bibfield
			{journal} {\bibinfo  {journal} {Comput. Phys. Commun.}\ }\textbf {\bibinfo
				{volume} {261}},\ \bibinfo {pages} {107760} (\bibinfo {year}
			{2021})}\BibitemShut {NoStop}%
		\bibitem [{\citenamefont {Turbil}\ \emph {et~al.}(1969)\citenamefont {Turbil},
			\citenamefont {Billiet},\ and\ \citenamefont {Michel}}]{Turbil1969Fe3Ge}%
		\BibitemOpen
		\bibfield  {author} {\bibinfo {author} {\bibfnamefont {J.~P.}\ \bibnamefont
				{Turbil}}, \bibinfo {author} {\bibfnamefont {Y.}~\bibnamefont {Billiet}},\
			and\ \bibinfo {author} {\bibfnamefont {A.}~\bibnamefont {Michel}},\
		}\bibfield  {title} {\bibinfo {title} {Contribution {\`a} l'{\'e}tude
				cristallographique de la phase {Fe$_3$Ge} de structure {D0$_{19}$}},\ }\href
		{https://rruff.geo.arizona.edu/AMS/result.php?key=_database_code_amcsd+0012279&viewing=html}
		{\bibfield  {journal} {\bibinfo  {journal} {Comptes Rendus Hebdomadaires des
					Seances de l'Academie des Sciences, Serie C}\ }\textbf {\bibinfo {volume}
				{269}},\ \bibinfo {pages} {309} (\bibinfo {year} {1969})}\BibitemShut
		{NoStop}%
		\bibitem [{\citenamefont {Nial}(1943)}]{Nial1943Fe3Sn}%
		\BibitemOpen
		\bibfield  {author} {\bibinfo {author} {\bibfnamefont {O.}~\bibnamefont
				{Nial}},\ }\bibfield  {title} {\bibinfo {title} {X-ray analysis of {FeSn$_2$}
				and {Fe$_3$Sn}},\ }\href@noop {} {\bibfield  {journal} {\bibinfo  {journal}
				{Arkiv f{\"o}r Kemi, Mineralogi och Geologi B}\ }\textbf {\bibinfo {volume}
				{17}} (\bibinfo {year} {1943})}\BibitemShut {NoStop}%
		\bibitem [{\citenamefont {Nial}(1947)}]{Nial1947FeSnAlloys}%
		\BibitemOpen
		\bibfield  {author} {\bibinfo {author} {\bibfnamefont {O.}~\bibnamefont
				{Nial}},\ }\bibfield  {title} {\bibinfo {title} {X-ray studies on binary
				alloys of tin with transition metals},\ }\href
		{https://www.crystallography.net/cod/1538363.html} {\bibfield  {journal}
			{\bibinfo  {journal} {Svensk Kemisk Tidskrift}\ }\textbf {\bibinfo {volume}
				{59}},\ \bibinfo {pages} {165} (\bibinfo {year} {1947})}\BibitemShut
		{NoStop}%
		\bibitem [{\citenamefont {Singh}\ \emph {et~al.}(1968)\citenamefont {Singh},
			\citenamefont {Pal}, \citenamefont {Chandrasekaran},\ and\ \citenamefont
			{Gupta}}]{Singh1968Mn3Sn}%
		\BibitemOpen
		\bibfield  {author} {\bibinfo {author} {\bibfnamefont {U.~P.}\ \bibnamefont
				{Singh}}, \bibinfo {author} {\bibfnamefont {A.~K.}\ \bibnamefont {Pal}},
			\bibinfo {author} {\bibfnamefont {L.}~\bibnamefont {Chandrasekaran}},\ and\
			\bibinfo {author} {\bibfnamefont {K.~P.}\ \bibnamefont {Gupta}},\ }\bibfield
		{title} {\bibinfo {title} {Study of the manganese-rich end of the {Mn--Sn}
				system},\ }\href
		{https://onemine.org/documents/part-viii-august-1968-papers-study-of-the-manganese-rich-end-of-mn-sn-system}
		{\bibfield  {journal} {\bibinfo  {journal} {Transactions of the Metallurgical
					Society of AIME}\ }\textbf {\bibinfo {volume} {242}},\ \bibinfo {pages}
			{1661} (\bibinfo {year} {1968})}\BibitemShut {NoStop}%
		\bibitem [{\citenamefont {Ye}\ \emph {et~al.}(2024)\citenamefont {Ye},
			\citenamefont {Fang}, \citenamefont {Kang}, \citenamefont {Kaufmann},
			\citenamefont {Lee}, \citenamefont {John}, \citenamefont {Neves},
			\citenamefont {Zhao}, \citenamefont {Denlinger}, \citenamefont {Jozwiak},
			\citenamefont {Bostwick}, \citenamefont {Rotenberg}, \citenamefont {Kaxiras},
			\citenamefont {Bell}, \citenamefont {Janson}, \citenamefont {Comin},\ and\
			\citenamefont {Checkelsky}}]{Ye2024Ni3In}%
		\BibitemOpen
		\bibfield  {author} {\bibinfo {author} {\bibfnamefont {L.}~\bibnamefont
				{Ye}}, \bibinfo {author} {\bibfnamefont {S.}~\bibnamefont {Fang}}, \bibinfo
			{author} {\bibfnamefont {M.}~\bibnamefont {Kang}}, \bibinfo {author}
			{\bibfnamefont {J.}~\bibnamefont {Kaufmann}}, \bibinfo {author}
			{\bibfnamefont {Y.}~\bibnamefont {Lee}}, \bibinfo {author} {\bibfnamefont
				{C.}~\bibnamefont {John}}, \bibinfo {author} {\bibfnamefont {P.~M.}\
				\bibnamefont {Neves}}, \bibinfo {author} {\bibfnamefont {S.~Y.~F.}\
				\bibnamefont {Zhao}}, \bibinfo {author} {\bibfnamefont {J.}~\bibnamefont
				{Denlinger}}, \bibinfo {author} {\bibfnamefont {C.}~\bibnamefont {Jozwiak}},
			\bibinfo {author} {\bibfnamefont {A.}~\bibnamefont {Bostwick}}, \bibinfo
			{author} {\bibfnamefont {E.}~\bibnamefont {Rotenberg}}, \bibinfo {author}
			{\bibfnamefont {E.}~\bibnamefont {Kaxiras}}, \bibinfo {author} {\bibfnamefont
				{D.~C.}\ \bibnamefont {Bell}}, \bibinfo {author} {\bibfnamefont
				{O.}~\bibnamefont {Janson}}, \bibinfo {author} {\bibfnamefont
				{R.}~\bibnamefont {Comin}},\ and\ \bibinfo {author} {\bibfnamefont {J.~G.}\
				\bibnamefont {Checkelsky}},\ }\bibfield  {title} {\bibinfo {title} {Hopping
				frustration-induced flat band and strange metallicity in a kagome metal},\
		}\href {https://doi.org/10.1038/s41567-023-02360-5} {\bibfield  {journal}
			{\bibinfo  {journal} {Nature Physics}\ }\textbf {\bibinfo {volume} {20}},\
			\bibinfo {pages} {610} (\bibinfo {year} {2024})}\BibitemShut {NoStop}%
		\bibitem [{\citenamefont {Lyubimtsev}\ \emph {et~al.}(2002)\citenamefont
			{Lyubimtsev}, \citenamefont {Baranov}, \citenamefont {Fischer}, \citenamefont
			{Kloo},\ and\ \citenamefont {Popovkin}}]{Lyubimtsev2002Ni3Sn}%
		\BibitemOpen
		\bibfield  {author} {\bibinfo {author} {\bibfnamefont {A.~L.}\ \bibnamefont
				{Lyubimtsev}}, \bibinfo {author} {\bibfnamefont {A.~I.}\ \bibnamefont
				{Baranov}}, \bibinfo {author} {\bibfnamefont {A.}~\bibnamefont {Fischer}},
			\bibinfo {author} {\bibfnamefont {L.}~\bibnamefont {Kloo}},\ and\ \bibinfo
			{author} {\bibfnamefont {B.~A.}\ \bibnamefont {Popovkin}},\ }\bibfield
		{title} {\bibinfo {title} {The structure and bonding of {Ni$_3$Sn}},\ }\href
		{https://doi.org/10.1016/S0925-8388(02)00047-6} {\bibfield  {journal}
			{\bibinfo  {journal} {Journal of Alloys and Compounds}\ }\textbf {\bibinfo
				{volume} {340}},\ \bibinfo {pages} {167} (\bibinfo {year}
			{2002})}\BibitemShut {NoStop}%
		\bibitem [{\citenamefont {Palenzona}\ \emph {et~al.}(1996)\citenamefont
			{Palenzona}, \citenamefont {Manfrinetti},\ and\ \citenamefont
			{Palenzona}}]{Palenzona1996ScIn}%
		\BibitemOpen
		\bibfield  {author} {\bibinfo {author} {\bibfnamefont {A.}~\bibnamefont
				{Palenzona}}, \bibinfo {author} {\bibfnamefont {P.}~\bibnamefont
				{Manfrinetti}},\ and\ \bibinfo {author} {\bibfnamefont {R.}~\bibnamefont
				{Palenzona}},\ }\bibfield  {title} {\bibinfo {title} {A reinvestigation of
				the {Sc--In} system and related {(Tm,Lu)--In} compounds},\ }\href
		{https://doi.org/10.1016/S0925-8388(96)02402-4} {\bibfield  {journal}
			{\bibinfo  {journal} {Journal of Alloys and Compounds}\ }\textbf {\bibinfo
				{volume} {243}},\ \bibinfo {pages} {182} (\bibinfo {year}
			{1996})}\BibitemShut {NoStop}%
		\bibitem [{\citenamefont {Raghavan}(2005)}]{Raghavan2005AlTi}%
		\BibitemOpen
		\bibfield  {author} {\bibinfo {author} {\bibfnamefont {V.}~\bibnamefont
				{Raghavan}},\ }\bibfield  {title} {\bibinfo {title} {{Al--Ti}
				(aluminum--titanium)},\ }\href {https://doi.org/10.1361/15477030523049}
		{\bibfield  {journal} {\bibinfo  {journal} {Journal of Phase Equilibria and
					Diffusion}\ }\textbf {\bibinfo {volume} {26}},\ \bibinfo {pages} {171}
			(\bibinfo {year} {2005})}\BibitemShut {NoStop}%
		\bibitem [{\citenamefont {Du}\ \emph {et~al.}(2019)\citenamefont {Du},
			\citenamefont {Cui},\ and\ \citenamefont {Liu}}]{Du2019Ti3Sn}%
		\BibitemOpen
		\bibfield  {author} {\bibinfo {author} {\bibfnamefont {M.}~\bibnamefont
				{Du}}, \bibinfo {author} {\bibfnamefont {L.}~\bibnamefont {Cui}},\ and\
			\bibinfo {author} {\bibfnamefont {F.}~\bibnamefont {Liu}},\ }\bibfield
		{title} {\bibinfo {title} {A new reversible phase transformation of
				intermetallic {Ti$_3$Sn}},\ }\href {https://doi.org/10.3390/ma12152484}
		{\bibfield  {journal} {\bibinfo  {journal} {Materials}\ }\textbf {\bibinfo
				{volume} {12}},\ \bibinfo {pages} {2484} (\bibinfo {year}
			{2019})}\BibitemShut {NoStop}%
		\bibitem [{\citenamefont {Fernandez}\ \emph {et~al.}(1976)\citenamefont
			{Fernandez}, \citenamefont {Tejedor},\ and\ \citenamefont
			{Bru}}]{Fernandez1976Fe3Ge}%
		\BibitemOpen
		\bibfield  {author} {\bibinfo {author} {\bibfnamefont {A.}~\bibnamefont
				{Fernandez}}, \bibinfo {author} {\bibfnamefont {L.}~\bibnamefont {Tejedor}},\
			and\ \bibinfo {author} {\bibfnamefont {L.}~\bibnamefont {Bru}},\ }\bibfield
		{title} {\bibinfo {title} {Electron microscopy study of the phase change
				{D0$_{19}$} to {L1$_2$} in the {Fe$_3$Ge} compound},\ }\href
		{https://doi.org/10.1002/pssa.2210340147} {\bibfield  {journal} {\bibinfo
				{journal} {Physica Status Solidi (a)}\ }\textbf {\bibinfo {volume} {34}},\
			\bibinfo {pages} {K17} (\bibinfo {year} {1976})}\BibitemShut {NoStop}%
		\bibitem [{\citenamefont {Chen}\ \emph {et~al.}(1998)\citenamefont {Chen},
			\citenamefont {Ngan},\ and\ \citenamefont {Duggan}}]{Chen1998Fe3GeL12DO19}%
		\BibitemOpen
		\bibfield  {author} {\bibinfo {author} {\bibfnamefont {Q.~Z.}\ \bibnamefont
				{Chen}}, \bibinfo {author} {\bibfnamefont {A.~H.~W.}\ \bibnamefont {Ngan}},\
			and\ \bibinfo {author} {\bibfnamefont {B.~J.}\ \bibnamefont {Duggan}},\
		}\bibfield  {title} {\bibinfo {title} {An in-situ tem study of the {L1$_2$}
				to {D0$_{19}$} phase transformation in the intermetallic compound
				{Fe$_3$Ge}},\ }\href {https://doi.org/10.1016/S0966-9795(97)00052-6}
		{\bibfield  {journal} {\bibinfo  {journal} {Intermetallics}\ }\textbf
			{\bibinfo {volume} {6}},\ \bibinfo {pages} {105} (\bibinfo {year}
			{1998})}\BibitemShut {NoStop}%
		\bibitem [{\citenamefont {Ochiai}\ \emph {et~al.}(1984)\citenamefont {Ochiai},
			\citenamefont {Oya},\ and\ \citenamefont {Suzuki}}]{Ochiai1984Ni3X}%
		\BibitemOpen
		\bibfield  {author} {\bibinfo {author} {\bibfnamefont {S.}~\bibnamefont
				{Ochiai}}, \bibinfo {author} {\bibfnamefont {Y.}~\bibnamefont {Oya}},\ and\
			\bibinfo {author} {\bibfnamefont {T.}~\bibnamefont {Suzuki}},\ }\bibfield
		{title} {\bibinfo {title} {Alloying behaviour of {Ni$_3$Al}, {Ni$_3$Ga},
				{Ni$_3$Si} and {Ni$_3$Ge}},\ }\href
		{https://doi.org/10.1016/0001-6160(84)90057-9} {\bibfield  {journal}
			{\bibinfo  {journal} {Acta Metallurgica}\ }\textbf {\bibinfo {volume} {32}},\
			\bibinfo {pages} {289} (\bibinfo {year} {1984})}\BibitemShut {NoStop}%
		\bibitem [{\citenamefont {Predel}(2016)}]{Predel2016PdSn}%
		\BibitemOpen
		\bibfield  {author} {\bibinfo {author} {\bibfnamefont {F.}~\bibnamefont
				{Predel}},\ }\bibinfo {title} {Phase diagram of {Pd--Sn} (palladium--tin)
			system},\ in\ \href {https://doi.org/10.1007/978-3-642-24977-8_102} {\emph
			{\bibinfo {booktitle} {Phase Equilibria, Crystallographic and Thermodynamic
					Data of Binary Alloys}}}\ (\bibinfo  {publisher} {Springer Berlin
			Heidelberg},\ \bibinfo {year} {2016})\ pp.\ \bibinfo {pages}
		{170--173}\BibitemShut {NoStop}%
		\bibitem [{\citenamefont {Cui}\ \emph {et~al.}(2021)\citenamefont {Cui},
			\citenamefont {Lynn},\ and\ \citenamefont {Napolitano}}]{Cui2021PtSn}%
		\BibitemOpen
		\bibfield  {author} {\bibinfo {author} {\bibfnamefont {S.}~\bibnamefont
				{Cui}}, \bibinfo {author} {\bibfnamefont {M.}~\bibnamefont {Lynn}},\ and\
			\bibinfo {author} {\bibfnamefont {R.~E.}\ \bibnamefont {Napolitano}},\
		}\bibfield  {title} {\bibinfo {title} {Experimental investigation and
				thermodynamic modeling of the binary {Pt--Sn} system},\ }\href
		{https://doi.org/10.1016/j.jallcom.2020.157064} {\bibfield  {journal}
			{\bibinfo  {journal} {Journal of Alloys and Compounds}\ }\textbf {\bibinfo
				{volume} {854}},\ \bibinfo {pages} {157064} (\bibinfo {year}
			{2021})}\BibitemShut {NoStop}%
		\bibitem [{\citenamefont {Murray}\ \emph {et~al.}(1992)\citenamefont {Murray},
			\citenamefont {Peruzzi},\ and\ \citenamefont {Abriata}}]{Murray1992AlZr}%
		\BibitemOpen
		\bibfield  {author} {\bibinfo {author} {\bibfnamefont {J.}~\bibnamefont
				{Murray}}, \bibinfo {author} {\bibfnamefont {A.}~\bibnamefont {Peruzzi}},\
			and\ \bibinfo {author} {\bibfnamefont {J.~P.}\ \bibnamefont {Abriata}},\
		}\bibfield  {title} {\bibinfo {title} {The {Al--Zr} (aluminum--zirconium)
				system},\ }\href {https://doi.org/10.1007/BF02667556} {\bibfield  {journal}
			{\bibinfo  {journal} {Journal of Phase Equilibria}\ }\textbf {\bibinfo
				{volume} {13}},\ \bibinfo {pages} {277} (\bibinfo {year} {1992})}\BibitemShut
		{NoStop}%
		\bibitem [{\citenamefont {Nial}(1938)}]{Nial1938CoSn}%
		\BibitemOpen
		\bibfield  {author} {\bibinfo {author} {\bibfnamefont {O.}~\bibnamefont
				{Nial}},\ }\bibfield  {title} {\bibinfo {title} {R{\"o}ntgenuntersuchung von
				kobalt--zinnlegierungen und ein vergleich des systems co--sn mit fe--sn und
				ni--sn},\ }\href {https://doi.org/10.1002/zaac.19382380213} {\bibfield
			{journal} {\bibinfo  {journal} {Zeitschrift f{\"u}r anorganische und
					allgemeine Chemie}\ }\textbf {\bibinfo {volume} {238}},\ \bibinfo {pages}
			{287} (\bibinfo {year} {1938})}\BibitemShut {NoStop}%
		\bibitem [{\citenamefont {Watanabe}\ and\ \citenamefont
			{Kunitomi}(1966)}]{Watanabe1966FeGe}%
		\BibitemOpen
		\bibfield  {author} {\bibinfo {author} {\bibfnamefont {H.}~\bibnamefont
				{Watanabe}}\ and\ \bibinfo {author} {\bibfnamefont {N.}~\bibnamefont
				{Kunitomi}},\ }\bibfield  {title} {\bibinfo {title} {On the neutron
				diffraction study of {FeGe}},\ }\href {https://doi.org/10.1143/JPSJ.21.1932}
		{\bibfield  {journal} {\bibinfo  {journal} {Journal of the Physical Society
					of Japan}\ }\textbf {\bibinfo {volume} {21}},\ \bibinfo {pages} {1932}
			(\bibinfo {year} {1966})}\BibitemShut {NoStop}%
		\bibitem [{\citenamefont {Richardson}\ \emph {et~al.}(1967)\citenamefont
			{Richardson}, \citenamefont {Ingri}, \citenamefont {Salomaa}, \citenamefont
			{Bloom},\ and\ \citenamefont {Hagen}}]{Richardson1967FeGe}%
		\BibitemOpen
		\bibfield  {author} {\bibinfo {author} {\bibfnamefont {M.}~\bibnamefont
				{Richardson}}, \bibinfo {author} {\bibfnamefont {N.}~\bibnamefont {Ingri}},
			\bibinfo {author} {\bibfnamefont {P.}~\bibnamefont {Salomaa}}, \bibinfo
			{author} {\bibfnamefont {G.}~\bibnamefont {Bloom}},\ and\ \bibinfo {author}
			{\bibfnamefont {G.}~\bibnamefont {Hagen}},\ }\bibfield  {title} {\bibinfo
			{title} {The partial equilibrium diagram of the {Fe--Ge} system in the range
				40--72 at.\% {Ge}, and the crystallisation of some iron germanides by
				chemical transport reactions},\ }\href
		{https://doi.org/10.3891/acta.chem.scand.21-2305} {\bibfield  {journal}
			{\bibinfo  {journal} {Acta Chemica Scandinavica}\ }\textbf {\bibinfo {volume}
				{21}},\ \bibinfo {pages} {2305} (\bibinfo {year} {1967})}\BibitemShut
		{NoStop}%
		\bibitem [{\citenamefont {Ehret}\ and\ \citenamefont
			{Westgren}(1933)}]{Ehret1933FeSn}%
		\BibitemOpen
		\bibfield  {author} {\bibinfo {author} {\bibfnamefont {W.~F.}\ \bibnamefont
				{Ehret}}\ and\ \bibinfo {author} {\bibfnamefont {A.~F.}\ \bibnamefont
				{Westgren}},\ }\bibfield  {title} {\bibinfo {title} {X-ray analysis of
				iron-tin alloys},\ }\href {https://doi.org/10.1021/ja01331a006} {\bibfield
			{journal} {\bibinfo  {journal} {Journal of the American Chemical Society}\
			}\textbf {\bibinfo {volume} {55}},\ \bibinfo {pages} {1339} (\bibinfo {year}
			{1933})}\BibitemShut {NoStop}%
		\bibitem [{\citenamefont {Yamaguchi}\ and\ \citenamefont
			{Watanabe}(1967)}]{Yamaguchi1967FeSn}%
		\BibitemOpen
		\bibfield  {author} {\bibinfo {author} {\bibfnamefont {K.}~\bibnamefont
				{Yamaguchi}}\ and\ \bibinfo {author} {\bibfnamefont {H.}~\bibnamefont
				{Watanabe}},\ }\bibfield  {title} {\bibinfo {title} {Neutron diffraction
				study of {FeSn}},\ }\href {https://doi.org/10.1143/JPSJ.22.1210} {\bibfield
			{journal} {\bibinfo  {journal} {Journal of the Physical Society of Japan}\
			}\textbf {\bibinfo {volume} {22}},\ \bibinfo {pages} {1210} (\bibinfo {year}
			{1967})}\BibitemShut {NoStop}%
		\bibitem [{\citenamefont {Meier}\ \emph {et~al.}(2020)\citenamefont {Meier},
			\citenamefont {Du}, \citenamefont {Okamoto}, \citenamefont {Mohanta},
			\citenamefont {May}, \citenamefont {McGuire}, \citenamefont {Bridges},
			\citenamefont {Samolyuk},\ and\ \citenamefont {Sales}}]{Meier2020CoSnType}%
		\BibitemOpen
		\bibfield  {author} {\bibinfo {author} {\bibfnamefont {W.~R.}\ \bibnamefont
				{Meier}}, \bibinfo {author} {\bibfnamefont {M.-H.}\ \bibnamefont {Du}},
			\bibinfo {author} {\bibfnamefont {S.}~\bibnamefont {Okamoto}}, \bibinfo
			{author} {\bibfnamefont {N.}~\bibnamefont {Mohanta}}, \bibinfo {author}
			{\bibfnamefont {A.~F.}\ \bibnamefont {May}}, \bibinfo {author} {\bibfnamefont
				{M.~A.}\ \bibnamefont {McGuire}}, \bibinfo {author} {\bibfnamefont {C.~A.}\
				\bibnamefont {Bridges}}, \bibinfo {author} {\bibfnamefont {G.~D.}\
				\bibnamefont {Samolyuk}},\ and\ \bibinfo {author} {\bibfnamefont {B.~C.}\
				\bibnamefont {Sales}},\ }\bibfield  {title} {\bibinfo {title} {Flat bands in
				the {CoSn}-type compounds},\ }\href
		{https://doi.org/10.1103/PhysRevB.102.075148} {\bibfield  {journal} {\bibinfo
				{journal} {Physical Review B}\ }\textbf {\bibinfo {volume} {102}},\ \bibinfo
			{pages} {075148} (\bibinfo {year} {2020})}\BibitemShut {NoStop}%
		\bibitem [{\citenamefont {Malaman}\ \emph {et~al.}(1976)\citenamefont
			{Malaman}, \citenamefont {Roques}, \citenamefont {Courtois},\ and\
			\citenamefont {Protas}}]{Malaman1976Fe3Sn2}%
		\BibitemOpen
		\bibfield  {author} {\bibinfo {author} {\bibfnamefont {B.}~\bibnamefont
				{Malaman}}, \bibinfo {author} {\bibfnamefont {B.}~\bibnamefont {Roques}},
			\bibinfo {author} {\bibfnamefont {A.}~\bibnamefont {Courtois}},\ and\
			\bibinfo {author} {\bibfnamefont {J.}~\bibnamefont {Protas}},\ }\bibfield
		{title} {\bibinfo {title} {Structure cristalline du stannure de fer
				{Fe$_3$Sn$_2$}},\ }\href {https://doi.org/10.1107/S0567740876005323}
		{\bibfield  {journal} {\bibinfo  {journal} {Acta Crystallographica Section
					B}\ }\textbf {\bibinfo {volume} {32}},\ \bibinfo {pages} {1348} (\bibinfo
			{year} {1976})}\BibitemShut {NoStop}%
		\bibitem [{\citenamefont {Wang}\ \emph {et~al.}(2022)\citenamefont {Wang},
			\citenamefont {Gu}, \citenamefont {McGuire}, \citenamefont {Yan},
			\citenamefont {Shi}, \citenamefont {Cui}, \citenamefont {Chen}, \citenamefont
			{Wang}, \citenamefont {Zhang}, \citenamefont {Yang}, \citenamefont {Dong},
			\citenamefont {Jiang}, \citenamefont {Hu}, \citenamefont {Wang},
			\citenamefont {Sun},\ and\ \citenamefont {Cheng}}]{Wang2022V3Sb2}%
		\BibitemOpen
		\bibfield  {author} {\bibinfo {author} {\bibfnamefont {N.}~\bibnamefont
				{Wang}}, \bibinfo {author} {\bibfnamefont {Y.}~\bibnamefont {Gu}}, \bibinfo
			{author} {\bibfnamefont {M.~A.}\ \bibnamefont {McGuire}}, \bibinfo {author}
			{\bibfnamefont {J.}~\bibnamefont {Yan}}, \bibinfo {author} {\bibfnamefont
				{L.}~\bibnamefont {Shi}}, \bibinfo {author} {\bibfnamefont {Q.}~\bibnamefont
				{Cui}}, \bibinfo {author} {\bibfnamefont {K.}~\bibnamefont {Chen}}, \bibinfo
			{author} {\bibfnamefont {Y.}~\bibnamefont {Wang}}, \bibinfo {author}
			{\bibfnamefont {H.}~\bibnamefont {Zhang}}, \bibinfo {author} {\bibfnamefont
				{H.}~\bibnamefont {Yang}}, \bibinfo {author} {\bibfnamefont {X.}~\bibnamefont
				{Dong}}, \bibinfo {author} {\bibfnamefont {K.}~\bibnamefont {Jiang}},
			\bibinfo {author} {\bibfnamefont {J.}~\bibnamefont {Hu}}, \bibinfo {author}
			{\bibfnamefont {B.}~\bibnamefont {Wang}}, \bibinfo {author} {\bibfnamefont
				{J.}~\bibnamefont {Sun}},\ and\ \bibinfo {author} {\bibfnamefont
				{J.}~\bibnamefont {Cheng}},\ }\bibfield  {title} {\bibinfo {title} {A
				density-wave-like transition in the polycrystalline {V$_3$Sb$_2$} sample with
				bilayer kagome lattice},\ }\href {https://doi.org/10.1088/1674-1056/ac4227}
		{\bibfield  {journal} {\bibinfo  {journal} {Chinese Physics B}\ }\textbf
			{\bibinfo {volume} {31}},\ \bibinfo {pages} {017106} (\bibinfo {year}
			{2022})}\BibitemShut {NoStop}%
		\bibitem [{\citenamefont {Sun}\ \emph {et~al.}(2022)\citenamefont {Sun},
			\citenamefont {Zhou}, \citenamefont {Wang}, \citenamefont {Kumar},
			\citenamefont {Geng}, \citenamefont {Yue}, \citenamefont {Han}, \citenamefont
			{Haraguchi}, \citenamefont {Shimada}, \citenamefont {Cheng}, \citenamefont
			{Chen}, \citenamefont {Shi}, \citenamefont {Wu}, \citenamefont {Meng},\ and\
			\citenamefont {Feng}}]{RN327}%
		\BibitemOpen
		\bibfield  {author} {\bibinfo {author} {\bibfnamefont {Z.}~\bibnamefont
				{Sun}}, \bibinfo {author} {\bibfnamefont {H.}~\bibnamefont {Zhou}}, \bibinfo
			{author} {\bibfnamefont {C.}~\bibnamefont {Wang}}, \bibinfo {author}
			{\bibfnamefont {S.}~\bibnamefont {Kumar}}, \bibinfo {author} {\bibfnamefont
				{D.}~\bibnamefont {Geng}}, \bibinfo {author} {\bibfnamefont {S.}~\bibnamefont
				{Yue}}, \bibinfo {author} {\bibfnamefont {X.}~\bibnamefont {Han}}, \bibinfo
			{author} {\bibfnamefont {Y.}~\bibnamefont {Haraguchi}}, \bibinfo {author}
			{\bibfnamefont {K.}~\bibnamefont {Shimada}}, \bibinfo {author} {\bibfnamefont
				{P.}~\bibnamefont {Cheng}}, \bibinfo {author} {\bibfnamefont
				{L.}~\bibnamefont {Chen}}, \bibinfo {author} {\bibfnamefont {Y.}~\bibnamefont
				{Shi}}, \bibinfo {author} {\bibfnamefont {K.}~\bibnamefont {Wu}}, \bibinfo
			{author} {\bibfnamefont {S.}~\bibnamefont {Meng}},\ and\ \bibinfo {author}
			{\bibfnamefont {B.}~\bibnamefont {Feng}},\ }\bibfield  {title} {\bibinfo
			{title} {{Observation of Topological Flat Bands in the Kagome Semiconductor
					{Nb}$_{3}${Cl}$_{8}$}},\ }\href
		{https://doi.org/10.1021/acs.nanolett.2c00778} {\bibfield  {journal}
			{\bibinfo  {journal} {Nano Lett.}\ }\textbf {\bibinfo {volume} {22}},\
			\bibinfo {pages} {4596} (\bibinfo {year} {2022})}\BibitemShut {NoStop}%
		\bibitem [{\citenamefont {Regmi}\ \emph {et~al.}(2022)\citenamefont {Regmi},
			\citenamefont {Fernando}, \citenamefont {Zhao}, \citenamefont {Sakhya},
			\citenamefont {Dhakal}, \citenamefont {Bin~Elius}, \citenamefont {Vazquez},
			\citenamefont {Denlinger}, \citenamefont {Yang}, \citenamefont {Chu},
			\citenamefont {Xu}, \citenamefont {Cao},\ and\ \citenamefont
			{Neupane}}]{RN1260}%
		\BibitemOpen
		\bibfield  {author} {\bibinfo {author} {\bibfnamefont {S.}~\bibnamefont
				{Regmi}}, \bibinfo {author} {\bibfnamefont {T.}~\bibnamefont {Fernando}},
			\bibinfo {author} {\bibfnamefont {Y.}~\bibnamefont {Zhao}}, \bibinfo {author}
			{\bibfnamefont {A.~P.}\ \bibnamefont {Sakhya}}, \bibinfo {author}
			{\bibfnamefont {G.}~\bibnamefont {Dhakal}}, \bibinfo {author} {\bibfnamefont
				{I.}~\bibnamefont {Bin~Elius}}, \bibinfo {author} {\bibfnamefont
				{H.}~\bibnamefont {Vazquez}}, \bibinfo {author} {\bibfnamefont {J.~D.}\
				\bibnamefont {Denlinger}}, \bibinfo {author} {\bibfnamefont {J.}~\bibnamefont
				{Yang}}, \bibinfo {author} {\bibfnamefont {J.-H.}\ \bibnamefont {Chu}},
			\bibinfo {author} {\bibfnamefont {X.}~\bibnamefont {Xu}}, \bibinfo {author}
			{\bibfnamefont {T.}~\bibnamefont {Cao}},\ and\ \bibinfo {author}
			{\bibfnamefont {M.}~\bibnamefont {Neupane}},\ }\bibfield  {title} {\bibinfo
			{title} {{Spectroscopic evidence of flat bands in breathing kagome
					semiconductor {Nb}$_{3}${I}$_{8}$}},\ }\href
		{https://doi.org/10.1038/s43246-022-00318-3} {\bibfield  {journal} {\bibinfo
				{journal} {Commun. Mater.}\ }\textbf {\bibinfo {volume} {3}},\ \bibinfo
			{pages} {100} (\bibinfo {year} {2022})}\BibitemShut {NoStop}%
	\end{thebibliography}

	%
	
\end{document}